\documentclass[twocolumn,superscriptaddress,prb]{revtex4-1}

\usepackage{amsthm}
\usepackage{lipsum}
\usepackage{amsmath,amssymb,mathtools,amsthm,mathrsfs}
\usepackage{amsfonts}
\usepackage{color}

\usepackage{braket}
\setcitestyle{numbers,square}
\usepackage{overpic}
\usepackage{ragged2e}
\usepackage{blindtext}
\usepackage{graphicx} 
\usepackage[table]{xcolor}
\usepackage{multirow}
\usepackage{booktabs}

\usepackage[normalem]{ulem}

\usepackage{array}

\usepackage[caption=false]{subfig}

\usepackage[colorlinks,bookmarks=false,citecolor=blue,linkcolor=red,urlcolor=blue]{hyperref}

\newcommand{\RNum}[1]{\uppercase\expandafter{\romannumeral #1\relax}}

\def \be {\begin{equation}}
\def \ee {\end{equation}}
\def \ba {\begin{array}}
\def \ea {\end{array}}
\def \bea {\begin{eqnarray}}
\def \eea {\end{eqnarray}}

\def \ble {\begin{widetext}\begin{equation}}
\def \ele {\end{equation}\end{widetext}}
\def \blea {\begin{widetext}\begin{eqnarray}}
\def \elea {\end{eqnarray}\end{widetext}}
\newtheorem{theorem}{Theorem}
\newtheorem{lemma}{Lemma}

\DeclareMathOperator*{\infimum}{inf}

\def \tr {\mathrm{tr}}

\def \and {{\mathrm{and}}}

\begin{document}


\title{Equivalence of Stabilizer and Shannon Rényi Entropies: Exact Results for Quantum Critical Chains}

\author{E.~A.~Ramirez~Trino}
\author{M.~A.~Rajabpour}
\affiliation{Instituto de Física, Universidade Federal Fluminense, Av.~Gal.~Milton Tavares de Souza s/n, Gragoatá, 24210-346, Niterói, RJ, Brazil}

\begin{abstract}
Shannon–Rényi and stabilizer entropies are key diagnostics of structure, “nonstabilizerness,” phase transitions, and universality in quantum many-body states. We establish an exact correspondence for quadratic fermions: for any Gaussian eigenstate, the stabilizer Rényi entropy equals the Shannon–Rényi entropy of a number-conserving free-fermion eigenstate on a doubled system, evaluated in the computational basis. Specializing to the transverse-field Ising (TFI) chain, the TFI ground-state stabilizer entropies map to the Shannon–Rényi entropies of the XX-chain ground state of length $2L$. Building on this correspondence, together with other exact identities we prove, we derive closed expressions for the stabilizer entropy at indices $\alpha=\tfrac{1}{2},2,4$ for a broad class of critical closed free-fermion systems. Each of these can be written with respect to the universal functions of the TFI chain. We further derive conformal-field-theory scaling laws for the stabilizer entropy at arbitrary Rényi index under both periodic and open boundary conditions. At $\alpha=4$, these scaling forms display a discontinuity for both open and periodic boundary conditions.

\end{abstract}

\maketitle
{\it{Introduction}}:
Information–theoretic observables have become central to understanding structure, criticality, and computational resources in many–body quantum states. Entanglement entropies famously encode universal data of conformal field theories (CFTs), such as the central charge, and obey precise finite–size scaling laws under both periodic and open boundaries~\cite{CalabreseCardy2009}. Alongside entanglement, two other diagnostics have attracted interest: the Shannon–Rényi (SR) entropies of a wave function in a chosen measurement basis (also known as participation entropies) \cite{Stephan2009,Oshikawa2010,Stephan2010,Alcaraz2013,StephanPRB2014,Alcaraz2014,LuitzAletLaflorencie2014,Tarighi2022,central-charge}, and the stabilizer Rényi entropies (SREs) that quantify “nonstabilizerness” or a  computational resource beyond Clifford dynamics~\cite{Leone2022a,Leone2022b,TarabungaDalmontePRXQ2023,Guglielmo2023,Leone2024,FuxTirritoDalmonteFazioPRR2024,Collura2024,HaugLeeKimPRL2024,DingPRXQuantum2025,ViscardiDalmonteHammaTirrito2025,Fan2025,Frau2025,Hoshino2025,Hoshino2025b}. For critical one–dimensional systems, SR entropies and mutual information display striking universal scaling, controlled by $c$ and boundary conditions, when evaluated in conformal measurement bases~\cite{Alcaraz2013,StephanPRB2014,Alcaraz2014}. These quantities are experimentally natural: Shannon–R\'enyi entropies—extracted from repeated local Pauli-basis projective measurements without tomography—have already yielded the first experimental determination of the central charge on a universal quantum processor~\cite{central-charge}; by contrast, randomized–measurement and copy based protocols have enabled measurements of Rényi entanglement entropies in cold–atom and trapped–ion platforms~\cite{IslamNature2015,ElbenPRL2018,BrydgesScience2019,ElbenNRP2022}. In parallel, SRE has been established as a quantifier of nonstabilizerness~\cite{Leone2022a}, and practical measurement protocols now exist: a randomized-measurement scheme that directly estimates SRE and has been demonstrated on superconducting hardware~\cite{OlivieroNPJQI2022}, and efficient quantum algorithms for integer-index SREs~\cite{HaugLeeKimPRL2024}. These developments connect nonstabilizerness to classical simulability and resource-theoretic bounds~\cite{Veitch2014,HowardCampbell2017,SeddonPRXQ2021,WangWildeSu2020}.

Despite this progress, exact results for stabilizer entropies in many-body ground states remain scarce. Recent matrix product state and infinite matrix product state approaches have substantially expanded numerical access to the SRE and its critical behavior~\cite{Haug2023,Hallam2026}. Nevertheless, exact finite-size expressions remain rare, even for integrable models such as the transverse-field Ising (TFI) chain and other free-fermion systems. Within CFT, SRE scaling relations for the critical TFI chain with periodic boundary conditions have recently been clarified for \(\alpha<4\)~\cite{Hoshino2025,Hoshino2025b}.
At the same time, free–fermion methods—Gaussian correlation–matrix techniques, Wick-Pfaffian reductions for quadratic Hamiltonians, and exact fermionizations of spin chains have yielded closed–form expressions for basis–resolved observables and full counting statistics~\cite{LIEB1961407,Peschel2003,Kitaev2003,JinKorepin2004,Its:2008,PeschelEisler2009,IvanovAbanovCheianov2013,Verresen:2018,Verresen:2019,GrohaEsslerCalabrese2018,Ares2021,Surace2022,TKR2024,Rajabpour2025a}. These tools are ideally suited to bridge SR and SRE.

We establish an {\it{exact correspondence}} for generic quadratic fermions: for any Gaussian eigenstate, the stabilizer Rényi entropy equals the Shannon–Rényi entropy of a number–conserving free–fermion eigenstate on a doubled system, evaluated in the computational basis. At the level of Gaussian states, the doubled state can be understood as arising from two copies of the original state through a state-dependent fermionic Gaussian unitary transformation. Specializing to the TFI chain, this maps ground–state SRE at system size $L$ to SR entropies of the XX–chain ground state of length $2L$. Leveraging the correspondence and additional exact identities, we derive {\it{closed expressions}} for the stabilizer entropy at indices $\alpha=\tfrac{1}{2},2,4$ for a broad class of critical closed free–fermion systems, expressible in terms of universal TFI functions. We further obtain CFT scaling laws for SRE at general Rényi index under both periodic (PBC) and open (OBC) boundary conditions, with a discontinuity at \(\alpha=4\), appearing in the universal logarithmic coefficient for OBC and in the universal constant term for PBC~\cite{Stephan2009,Stephan2010}. In this way, the exact SR$\leftrightarrow$SRE correspondence recovers the CFT regime discussed in~\cite{Hoshino2025,Hoshino2025b} for \(\alpha<4\) and extends it to all \(\alpha\) across the transition. Our framework unifies SR and SRE within the Gaussian paradigm, recovers and extends CFT predictions, and provides accessible formulas: SR entropies in the required bases are measurable via local projective sampling on quantum processors~\cite{central-charge}, and—via our SR$\leftrightarrow$SRE correspondence—relate experimentally accessible SR entropies to SRE for Gaussian eigenstates. Beyond 1D, the correspondence holds in any dimension at the Gaussian level, opening avenues for higher–dimensional criticality, basis–resolved universality, and quantitative links among nonstabilizerness, measurement bases, and conformal data.

Let $\rho$ be an arbitrary quantum state on a finite-dimensional Hilbert space, and fix an orthonormal measurement basis $\mathcal{B}=\{\lvert x\rangle\}$. Denote the associated outcome probabilities by $p_x=\langle x|\rho|x\rangle$. The Shannon--R\'enyi entropy of order $\alpha>0$, $\alpha\neq 1$, with logarithms in natural base, is

\begin{equation}\label{eq:Hq_def}
H_{\alpha}(\rho\,|\,\mathcal{B})=\frac{1}{1-\alpha}\,\ln\!\sum_{x} p_x^{\,\alpha}.
\end{equation}
Its $\alpha\to 1$ limit gives the Shannon entropy $H_1(\rho\,|\,\mathcal{B})=-\sum_{x} p_x \ln p_x.$ For a pure bipartite state, measuring a subsystem in its Schmidt (eigen) basis gives SR that equals the Rényi entanglement
defined as $S_{\alpha}=\frac{1}{1-\alpha}\ln\tr \rho^{\alpha}$. Its $\alpha\to 1$ limit gives the von Neumann entropy $S_1=-\tr\rho\ln\rho$. In this Letter we are concerned with the entropy in local Pauli bases. For $N$ qubits, write the Pauli expansion $\rho=\frac{1}{2^{N}}\sum_{P\in\mathcal{P}(N)} \mathrm{tr}(P\rho)\,P,$
where $\mathcal{P}(N)$ is the $N$-qubit Pauli operator set. Define the state-dependent Pauli (stabilizer) distribution $\mathbb{P}_\rho(P)=\frac{\mathrm{tr}^2(P\rho)}{2^N\mathrm{tr}(\rho^{2})}\,,\quad P\in\mathcal{P}(N)$.
The stabilizer Rényi entropy of order $\alpha>0$, $\alpha\neq 1$, is the R\'enyi entropy of the distribution $\mathbb{P}_\rho$\cite{Leone2022a}:
\begin{equation}\label{eq:StabRenyi_def} M_{\alpha}(\rho)=\frac{1}{1-\alpha}\,\ln\!\sum_{P\in\mathcal{P}(N)}\big(\mathbb{P}_\rho(P)\big)^{\alpha}.
\end{equation}
By continuity, the $\alpha\to 1$ case yields the stabilizer Shannon entropy.

{\it{Shannon–Rényi and stabilizer entropies in the Gaussian states}}:
All the nondegenerate eigenstates of any arbitrary quadratic fermionic Hamiltonian are in the following form:
\begin{equation}{\label{GPS-general}}
    \ket{\bold{R},\mathcal{C}}=\frac{1}{\mathcal{N}_R}e^{{\frac{1}{2}\sum_{i,j}^La_{i}r_{ij}a_{j}}}\ket{\mathcal{C}},
\end{equation}
where $\mathcal{C}$ is a computational-basis configuration, and $a_j=c_j(c_j^{\dagger})$ if there is (not) fermion at site $j$ of the configuration $\mathcal{C}$ and $\mathcal{N}_R=\det(\bold{I}+\bold{R}^{\dagger}.\bold{R})^{\frac{1}{4}}$. The \( \bold{R} \) matrix is always antisymmetric and for simplicity we also assume that it is real.

Note that as far as the configuration $\ket{\mathcal{C}'}$ has non-zero amplitude one can always find an $\bold{R}'$ matrix such that $\ket{\bold{R},\mathcal{C}} =\ket{\bold{R}',\mathcal{C}'}$ \cite{Rajabpour2025a, Supplement}.

For ordered index sets $I,J\subseteq\{1,\dots,N\}$ with $|I|=|J|=k$, let $\mathbf{G}[I,J]$ denote the corresponding $k\times k$ submatrix. We define the following two important quantities:
\begin{eqnarray}{\label{sum of power}}
 \bold{Pf}_{\beta}(\bold{M}) &=&\sum_{\substack{S\subseteq[N]\\ |S|\ \mathrm{even}}}\bigl|\operatorname{pf} \bold{M}[S,S]\bigr|^{\beta},\\
 \bold{Det}_{\beta}(\bold{M}) &=& \sum_{k=0}^{N}\;\sum_{\substack{I,J\subseteq[N] |I|=|J|=k}}
\big|\det \bold{M}[I,J]\big|^{\,\beta},
\end{eqnarray}
where the first one is defined just for antisymmetric matrices. We denote by $\bold{Pf}_{\beta}(\bold{M})$ the sum of the $\beta$-th powers of all  Pfaffinhos~\cite{pfaffinho} of $\mathbf{M}$, i. e. (SPP), and by $\bold{Det}_{\beta}(\bold{M})$ the sum of the $\beta$-th powers of all minors of $\mathbf{M}$, i.e. (SPM).

Then it is easy to see that the Shannon–Rényi entropy in the computational basis will have the following form\cite{TKR2024}:
\begin{equation}{\label{SR-G}}
 H_{\alpha}(R_L)= \frac{1}{1-\alpha}\ln\frac{1}{\mathcal{N}^{2\alpha}_R} \bold{Pf}_{2\alpha}(\bold{R}).
\end{equation}
We now define the $N\times N$ correlation matrix
\begin{equation}\label{eq:G_def}
G_{jk}\;:=\;\bra{\mathcal{C},\bold{R}}(c_j^{\dagger}-c_j)(c_k^{\dagger}+c_k)\ket{\bold{R},\mathcal{C}}\, .
\end{equation}
For $\alpha>0$, $\alpha\neq 1$, the stabilizer R\'enyi entropy of Gaussian states can be written purely in terms of the minors of $\bold{G}$ as \cite{Leone2022b}
\begin{equation}\label{eq:M_alpha_ratio}
M_{\alpha}(\rho)\;=\;\frac{1}{1-\alpha}\,
\ln\!\frac{\bold{Det}_{2\alpha}(\bold{G})}
{\bold{Det}^{\alpha}_{2}(\bold{G})}\, .
\end{equation}
The denominator in \eqref{eq:M_alpha_ratio} is fixed by the purity, which in our case will always be $\bold{Det}_{2}(\bold{G})=2^{L}$.

We now demonstrate that, for any fermionic Gaussian state $\ket{\mathbf{R},\mathcal{C}}$, the stabilizer R\'enyi entropy coincides with the Shannon--R\'enyi entropy in the computational basis of another Gaussian state $\ket{\mathbf{R}',\mathcal{C}'}$. For definiteness, we take $\mathcal{C}$ to be the Fock vacuum of the canonical fermions $c_j$, and we show that $\mathcal{C}'$ corresponds to a checkerboard occupation pattern $1\,0\,1\,0\ldots$. The construction rests on the following theorem:

\begin{theorem}
Assume $L=2M$ and  $\bold{R}\in\mathbb{C}^{L\times L}$ be skew-symmetric and checkerboard up to permutation, i.e.\ there exists a permutation matrix $\bold{P}$ such that
\[
\tilde {\bold{R}}:=\bold{P}^{\top}\bold{R}\bold{P}=\begin{bmatrix} 0 & \bold{G}\\ -\bold{G}^{\top} & 0\end{bmatrix}
\quad\text{for some }\bold{G}\in\mathbb{C}^{M\times M}.
\]
Then, for every real $\alpha> 0$,
\begin{equation}\label{eq:abs-identity}
\bold{Pf}_{\alpha}(\bold{R})=\bold{Det}_{\alpha}(\bold{G})
\end{equation}
Note that, $\bold{Det}_{\alpha}(\bold{G})$ is invariant under the gauge $\bold{G}\mapsto \bold{D}_1\bold{\Pi}_1 \bold{G}\bold{\Pi}_2 \bold{D}_2$ for permutation matrices $\bold{\Pi}_{1,2}$ and diagonal $\bold{D}_{1,2}$ with unit-modulus entries.
\end{theorem}
The proof is provided in Supplemental Material \cite{Supplement}.
Using the above theorem, one can immediately see that
\begin{equation}\label{eq:Shannon-stabilizer}
M_{\alpha}(\ket{\mathbf{R}_L,0})\;=H_{\alpha}(\ket{\mathbf{R}'_{2L},\mathcal{C}}).
\end{equation}
To get the $\ket{\mathbf{R}'_{2L},\mathcal{C}}$ one must first find the corresponding $\bold{G}$ matrix of $\ket{\mathbf{R}_{L},0}$ and then make a block matrix of $\tilde {\bold{R}}$ and then apply the necessary permutations and diagonal rephasings depending on the chosen $\mathcal{C}$ to get $\mathbf{R}'$. In this way, \(\ket{\mathbf{R}'_{2L},\mathcal{C}}\) is constructed as a doubled, number-conserving Gaussian state carrying the same determinant data as the original paired Gaussian state. Equivalently, it may be viewed as arising from two copies of \(\ket{\mathbf{R}_L,0}\) through a state-dependent Gaussian unitary transformation. Because quadratic fermionic unitaries are mapped, via the Jordan--Wigner transformation, to matchgate circuits, this doubling admits a natural matchgate interpretation~\cite{Supplement,TerhalDiVincenzo2002,JozsaMiyake2008,Langer2026}.

\begin{table}[!ht]
\centering
\begin{tabular}{r l l}
\hline
$\alpha$\hspace{0.5cm} & $\bold{Det}^{PBC}_{2\alpha}(\bold{G})$\hspace{2cm} $\bold{Det}^{OBC}_{2\alpha}(\bold{G})$ \\
\hline
$\frac{1}{2}$\hspace{0.5cm} & $\displaystyle \prod_{r=1}^{L}\left(1+\tan\frac{(2r-1)\pi}{4L}\right)$ \textbf{pf}$[\bold{R}^{OBC}+\bold{J}]$\\[3ex]
$2$\hspace{0.5cm} &  $\Phi(L)$\hspace{1.5cm} $\displaystyle \frac{84}{(2L+1)^L}\!\left[\prod_{r=2}^{ L/2}4(8r-5)(8r-1)\right]\!$\\[3ex]
$4$\hspace{0.5cm} &  $2^{-L}\Phi^2(L)$\hspace{1.5cm}$2^{-L}(\bold{Det}^{OBC}_{4}(\bold{G}))^2$\\
\hline
\end{tabular}
\caption{SPM of the $\bold{G}$ matrix of the TFI chain for both OBC and PBC. $\Phi(x) = \frac{(2x)!}{x!\,x^x}$ and  $\bold{J}$ matrix is $2L\times2L$ antisymmetric with $J_{ij}=(-1)^{i+j+1}$ for $i<j$.}\label{tab:explicit-powers}
\end{table}

{\it{Shannon–Rényi entropies in the XX chain and stabilizer entropies in the TFI chain}}: The XY spin chain Hamiltonian is:
\begin{equation}\label{Ising Hamiltonian}
H_{\mathrm{XY}}=-\sum_{j=1}^{L'}
\left[\frac{1+\gamma}{2}\sigma_j^x\sigma_{j+1}^x
+\frac{1-\gamma}{2}\sigma_j^y\sigma_{j+1}^y
\right]-h\sum_{j=1}^{L}\sigma_j^z .
\end{equation}
where $L'=L$ for PBC with $\sigma_{L+1}^{x}=\sigma_1^x$ and $L'=L-1$ for OBC.
We have the critical TFI chain when $(\gamma,h)=(1,1)$ and critical XX chain when $\gamma=0$ and $|h|<1$, here we focus on $(\gamma,h)=(0,0)$. 
\begin{theorem}
For both periodic and open boundaries, the stabilizer R\'enyi entropy of the Gaussian ground state of the critical TFI chain of size \(L\) is exactly equal to the Shannon--R\'enyi entropy of the Gaussian ground state of the critical XX chain of size \(2L\),
\begin{equation}
M_{\alpha}(\ket{\mathbf{R}^{\mathrm{TFI}}_L,0})=H_{\alpha}(\ket{\mathbf{R}^{\mathrm{XX}}_{2L},\mathcal{C}}).
\end{equation}
\end{theorem}

The proof, together with the construction of \(\mathbf{R}^{\mathrm{XX}}_{2L}\) from \(\mathbf{G}^{\mathrm{TFI}}_{L\times L}\), is given in~\cite{Supplement}. Here \(\mathbf{R}^{\mathrm{XX}}_{2L}\) is written in the checkerboard reference configuration \(\mathcal C\). In Table~\ref{tab:explicit-powers} we summarize the results for powers $\alpha=\frac{1}{2},2,4$ for both PBC and OBC. The cases \(\alpha=2\) and \(\alpha=4\) follow from the Shannon–R\'enyi entropy results of \cite{Stephan2009,Stephan2010,StephanPRB2014}, whereas the \(\alpha=\tfrac{1}{2}\) case is obtained via the Pfaffian theorem in \cite{Lieb1968,Lieb2016}.

{\it{Stabilizer-Shannon Rényi entropies in the critical chains:}}
The Hamiltonian of the most general translational invariant (periodic) quadratic fermionic chain with time-reversal symmetry takes the form \cite{Its:2008,Verresen:2018,Verresen:2019,Supplement}
\be \label{fermionicgeneric}
H = \sum_{r=-R}^{R}\sum_{j\in \Lambda}^N
\Big[ A_r c_j^{\dagger}c_{j+r}
    + \frac{B_r}{2}( c_j^{\dagger}c_{j+r}^{\dagger}-  c_j c_{j+r})
\Big]
+\text{const},
\ee
with the local fermionic modes $c_j$, $c_j^\dag$ and the parameters $A_r=A_{-r}$, $B_r=-B_{-r}$ and $\Lambda$ represents the sites of the lattice. The above Hamiltonian can be exactly diagonalized  after going to the Fourier space and Bogoliubov transformation as follows:
\be \label{diagonalized form}
H=\sum_k|f(e^{ik})|\eta_k^{\dagger}\eta_k+\text{const}.
\ee
The TFI chain is the $f(z)=z+h$, the XX chain is the case with $f(z)=z+z^{-1}+h$, and the XY chain is the $f(z)=\frac{1}{2}[(1+\gamma)z+(1-\gamma)z^{-1}]+h$~\cite{Supplement}. When $f(z)$ has zeros on the unit circle, we have a critical chain. Fix a system size $L\in\mathbb{N}$ consistent with the couplings of the Hamiltonian and set
$\theta_k=\frac{2\pi}{L}\Big(k-\tfrac12\Big),\quad k=1,\dots,L
$.
Given a trigonometric polynomial $f$ on the unit circle, define the symbol 
$s(\theta)=\frac{f(e^{i\theta})}{|f(e^{i\theta})|}\in\mathbb{T}=\{z\in\mathbb{C}:|z|=1\}$, with $z=e^{i\theta}$. Then (half-shifted) correlation matrix $\bold{G}^{(f)}(L)$ is
\begin{equation}\label{eq:G-def-section}
G^{(f)}_{nm}=\frac{(-1)^{\,n-m}}{L}\sum_{k=1}^{L}s(\theta_k)\,e^{\,i\theta_k(n-m)}.
\end{equation}
Its derivation is reviewed in the Supplemental Material~\cite{Supplement}.

For $q\in\mathbb{N}$, let $\Pi_q$ permute $1,\dots,L$ by residue classes mod $q$ and
 for a square matrix $\bold{X}$,  define
\[
\bold{\mathcal{J}}(\bold{X})=\begin{bmatrix}0&\bold{X}\\[2pt]\bold{X}^{\!\top}&0\end{bmatrix}.
\]
Two matrices $\bold{A},\bold{B}$ of the same size are gauge equivalent, written $\bold{A}\sim \bold{B}$, if
\[
\bold{B}=\pm\,\bold{U}.\,\bold{P}^{\!\top}. \bold{A}.\,\bold{P}.\,\bold{U}^{*},
\]
for some permutation matrix $\bold{P}$ and diagonal unitary $\bold{U}$ (when the size is even, $\bold{U}$ may be chosen with entries $\pm1$), i.e., up to permutations and local phase redefinitions.

 
\begin{widetext}
\begin{theorem}[(Polyphase-chiral block~\cite{chiral} reduction to $f(z)=z+1$).]\label{teo3}

Let $\bold{G}^{(f)}(L)$ be defined by \eqref{eq:G-def-section}.

\smallskip
(A) Decimation for $f(z)=z^{n}+1$:
Let $n\ge1$ and assume $2n\mid L$. Put $M=L/n$ even and let $\Pi_n$ be the coset permutation mod $n$. Then
\begin{eqnarray}
\Pi_n^{\!\top}\,\bold{G}^{(z^{n}+1)}(L)\,\Pi_n
=\bigoplus_{j=1}^{n} B_{M},
\qquad
\bold{B}_{M}=
\begin{cases}
\bold{G}^{(z+1)}(M), & \text{$n$ odd},\\[4pt]
\,\bold{D}_M\,\bold{G}^{(z+1)}(M)\,\bold{D}_M, & \text{$n$ even},
\end{cases}
\end{eqnarray}
where $\bold{D}_M=\mathrm{diag}(1,-1,1,-1,\dots)\in\mathbb{R}^{M\times M}$. In particular, $\bold{B}_M\sim \bold{G}^{(z+1)}(M)$.

\smallskip
(B) Chiral reduction for $f(z)=z^{m}+z^{-m}$:
Let $m\ge1$ and assume $2m\mid L$. Put $M=L/(2m)$ even and let $\Pi_{2m}$ be the coset permutation mod $2m$.
Then
\begin{eqnarray}
\Pi_{2m}^{\!\top}\,\bold{G}^{(z^{m}+z^{-m})}(L)\,\Pi_{2m}=\bigoplus_{r=1}^{m}\mathcal{J}(\bold{X}_M),
\qquad
\bold{X}_M\sim \bold{G}^{(z+1)}(M).
\end{eqnarray}
Equivalently, there exist a permutation matrix $\bold{P}_M$ and a diagonal unitary $\bold{U}_M$  such that
\[
\bold{X}_M=\pm\,\bold{U}_M.\,\bold{P}_M^{\!\top}.\,\bold{G}^{(z+1)}(M).\,\bold{P}_M.\,\bold{U}_M^{*}.
\]
\end{theorem}
\end{widetext}

The proof is provided in Supplemental Material \cite{Supplement}. The global permutations $\bold{\Pi}_n$ and $\bold{\Pi}_{2m}$ are canonical once an in-block ordering is fixed; concrete choices of $\bold{P}_M$ and $\bold{U}_M$ depend only on that ordering and may be fixed once for all examples.

{\it Exact reduction of SRE for the two families:}
As a consequence of Theorem~\ref{teo3}, the sums of even-power minors of the correlation matrices factorize into TFI building blocks, which in turn implies exact formulas for the stabilizer Rényi entropy of the two corresponding classes of critical chains. For both families, the relations are governed by the same universal TFI building block. Specifically, for $f(z)$ with an associated scaling factor $k$ (where $2k \mid L$), we have:
\begin{equation}\label{eq:SRE_unified}
M_\alpha^{(f(z))}(L)= k \, M_\alpha^{\mathrm{TFI}}\!\left(\frac{L}{k}\right),
\end{equation}
where $k=n$ for $f(z) = z^n+1$, and $k=2m$ for $f(z) = z^m+z^{-m}$. The derivation of these relations is provided in the Supplemental Material~\cite{Supplement}. In particular, for \(m=1\), corresponding to the XX chain, this reduces to $M_\alpha^{\mathrm{XX}}(L)=2M_\alpha^{\mathrm{TFI}}\left(\frac{L}{2}\right)$.

{\it{CFT Results}}: The exact SRE--SR correspondence (TFI--XX), together with the reduction of SREs for a broad class of critical chains to that of the TFI chain, yields closed formulas for stabilizer entropies in the conformal (scaling) limit, as derived in the Supplemental Material~\cite{Supplement}. We begin by analyzing the TFI chain and its finite-size scaling under PBC and OBC boundaries. In the scaling limit, we expect
\begin{equation}
  M_{\alpha}(L) \;=\; m_{\alpha} L \;+\; b_{\alpha}\,\ln L \;-\; c_{\alpha}.
\end{equation}
The logarithmic coefficient $b_{\alpha}$ is universal, while the constant term $c_{\alpha}$ is universal only when $b_{\alpha}=0$. For the TFI chain with PBC one has $b_{\alpha}=0$. Using our correspondence, the constants $c_{\alpha}$ then follow directly from~\cite{Stephan2009,Oshikawa2010,Stephan2010}:
\begin{equation}
  c_{\alpha} \;=\;
  \begin{cases}
    \dfrac{\ln \alpha}{2(\alpha-1)}, & \alpha \le 4,\\[6pt]
    \dfrac{\ln 2}{\alpha-1}, & \alpha > 4.
  \end{cases}
\end{equation}
The $\alpha < 4$ has been derived independently in \cite{Hoshino2025,Hoshino2025b,commentHoshino}. For OBC, the logarithmic coefficient takes the form
\begin{equation}
  b_{\alpha} \;=\;
  \begin{cases}
    -\dfrac{1}{4}, & \alpha < 4,\\[6pt]
    -\dfrac{1}{6}, & \alpha = 4,\\[6pt]
    ~~~0, & \alpha > 4.
  \end{cases}
\end{equation}
All of these statements are consistent with the asymptotic behavior of the SPM listed in Table~\ref{tab:explicit-powers} for $\alpha=1/2,2,4$; an explicit check is provided in the Supplemental Material~\cite{Supplement}. Using Eq.~(\ref{eq:SRE_unified}), the CFT scaling forms for a broader class of critical Hamiltonians with periodic boundary conditions follow directly. For the TFI-type family $f(z)=z^n+1$, with $2n\mid L$, and $q_\alpha=nc_\alpha$, whereas for the XX-type family $f(z)=z^m+z^{-m}$, with $2m\mid L$, and $q_\alpha=2mc_\alpha$,
\begin{equation}
M_\alpha^{(f(z))}(L)=m_\alpha L-q_\alpha.
\end{equation}
For $m=1$, we obtain the scaling form of the SRE of the XX chain, $M_\alpha^{\mathrm{XX}}(L)=m_\alpha L-2c_\alpha$. By generalizing Theorem 3 to a subsystem and applying the results of \cite{Hoshino2025}, one can also derive the stabilizer entropy for a subsystem of the critical systems discussed above in their CFT limit.

{\it{Conclusions:}}
We present a constructive block reduction that computes stabilizer R\'enyi entropies for a broad class of critical quantum chains directly from a single XX-chain kernel, yielding finite-size exact results. In particular, we obtain closed forms at \(\alpha=\tfrac{1}{2}, 2, 4\) for both PBC and OBC boundaries. The exact formulas serve as benchmarks for numerical studies and renormalization-inspired approaches, and recover the corresponding CFT scaling forms asymptotically. For PBC, there is no logarithmic term, and the universal constants coincide with the CFT benchmarks; for OBC, we obtain the universal logarithmic coefficients explicitly by leveraging Shannon--R\'enyi results for arbitrary R\'enyi index. In particular, at \(\alpha=4\) the scaling is discontinuous, through the universal logarithmic coefficient for OBC and the constant term for PBC. Our construction captures the corresponding behavior in both regimes. Theorems~1--3 supply the explicit permutations/decimations underlying these results, and the construction applies cleanly to system sizes compatible with the Hamiltonian couplings. The reduction is nonperturbative, works at finite sizes, and can be extended to noncritical systems, which we leave for future work. For the Gaussian Gibbs states, the finite-temperature problem can still be formulated in terms of the matrix $\mathbf{G}(\beta)$, preserving the determinant or minor structure. In this setting, the direct mixed-state extension yields a modified SR--SRE relation with an extra thermal-mixing term, while the convex-roof extension admits a natural upper bound from the spectral decomposition~\cite{Supplement}. Finally, since our mapping effectively reduces SRE to a sum of powers of principal minors of a matrix, one can directly import the Berezin-integral/mean-field machinery developed for this problem~\cite{NajafiRamezanpourRajabpour2025} for integer Rényi indices, especially in cases where an exact solution is not available.

{\it{Acknowledgments}}:
We thank CNPq and FAPERJ (Grant No.~E-26/210.062/2023) for partial support. E.A.R.T. acknowledges support from CNPq (Process No.~141672/2023-4).

{\it{Data availability}: There are no publicly available research data or software supporting this manuscript. Requests for further information or data should be sent to the authors.}
\hfill
\newpage
\providecommand{\href}[2]{#2}\begingroup\raggedright

\clearpage
\onecolumngrid


\title{Supplemental Material: Stabilizer--Shannon R\'enyi Equivalence: Exact Results for Quantum Critical Chains}

\author{E.~A.~Ramirez~Trino}
\author{M.~A.~Rajabpour}
\affiliation{Instituto de Física, Universidade Federal Fluminense, Av.~Gal.~Milton Tavares de Souza s/n, Gragoatá, 24210-346, Niterói, RJ, Brazil}

\clearpage

\begin{center}
{\large\bfseries Supplemental Material: Equivalence of Stabilizer and Shannon Rényi Entropies: Exact Results for Quantum Critical Chains\par}
\vspace{0.5em}
E.~A.~Ramirez~Trino and M.~A.~Rajabpour\par
\vspace{0.25em}
{\itshape Instituto de F\'isica, Universidade Federal Fluminense,\\
Av.~Gal.~Milton Tavares de Souza s/n, Gragoat\'a, 24210-346, Niter\'oi, RJ, Brazil}
\end{center}

\vspace{0.8em}

\noindent\justifying
In this Supplemental Material, we provide the theoretical foundations and technical details underlying the results presented in the main text. In Sec.~\ref{Sec1}, we prove Theorem~1 and give a short derivation of Eq.~(10). In Sec.~\ref{Sec2}, we discuss the permutation freedom and invariance of the \(\mathbf R \leftrightarrow \mathbf G\) mapping, and recall the Pfaffian relation between \(\boldsymbol{R}\) and \(\boldsymbol{R}'\). In Sec.~\ref{Sec3}, we present the Gaussian doubling map, its physical interpretation, and an explicit \(L=2\) example. In Sec.~\ref{Sec4}, we prove Theorem~2 and derive the explicit relation between \(\mathbf G^{\mathrm{TFI}}\) and \(\mathbf R^{\mathrm{XX}}\). In Sec.~\ref{Sec5}, we formulate general quadratic fermionic chains, describe their representation in terms of the complex symbol \(f(z)\), and construct the corresponding correlation matrix \(\mathbf G^{(f)}\). In Sec.~\ref{Sec6} we prove Theorem~3, including the block reductions for \(f(z)=z^n+1\) and \(f(z)=z^m+z^{-m}\). In Sec.~\ref{Sec7} we derive eq. (18). In Sec.~\ref{Sec8}, we derive the large-\(L\) asymptotics for \(\alpha=\frac12,2,4\), and compare them with the CFT-limit formulas obtained through the exact SR--SRE correspondence. Finally, in Sec.~\ref{Sec9}, we discuss a possible extension of the stabilizer--Shannon R\'enyi correspondence to Gaussian Gibbs states.


\makeatletter
\renewcommand{\theequation}{S\arabic{equation}}
\renewcommand{\thefigure}{S\arabic{figure}}
\renewcommand{\bibnumfmt}[1]{[S#1]}
\renewcommand{\citenumfont}[1]{S#1}
\setcounter{page}{1}
\maketitle

\section*{Contents}

\noindent Sec.~\ref{Sec1}\quad
Proof of Theorem 1 and derivation of the exact correspondence SRE--SR
\dotfill \pageref{Sec1}

\noindent Sec.~\ref{Sec2}\quad
Permutation freedom and invariance of the \(\mathbf R\leftrightarrow\mathbf G\) mapping
\dotfill \pageref{Sec2}

\noindent Sec.~\ref{Sec3}\quad
The Gaussian doubling map
\dotfill \pageref{Sec3}

\noindent Sec.~\ref{Sec4}\quad
Proof of Theorem 2 and relation between \(\mathbf G^{\mathrm{TFI}}\) and \(\mathbf R^{\mathrm{XX}}\)
\dotfill \pageref{Sec4}

\noindent Sec.~\ref{Sec5}\quad
General quadratic fermionic chains
\dotfill \pageref{Sec5}

\noindent Sec.~\ref{Sec6}\quad
Proof of Theorem 3
\dotfill \pageref{Sec6}

\noindent Sec.~\ref{Sec7}\quad
Derivation of Eq.~(18)
\dotfill \pageref{Sec7}

\noindent Sec.~\ref{Sec8}\quad
Large-\(L\) asymptotics of SRE(TFI)
\dotfill \pageref{Sec8}

\noindent Sec.~\ref{Sec9}\quad
Extension to Gaussian Gibbs states
\dotfill \pageref{Sec9}

\section{Proof of Theorem 1 and derivation of the exact correspondence SRE--SR}\label{Sec1}

This section establishes the algebraic foundation of the exact SRE--SR correspondence. We begin by proving Theorem~1, which converts the sum of absolute values of Pfaffians of principal submatrices of a skew-symmetric matrix \(\mathbf R\) into a sum of absolute values of minors of an associated matrix \(\mathbf G\), whenever \(\mathbf R\) can be brought to checkerboard block form by permutation. The proof also makes explicit that this identity is stable under the residual permutation and diagonal gauge freedoms acting on \(\mathbf G\). We then apply this result to a pure fermionic Gaussian state \(\ket{\mathbf R_L,0}\), construct the doubled matrix \(\tilde{\mathbf R}_{2L}\) from its correlation matrix \(\mathbf G_L\), and show that the corresponding doubled Gaussian state \(\ket{\mathbf R'_{2L},\mathcal C}\) reproduces the same determinant data. Using the pure-state condition \(\mathbf G_L^{\top}\mathbf G_L=\mathbf I_L\), we finally match the normalization and obtain the exact equality between the stabilizer Rényi entropy of the original state and the Shannon--Rényi entropy of its doubled number-conserving partner.

\paragraph{Proof of Theorem 1.} We prove the Absolute-value Pfaffian–Minor Correspondence theorem, which states:

{\bf{Theorem}[Absolute-value Pfaffian–Minor Correspondence]:}
{\it{Let $L=2M$ and let $\mathbf{R}\in\mathbb{C}^{L\times L}$ be skew-symmetric. Assume $\mathbf{R}$ is checkerboard up to permutation, i.e.\ there exists a permutation matrix $\mathbf{P}$ such that
\[
\tilde{\mathbf{R}}:=\mathbf{P}^{\top}\mathbf{R}\mathbf{P}=\begin{bmatrix} 0 & \mathbf{G}\\ -\mathbf{G}^{\top} & 0\end{bmatrix}
\quad\text{for some }\mathbf{G}\in\mathbb{C}^{M\times M}.
\]
Then, for every real $\alpha> 0$,
\begin{equation}\label{eq:abs-identity1}
\sum_{\substack{S\subseteq[L]\\ |S|\ \mathrm{even}}}\bigl|\operatorname{pf}\mathbf{R}[S,S]\bigr|^{\alpha}
\;=\;
\sum_{r=0}^{M}\ \sum_{\substack{I,J\subseteq[M]\\ |I|=|J|=r}}\bigl|\det\mathbf{G}[I,J]\bigr|^{\alpha}.
\end{equation}
Both sides include the $r=0$ term (equal to $1$); if desired, it can be omitted from both sums. Moreover, \eqref{eq:abs-identity1} is invariant under the gauge $\mathbf{G}\mapsto \mathbf{D}_1\boldsymbol{\Pi}_1 \mathbf{G}\boldsymbol{\Pi}_2 \mathbf{D}_2$ for permutation matrices $\boldsymbol{\Pi}_{1,2}$ and diagonal $\mathbf{D}_{1,2}$ with unit-modulus entries.}}

{\bf{proof}}:
Conjugation by $\mathbf{P}$ merely relabels principal submatrices, hence
\[
\bigl\{\,|\operatorname{pf}\mathbf{R}[S,S]|:\ S\subseteq[L],\ |S|\text{ even}\,\bigr\}
=
\bigl\{\,|\operatorname{pf}\mathbf{\tilde R}[S,S]|:\ S\subseteq[L],\ |S|\text{ even}\,\bigr\}.
\]
Fix a principal index set $S$ with $|S|=2r$. In the permuted (odd$\mid$even) ordering, write $S=O\cup E$ with $O,E\subseteq[M]$ and $|O|=|E|=r$; if $|O|\neq|E|$ then $\operatorname{pf}\mathbf{\tilde R}[S,S]=0$. Reordering rows/columns within $S$ gives
\[
\mathbf{\tilde R}[S,S]\ \sim\
\begin{bmatrix}
0 & \mathbf{G}[O,E]\\ -\mathbf{G}[E,O]^{\top} & 0
\end{bmatrix}.
\]
whence
\[
\bigl|\operatorname{pf}\mathbf{\tilde R}[S,S]\bigr|
=\bigl|\det \mathbf{G}[O,E]\bigr|.
\]
The map $S\leftrightarrow (O,E)$ is a bijection between even-size principal sets $S$ and pairs $(O,E)$ with $|O|=|E|$, so that we have:
\[
\bigl\{\,|\operatorname{pf}\mathbf{\tilde R}[S,S]|\,\bigr\}_{|S|\ \mathrm{even}}=\bigcup_{r=0}^{M}\bigl\{\,|\det \mathbf{G}[I,J]|:\ |I|=|J|=r\,\bigr\}.
\]
Summing $|\cdot|^{\alpha}$ over these equal multisets yields \eqref{eq:abs-identity1}.

For the gauge invariance, if $\widehat {\mathbf{G}}=\mathbf{D}_1\boldsymbol{\Pi}_1 \mathbf{G}\boldsymbol{\Pi}_2 \mathbf{D}_2$ with $|\mathbf{D}_{1,2,ii}|=1$, then
\[
|\det \mathbf{\widehat G}[I,J]| = |\det \mathbf{D}_{1,II}|\ |\det \mathbf{G}[\bold \Pi_1(I),\bold\Pi_2(J)]|\ |\det\mathbf{D}_{2,JJ}|
= |\det \mathbf{G}[\bold\Pi_1(I),\bold\Pi_2(J)]|,
\]
and permutations merely relabel the index sets in the sum, leaving the right-hand side of \eqref{eq:abs-identity1} unchanged.
\paragraph{Derivation of the exact correspondence SRE--SR.} Let \(\ket{\mathbf R_L,0}\) be a pure fermionic Gaussian state, and let \(\boldsymbol G\) be its associated correlation matrix. For pure Gaussian states, one has $\mathbf{G}^{\top}_L\mathbf{G}_L=\mathbf I_L$.

Now construct the doubled skew-symmetric matrix
\begin{equation*}
\tilde{\mathbf R}_{2L}=\begin{bmatrix}
0 & \mathbf{G}_L\\
-\mathbf{G}^{\top}_L& 0
\end{bmatrix},
\end{equation*}
from which, after suitable permutations and diagonal rephasings depending on the chosen configuration \(\mathcal C\), one obtains the matrix \(\mathbf R'_{2L}\) defining the doubled Gaussian state \(\ket{\mathbf R'_{2L},\mathcal C}\).

By construction, the permutations and diagonal sign changes relating \(\tilde{\mathbf R}_{2L}\) and \(\mathbf R'_{2L}\) do not affect the absolute values of Pfaffians of principal submatrices. Hence
\begin{equation}
\mathrm{Pf}_{2\alpha}(\mathbf R'_{2L})
=
\mathrm{Pf}_{2\alpha}(\tilde{\mathbf R}_{2L})
=
\mathrm{Det}_{2\alpha}(\boldsymbol G),
\end{equation}
where in the second equality we used Theorem~1.

It remains only to match the normalization factor in the Shannon--R\'enyi entropy. Since for a pure Gaussian state one has \(\boldsymbol G^{\top}\boldsymbol G=\boldsymbol G\boldsymbol G^{\top}=\mathbf I_L\), the doubled matrix \(\tilde{\mathbf R}_{2L}\) satisfies
\begin{equation}
N_{R'}^2
=
\det(\mathbf I_{2L}+\mathbf R_{2L}^{\prime\top}\mathbf R'_{2L})^{1/2}
=
\det(\mathbf I_{2L}+\tilde{\mathbf R}_{2L}^{\top}\tilde{\mathbf R}_{2L})^{1/2}
=
2^L.
\end{equation}
Therefore,
\begin{equation*}
H_\alpha(\ket{\mathbf R'_{2L},\mathcal C})=\frac{1}{1-\alpha}\ln\!\left[\frac{\mathrm{Pf}_{2\alpha}(\mathbf R'_{2L})}{(N_{R'}^2)^\alpha}\right]=\frac{1}{1-\alpha}
\ln\!\left[\frac{\mathrm{Det}_{2\alpha}(\boldsymbol G)}{(2^L)^\alpha}\right]=M_\alpha(\ket{\mathbf R_L,0}).
\end{equation*}
Above, we use the formulas for Shannon and Stabilizer Rényi entropies in function of Pfaffians and minors derived in~\cite{TKR20241} for the Shannon Rényi entropy and~\cite{Leone2022b1} for the Stabilizer Rényi entropy. Finally, we have Eq.~\eqref{eq:Shannon-stabilizer}
\begin{equation*}
\boxed{H_\alpha(\ket{\mathbf R'_{2L},\mathcal C})=M_\alpha(\ket{\mathbf R_L,0}).}
\end{equation*}

\section{Permutation freedom and invariance of the \texorpdfstring{$\mathbf{R}\leftrightarrow\mathbf{G}$}{R-G} mapping}\label{Sec2}

This section is a direct continuation of the previous one. After establishing in Sec.~\ref{Sec1} the Pfaffian--minor correspondence and the resulting exact SRE--SR relation, we now clarify the status of the apparent non-uniqueness in the \(\mathbf R \leftrightarrow \mathbf G\) mapping. We show that this non-uniqueness is purely representational: the permutation bringing \(\mathbf R\) to checkerboard form can be chosen canonically, different choices of the reference configuration \(\mathcal C\) correspond to equivalent parametrizations of the same Gaussian state, and the remaining freedom in \(\mathbf G\) reduces to permutations and diagonal rephasings of fermionic modes. Since the relevant entropy depends only on the absolute values of Pfaffians or minors, all such transformations leave the final result unchanged. In this way, the section makes explicit that the correspondence derived in Sec.~\ref{Sec1} is exact at the level of the physical Gaussian state and is not affected by gauge-like ambiguities in the intermediate matrix representation.

\paragraph{Canonical permutation.} The permutation matrix $\mathbf{P}$ used in our construction is fixed explicitly by the ordering of odd and even indices,
\[
(1,2,\dots,L)\;\longmapsto\;(L-1,L-3,\dots,3,1,\;L,L-2,\dots,4,2).
\]
With this choice, the reordered matrix $\tilde{\mathbf{R}}=\mathbf{P}^\top \mathbf{R}\mathbf{P}$ takes the checkerboard form 
\[
\tilde{\mathbf{R}}=
\begin{bmatrix}
0 & \mathbf{G} \\
-\mathbf{G}^\top & 0
\end{bmatrix}.
\]
Thus, the reduction of $\mathbf{R}$ to block-off-diagonal form is canonical.

\paragraph{Equivalent parametrizations of the same Gaussian state.} The parametrization $|\mathbf{R},\mathcal{C}\rangle$ is redundant. As discussed around Eq.~\eqref{GPS-general} and in Ref.~\cite{Rajabpour2025a1}, whenever a configuration $|\mathcal{C}'\rangle$ has nonzero amplitude, one may equivalently write the same Gaussian state as
\[
|\mathbf{R},\mathcal{C}\rangle = |\mathbf{R}',\mathcal{C}'\rangle
\]
for a suitable antisymmetric matrix $\mathbf{R}'$. In the doubled construction, the checkerboard occupation pattern $1010\ldots$ is simply a convenient representative of this equivalence class. Hence, changing $\mathcal{C}$ modifies only the representation, not the physical state.

To obtain $\mathbf R'$ from $\mathbf R$, let $\ket{\mathcal I'}$ be a configuration obtained from $\ket{\mathcal C'}$ by flipping two occupations. Then~\cite{Rajabpour2025a1}:
\begin{equation}
  r'_{ij}
  = \mathrm{sgn}(\mathcal C,\mathcal C')
    \frac{\mathrm{sgn}(\mathcal C,\mathcal I')}{\mathrm{sgn}(\mathcal C',\mathcal I')}
    \frac{\mathrm{pf}\,\mathbf R[\mathcal C,\mathcal I']}
         {\mathrm{pf}\,\mathbf R[\mathcal C,\mathcal C']},
\end{equation}
with 
\[
\mathrm{sgn}(\mathcal C,\mathcal I)=\prod_{i=2}^L (-1)^{|n_i-m_i|\sum_{j<i} n_j},
\]
for $\mathcal C=(n_1,\dots,n_L)$ and $\mathcal I=(m_1,\dots,m_L)$.

\paragraph{Residual freedom in $\mathbf{G}$.} Once $\mathbf{G}$ is obtained, there remains a freedom of the form
\[
\mathbf{G}\;\longmapsto\; \mathbf{D}_1\bold{\Pi}_1\,\mathbf{G}\bold{\Pi}_2\,\mathbf{D}_2,
\]
where $\bold{\Pi}_{1,2}$ are permutation matrices and $\mathbf{D}_{1,2}$ are diagonal unitary matrices. These transformations correspond to relabeling rows and columns and rephasing fermionic modes. They can only change minors of $\mathbf{G}$ by permutation signs and overall phase factors.

\paragraph{Why the entropy is unchanged.} Since $\mathrm{Det}_\alpha(\mathbf{G})$ is built from the absolute values of these minors, such signs and phases drop out. Consequently, the Shannon--Rényi entropy is invariant under the allowed permutations and phase redefinitions, even at finite size. The non-uniqueness in the intermediate matrices is therefore a gauge freedom of the representation, not a physical ambiguity.

\paragraph{Scope of validity.} The correspondence is exact for the Gaussian states considered in this work. Physically, this is because the mapping is constructed at the level of the Gaussian state itself, rather than for a particular matrix representation. In the doubled description, particle and hole degrees of freedom are paired symmetrically, so the checkerboard structure is a built-in feature of the construction.

Technically, Theorem~1 applies to matrices $\mathbf{R}$ that can be brought into checkerboard (block-off-diagonal) form, possibly after a permutation of the fermionic modes. In our construction, this condition is always satisfied: starting from a number-conserving Gaussian state described by $\mathbf{G}$, we define
\[
\tilde{\mathbf{R}}=
\begin{bmatrix}
0 & \mathbf{G} \\
-\mathbf{G}^\top & 0
\end{bmatrix},
\]
which is explicitly of checkerboard type. Any further permutation brings this matrix to the canonical form used in the theorem.

Therefore, within the doubled representation considered here, the checkerboard condition is not an additional restriction on the physical states, but a property of the chosen parametrization. As a consequence, the correspondence applies exactly to the Gaussian states in our framework.

\section{The Gaussian doubling map}
\label{Sec3}

This section develops the state-level interpretation of the exact correspondence established in Sec.~\ref{Sec1}. There, Theorem~1 provided the algebraic relation between Pfaffians and minors, leading to the equality between the stabilizer R\'enyi entropy of a paired Gaussian state and the Shannon--R\'enyi entropy of an associated doubled state. Here we explain how this correspondence can be implemented through a state-dependent Gaussian doubling map. Starting from the original paired Gaussian state \(\ket{\mathbf{R}_L,0}\), we construct a number-conserving Gaussian state on \(2L\) modes whose computational-basis amplitudes are determined by the same determinant data encoded in the correlation matrix \(\mathbf G\). In this way, the doubling provides a concrete Gaussian-state representation of the correspondence at the level of the states themselves. We then present an explicit construction of the map \(U_R\), discuss the role of the doubled matrix \(\mathbf{R}'_{2L}\), and illustrate the procedure in detail in the simplest nontrivial example, \(L=2\).

\paragraph{Gaussian doubling map: construction, interpretation, and single-unitary form.} The original state \(\ket{\mathbf{R}_L,0}\) is a paired Gaussian state whose stabilizer distribution is encoded in the minors of the correlation matrix \(\mathbf{G}\). The doubled construction replaces this paired description by a \emph{number-conserving} free-fermion state on \(2L\) modes whose computational-basis amplitudes are governed by the same determinant data. Thus, the doubling is an embedding of the pairing structure into a Slater-determinant structure on a larger Hilbert space.

Accordingly, \(U_R\) may be viewed as a physically realizable Gaussian transformation that maps two copies of the original state to the doubled number-conserving one. The correspondence Eq.~\eqref{eq:Shannon-stabilizer} also admits a natural circuit interpretation in terms of \emph{matchgates}. Since \(U_R\) is a quadratic fermionic unitary, it is mapped under the Jordan--Wigner transformation to a matchgate circuit on qubits~\cite{TerhalDiVincenzo20021,JozsaMiyake20081}. This interpretation is consistent with modern treatments of fermionic Gaussian unitaries, where matchgate circuits provide the natural circuit representation of free-fermion dynamics~\cite{Heyraud20251}, and with recent results showing how arbitrary fermionic Gaussian states can be prepared within this framework~\cite{Langer20261}.

At the same time, the construction is state-dependent: the map \(U_R\) is determined by the correlation matrix \(\mathbf{G}\) of the input state, and should therefore be viewed as a Gaussian transport adapted to \(\ket{\mathbf{R}_L,0}\), rather than as a universal basis transformation.

We consider the even-parity Gaussian state
\begin{equation}
\ket{\mathbf{R}_L,0}=\mathcal{N}_R\,
\exp\!\left(
\frac{1}{2}\sum_{i,j=1}^{L}
(\mathbf{R}_L)_{ij}\,c_i^\dagger c_j^\dagger
\right)\ket{0},
\label{eq:R_state}
\end{equation}
where \(\mathbf{R}_L^\top=-\mathbf{R}_L\), \(\ket{0}\) is the fermionic vacuum, and \(\mathcal{N}_R\) is a normalization constant. Its associated correlation matrix is $\mathbf{G}=(\mathbf{I}-\mathbf{R}_L)(\mathbf{I}+\mathbf{R}_L)^{-1}$. For the critical chains considered here, the entries of \(\mathbf{G}\) or \(\mathbf{R}\) are known explicitly for both PBC and OBC.

From \(\mathbf{G}\) one constructs the \(2L\times2L\) antisymmetric matrix
\begin{equation}
\widetilde{\mathbf{R}}_{2L}=
\begin{bmatrix}
0 & \mathbf{G}\\
-\mathbf{G}^{\mathsf T} & 0
\end{bmatrix}.
\label{eq:Rtilde_checkerboard}
\end{equation}
This is the basic doubled object. It already contains the full determinant/Pfaffian structure behind the correspondence, although to interpret it as a Gaussian state in a standard Fock basis one must also specify a reference configuration and, if desired, apply a fixed permutation and diagonal rephasing. The final doubled state is denoted by \(\ket{\mathbf{R}'_{2L},\mathcal C}\), where \(\mathbf{R}'_{2L}\) is obtained from \(\widetilde{\mathbf{R}}_{2L}\) by a convenient choice of ordering and gauge. Since permutations and diagonal phase redefinitions do not affect probabilities, they do not change the Shannon--Rényi entropy.

A useful factorized representation of the map is
\begin{equation}
U_R=\Pi^\dagger\,V_G\,\Pi\,S\,(B_R^\dagger\otimes B_R^\dagger).
\label{eq:UR_factorized}
\end{equation}
Each factor has a clear interpretation. First,
\[
B_R=\exp\!\left[
\frac{1}{2}\sum_{i,j=1}^{L}
(\mathbf{R}_L)_{ij}\,c_i^\dagger c_j^\dagger
-\text{h.c.}\right],
\qquad
B_R\ket{0}=\ket{\mathbf{R}_L,0},
\]
so that \(B_R^\dagger\ket{\mathbf{R}_L,0}=\ket{0}\), and therefore
\[
(B_R^\dagger\otimes B_R^\dagger)
\bigl(\ket{\mathbf{R}_L,0}\otimes\ket{\mathbf{R}_L,0}\bigr)=\ket{0}^{\otimes2}.
\]
Thus, the first step simply removes the original Bogoliubov pairing in each copy.

Next, we introduce the doubled system in the interleaved ordering
\[
(c_1,c_2,\ldots,c_{2L})=(A_1,B_1,A_2,B_2,\ldots,A_L,B_L),
\]
where \(A_j\) and \(B_j\) denote the two checkerboard sublattices. We choose a Gaussian unitary \(S\) such that
\[
S\ket{0}^{\otimes2}=\ket{\mathcal C},
\qquad
\ket{\mathcal C}=\ket{1010\cdots10}.
\]
A convenient choice is
\[
S=
\prod_{j=1}^{L}
\exp\!\left[
\frac{\pi}{2}
\left(
c_{2j-1}^\dagger c_{2j-1+L}^\dagger
-
c_{2j-1+L}c_{2j-1}
\right)
\right],
\]
with the notation understood modulo the chosen ordering convention. This makes explicit that \(S\) is a product of elementary Bogoliubov rotations that fill the odd sites of the doubled chain and prepare the checkerboard configuration. 

The permutation \(\Pi\) then rearranges the interleaved ordering into the checkerboard basis
\[
(d_1,\ldots,d_{2L})=(A_1,\ldots,A_L,B_1,\ldots,B_L),
\]
in which the doubled matrix naturally takes the block form \eqref{eq:Rtilde_checkerboard}.

Finally, \(V_G\) is a number-conserving Gaussian unitary whose one-body action is fixed by \(\mathbf G\). A convenient general form is
\begin{equation}
V_G=\exp\!\left(\sum_{a,b=1}^{2L}(\mathbf K_G)_{ab}\,d_a^\dagger d_b\right),
\qquad
\mathbf W_G=e^{\mathbf K_G},
\label{eq:VG_general}
\end{equation}
with
\begin{equation}
\mathbf{W}_G=\begin{bmatrix}
(\mathbf{I}+\mathbf{G}^{\mathsf T}\mathbf{G})^{-1/2}&-\mathbf{G}^{\mathsf T}(\mathbf{I}+\mathbf{G}\mathbf{G}^{\mathsf T})^{-1/2}\\[2mm]
\mathbf{G}(\mathbf{I}+\mathbf{G}^{\mathsf T}\mathbf{G})^{-1/2}&
(\mathbf{I}+\mathbf{G}\mathbf{G}^{\mathsf T})^{-1/2}
\end{bmatrix}.
\label{eq:WG_general}
\end{equation}
Equivalently, \(\mathbf K_G=\log\mathbf W_G\). This choice guarantees that the occupied block has Thouless matrix \(\mathbf G\), so the output state is precisely the doubled checkerboard Gaussian state associated with \(\widetilde{\mathbf R}_{2L}\).

A particularly simple expression arises when
\[
\mathbf G\mathbf G^{\mathsf T}=\mathbf I.
\]
In this case \(\widetilde{\mathbf R}_{2L}^2=-\mathbf I_{2L}\), so \(\widetilde{\mathbf R}_{2L}\) behaves as a complex structure, and
\begin{equation}
\mathbf W_G=
\frac{1}{\sqrt2}
\begin{pmatrix}
\mathbf I & -\mathbf G^{\mathsf T}\\
\mathbf G & \mathbf I
\end{pmatrix}
=
\exp\!\left[
\frac{\pi}{4}
\begin{pmatrix}
0 & -\mathbf G^{\mathsf T}\\
\mathbf G & 0
\end{pmatrix}
\right].
\label{eq:WG_orthogonal_case}
\end{equation}
Accordingly,
\[
\mathbf K_G=
\frac{\pi}{4}
\begin{pmatrix}
0 & -\mathbf G^{\mathsf T}\\
\mathbf G & 0
\end{pmatrix},
\]
so in this orthogonal case the \(G\)-dependent rotation is generated directly by the same block structure that defines the doubled matrix \(\widetilde{\mathbf R}_{2L}\), up to the sign convention associated with the chosen \(A/B\) ordering.

Combining all steps, one obtains
\begin{equation}
\boxed{
U_R\bigl(\ket{\mathbf R_L,0}\otimes\ket{\mathbf R_L,0}\bigr)=
\ket{\mathbf R'_{2L},\mathcal C}.}
\label{eq:UR_final_action}
\end{equation}
In the checkerboard basis \((A_1,\ldots,A_L,B_1,\ldots,B_L)\), the output state is characterized by a doubled antisymmetric matrix of the form
\[
\mathbf R'_{2L}\sim
\begin{pmatrix}
0 & \mathbf G\\
-\mathbf G^{\mathsf T} & 0
\end{pmatrix},
\]
up to the fixed permutation and rephasing gauge discussed above. Its computational-basis amplitudes are therefore governed by the same determinant data that appear in the stabilizer distribution of \(\ket{\mathbf R_L,0}\). This is the content of the equivalence of Eq.~\eqref{eq:Shannon-stabilizer}: a paired Gaussian state on \(L\) modes is mapped to a number-conserving Gaussian state on \(2L\) modes whose ordinary computational-basis probabilities reproduce the stabilizer probabilities of the original state.

The map \(U_R\) is Gaussian and hence preserves the free-fermion structure, although it does not in general preserve strict short-range locality in the original site basis, since the matrix \(\mathbf G\) entering \eqref{eq:Rtilde_checkerboard} is typically long-ranged, especially at criticality. The relevant computational basis on the doubled side is the Fock basis associated with the canonical fermions in the interleaved ordering together with the checkerboard reference configuration \(\mathcal C\). Different admissible choices of ordering or phase convention merely relabel basis vectors or multiply amplitudes by phases, and therefore leave the probability distribution, and thus \(H_\alpha\), unchanged.

Although Eq.~\eqref{eq:UR_factorized} is convenient for the construction, the full doubling map is itself a \emph{single} Gaussian unitary. Since Gaussian unitaries form a group, the product
\[
U_R=\Pi^\dagger\,V_G\,\Pi\,S\,(B_R^\dagger\otimes B_R^\dagger)
\]
may be written as
\begin{equation}
U_R=\exp\!\left(\frac12\,\Xi^\dagger \mathcal K_R\,\Xi\right),
\label{eq:UR_single_gaussian}
\end{equation}
where
\[
\Xi=
\begin{pmatrix}
c_1\\
\vdots\\
c_{2L}\\
c_1^\dagger\\
\vdots\\
c_{2L}^\dagger
\end{pmatrix},
\]
and \(\mathcal K_R\) is the Bogoliubov--de Gennes generator of the full transformation. Equivalently, if \(\mathcal W_{U_R}\) denotes the one-body Bogoliubov matrix acting on \(\Xi\), then
\begin{equation}
\mathcal W_{U_R}
=
\mathcal W_{\Pi}^\dagger\,
\mathcal W_{V_G}\,
\mathcal W_{\Pi}\,
\mathcal W_S\,
\bigl(\mathcal W_{B_R}^\dagger\oplus\mathcal W_{B_R}^\dagger\bigr),
\qquad
\mathcal K_R=\log\mathcal W_{U_R}.
\label{eq:KR_from_product}
\end{equation}
Thus, the factorized representation should be viewed only as a convenient decomposition of a single quadratic fermionic unitary. In this form the map is manifestly Gaussian and state-dependent: all dependence on the input state is encoded in \(\mathbf R\), equivalently in \(\mathbf G\), through the matrix \(\mathcal K_R\). Therefore the doubling map is not merely a sequence of auxiliary steps, but a single Bogoliubov transformation sending
\[
\ket{\mathbf R_L,0}\otimes\ket{\mathbf R_L,0}
\longmapsto
\ket{\mathbf R'_{2L},\mathcal C}.
\]

\paragraph{Example: \texorpdfstring{$L=2$}{L=2}.}
\label{subsec:UR_example_L2}

For the \(L=2\) periodic TFI case, the correlation matrix is
\begin{equation*}
\mathbf G=\begin{pmatrix}
0.707107 & 0.707107\\
-0.707107 & 0.707107
\end{pmatrix},
\end{equation*}
and the original pairing matrix is
\begin{equation*}
\mathbf R=
\begin{pmatrix}
0 & -0.414214\\
0.414214 & 0
\end{pmatrix},
\qquad
r=1-\sqrt2\approx -0.414214.
\end{equation*}
The corresponding single-copy Gaussian state is:
\begin{equation*}
\ket{R,0}=\frac{1}{\sqrt{1+r^2}}\bigl(\ket{00}+r\,\ket{11}\bigr),
\end{equation*}
so that the doubled input state is
\begin{equation}\label{eq:PsiinL2decimal}
\begin{split}
\ket{\Psi_{\mathrm{in}}}&=\ket{R,0}_{12}\otimes\ket{R,0}_{34}=\frac{1}{1+r^2}\Bigl(\ket{0000}+r\,\ket{0011}+r\,\ket{1100}+r^2\,\ket{1111}\Bigr)\\
&\approx 0.853553\,\ket{0000}-0.353553\,\ket{0011}-0.353553\,\ket{1100}+0.146447\,\ket{1111}.
\end{split}
\end{equation}

We now write the doubling map as a single Gaussian unitary
\begin{equation}
U_R=\exp(\widehat K_R),
\end{equation}
where \(\widehat K_R\) is quadratic in the fermionic operators. In the present \(L=2\) example one obtains
\begin{align*}
\widehat K_R \approx\;&
0.475371\,(c_1^\dagger c_2-c_2^\dagger c_1)
+0.206686\,(c_1^\dagger c_3-c_3^\dagger c_1)
+0.653649\,(c_1^\dagger c_4-c_4^\dagger c_1)
\nonumber\\
&+0.653649\,(c_2^\dagger c_3-c_3^\dagger c_2)
-0.206686\,(c_2^\dagger c_4-c_4^\dagger c_2)
+0.475371\,(c_3^\dagger c_4-c_4^\dagger c_3)
\nonumber\\
&-0.710660\,(c_1^\dagger c_2^\dagger-c_2c_1)
-1.227373\,(c_1^\dagger c_3^\dagger-c_3c_1)
+0.384965\,(c_1^\dagger c_4^\dagger-c_4c_1)
\nonumber\\
&+0.384965\,(c_2^\dagger c_3^\dagger-c_3c_2)
-0.343423\,(c_2^\dagger c_4^\dagger-c_4c_2)
-0.710660\,(c_3^\dagger c_4^\dagger-c_4c_3).
\label{eq:KRhatL2decimal}
\end{align*}
 In matrix form we obtain
\[
\widehat K_R=\frac12\,\Xi^\dagger \mathcal K_R \Xi,
\qquad
\Xi=
\begin{pmatrix}
c_1& c_2& c_3& c_4& c_1^\dagger& c_2^\dagger& c_3^\dagger& c_4^\dagger
\end{pmatrix}^\top,
\]
with
\[
\mathcal K_R \approx
\begin{pmatrix}
0 & 0.475371 & 0.206686 & 0.653649 & 0 & -0.710660 & -1.227373 & 0.384965 \\
-0.475371 & 0 & 0.653649 & -0.206686 & 0.710660 & 0 & 0.384965 & -0.343423 \\
-0.206686 & -0.653649 & 0 & 0.475371 & 1.227373 & -0.384965 & 0 & -0.710660 \\
-0.653649 & 0.206686 & -0.475371 & 0 & -0.384965 & 0.343423 & 0.710660 & 0 \\
0 & -0.710660 & -1.227373 & 0.384965 & 0 & 0.475371 & 0.206686 & 0.653649 \\
0.710660 & 0 & 0.384965 & -0.343423 & -0.475371 & 0 & 0.653649 & -0.206686 \\
1.227373 & -0.384965 & 0 & -0.710660 & -0.206686 & -0.653649 & 0 & 0.475371 \\
-0.384965 & 0.343423 & 0.710660 & 0 & -0.653649 & 0.206686 & -0.475371 & 0
\end{pmatrix}.
\]

Applying \(U_R\) to the input state \eqref{eq:PsiinL2decimal}, one finds

\begin{equation}\label{eq:URPsiinL2decimal}
\begin{split}
U_R\ket{\Psi_{\mathrm{in}}}
\approx\;&0.353553\,\ket{0011}+0.500000\,\ket{0101}-0.353553\,\ket{0110}\\
&-0.353553\,\ket{1001}+0.500000\,\ket{1010}+0.353553\,\ket{1100}.    
\end{split}
\end{equation}
The state above is precisely the ground state of the critical XX chain with size $2L=4$.

Therefore, in this explicit \(L=2\) example, the doubled paired Gaussian state is mapped by the single Gaussian unitary \(U_R=\exp(\widehat K_R)\) into the number-conserving doubled state
\begin{equation*}
\ket{\mathbf{R}',\mathcal C}=U_R\ket{\Psi_{\mathrm{in}}},
\qquad\mathcal C=1010,
\end{equation*}
with amplitudes given in Eq.~\eqref{eq:URPsiinL2decimal}. 

We can also rewrite~\eqref{eq:URPsiinL2decimal} in terms of the $\mathbf{R}'_{2L}$ matrix:
sing the Jordan--Wigner mapping, the basis states can be written in terms of fermionic creation operators acting on $|1010\rangle$ (we have freedom to pick one of the vectors),
\begin{align*}
|0011\rangle &=c_1 c_4^\dagger |1010\rangle,\\
|0110\rangle &=c_2^\dagger c_1 |1010\rangle,\\
|1001\rangle &=c_4^\dagger c_3 |1010\rangle,\\
|1100\rangle &=c_2^\dagger c_3 |1010\rangle,\\
|0101\rangle &=c_2^\dagger c_4^\dagger c_3 c_1 |1010\rangle.
\end{align*}
Equation~\eqref{eq:URPsiinL2decimal} is of the form
\begin{equation}
\ket{\mathbf{R}',\mathcal C}=\frac{1}{\mathcal{N}}e^{\tfrac{1}{2}\sum_{i,j=1}^{4} R'_{ij}\,a_i a_j}
|1010\rangle,
\end{equation}
with $a=\begin{pmatrix}
    c_1&c^\dagger_2&c_3&c^\dagger_4
\end{pmatrix}$, and a $\mathbf{R}$ matrix
\begin{equation}
\mathbf{R}'_{2L}=
\begin{pmatrix}
0 & -0.707106 & 0 & 0.707106\\
0.707106 & 0 & 0.707106 & 0\\
0 & -0.707106 & 0 & -0.707106\\
-0.707106 & 0 & 0.707106 & 0
\end{pmatrix},
\end{equation}
and normalization constant $\mathcal{N}=2$.

\section{Proof of Theorem 2 and Relation between \texorpdfstring{$\mathbf{G}^{TFI}_{L\times L}$}{GTFIL} and \texorpdfstring{$\mathbf{R}^{XX}_{2L\times2L}$}{RXX2L}\label{Sec4}}

In this Appendix, we illustrate the relationship between the matrices $\mathbf{G}^{\mathrm{TFI}}_{L\times L}$ and $\mathbf{R}^{\mathrm{XX}}_{2L\times 2L}$, for both periodic (PBC) and open (OBC) boundary conditions, through explicit examples. The precise definitions of these matrices are given in Table~\ref{tab:GR_matricesTFI-X}.

\begin{table}[!ht]
\centering
\footnotesize
\setlength{\tabcolsep}{7pt}
\renewcommand{\arraystretch}{1.15}

\begin{tabular}{|l|l|l|}
\hline
Model & BC & Matrix entries \\ \hline
TFI & PBC &
\(\displaystyle
G^{\mathrm{PBC}}_{jk}
=\frac{(-1)^{\,j-k}}{L}\,
\csc\!\Big(\frac{\pi}{L}\big(j-k+\tfrac12\big)\Big)
\) \\[6pt]
TFI & OBC &
\(\displaystyle
G^{\mathrm{OBC}}_{jk}
=\frac{(-1)^{\,j-k}}{2L+1}\!\left[
\csc\!\Big(\frac{\pi}{2L+1}\big(j-k+\tfrac12\big)\Big)
+\csc\!\Big(\frac{\pi}{2L+1}\big(j+k-\tfrac12\big)\Big)
\right]
\) \\[8pt]
XX  & PBC &
\(\displaystyle
\text{for } j<k:\quad
R^{\mathrm{PBC}}_{jk}=
\begin{cases}
0, & j+k\ \text{even},\\[4pt]
(-1)^{\frac{j+k+1}{2}}\dfrac{2}{L}\,
\csc\!\Big(\dfrac{\pi}{L}(j-k)\Big), & j+k\ \text{odd}.
\end{cases}
\) \\[12pt]
XX  & OBC &
\(\displaystyle
\theta=\frac{\pi}{L+1},\quad
R^{\mathrm{OBC}}_{jk}=
\begin{cases}
\dfrac{1}{L+1}\!\left[\csc\!\big(\tfrac{k-j}{2}\,\theta\big)
+\csc\!\big(\tfrac{k+j}{2}\,\theta\big)\right], & j\ \text{odd},\ k\ \text{even},\\[6pt]
-\,R^{\mathrm{OBC}}_{kj}, & j\ \text{even},\ k\ \text{odd},\\[4pt]
0, & j+k\ \text{even}.
\end{cases}
\)
\\ \hline
\end{tabular}
\caption{Relevant matrices for the critical TFI and XX chains with periodic (PBC) and open (OBC) boundary conditions. \(\mathbf{R}\) is antisymmetric, \(L\) is even.} \label{tab:GR_matricesTFI-X}
\end{table}

Our procedure is as follows. We first compute the explicit form of the $\mathbf{G}$ matrix, and then construct the corresponding $\mathbf{R}$ matrix. Next, we apply the permutation $\mathbf{P}$ to $\mathbf{R}$, which reorganizes it into a block off-diagonal form. Up to this stage, the steps follow exactly the same construction as in Theorem~1.

The key observation is that the resulting off-diagonal blocks coincide, up to simple transformations, with the $\mathbf{G}$ matrices. More precisely, by applying suitable diagonal matrices $\mathbf{D}$, one can map these blocks directly to $\mathbf{G}^{\mathrm{TFI}}$.

This correspondence naturally leads to Theorem~2, which establishes a direct connection between the ground-state stabilizer Rényi entropy of the critical TFI chain of size $L$ and the Shannon entropy of the ground state of the critical XX chain of size $2L$. Finally, we give two explicit examples for this derivation. Figure~\ref{fig:placeholder} summarizes the exact correspondence established in this work for the particular TFI--XX case.

\begin{figure}[!ht]
\centering
\includegraphics[width=0.6\linewidth]{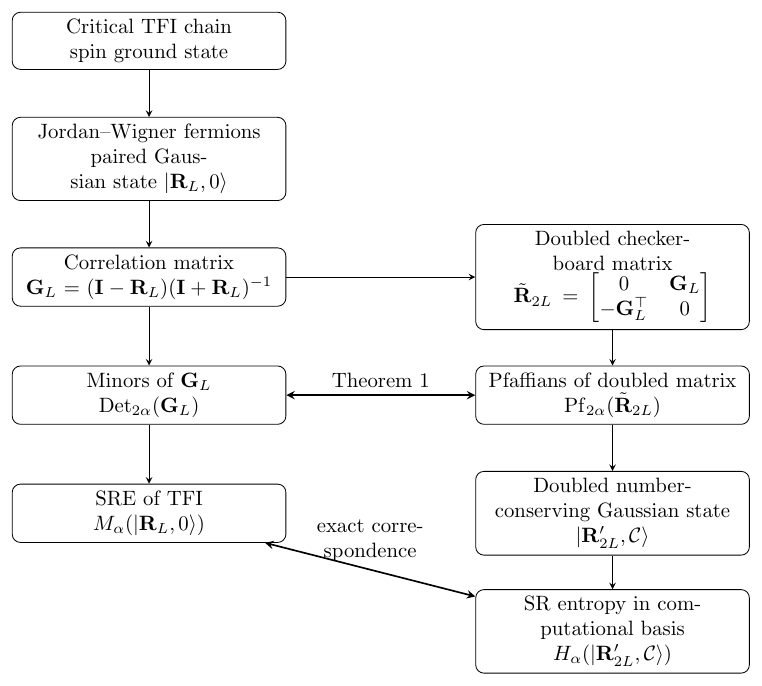}
\caption{Schematic mapping between the stabilizer R\'enyi entropy (SRE) of the critical TFI chain and the Shannon--R\'enyi entropy (SR) of the doubled critical XX system. The correspondence is mediated by Theorem~1, which establishes the equivalence between the Pfaffian structure of $\mathbf R_{2L}$ and the determinant structure of $\mathbf G_L$.}
    \label{fig:placeholder}
\end{figure}

\paragraph{Constructing $\mathbf{R}^{\mathrm{XX}}$ from $\mathbf{G}^{\mathrm{TFI}}$.} Let $L$ be even. We begin with the PBC case. We consider the permutation
\[
(1,2,\dots,L)\ \longrightarrow\ (L-1,L-3,\dots,3,1\,|\,L,L-2,\dots,4,2),
\]
that is, the odd indices appear first in decreasing order, followed by the even indices, also in decreasing order. In the permuted basis, the antisymmetric matrix takes the checkerboard form
\begin{equation}
\tilde{\mathbf{R}}^{\mathrm{PBC}}=\mathbf{P}^{\top}\mathbf{R}^{\mathrm{PBC}}\mathbf{P}=\begin{pmatrix}
0 &\mathbf{G}^{\mathrm{PBC}}\\[2mm]
-\big(\mathbf{G}^{\mathrm{PBC}}\big)^{\top} & 0
\end{pmatrix},
\end{equation}
where $G^{\mathrm{PBC}}\in\mathbb{C}^{(L/2)\times(L/2)}$ has entries
\begin{equation}
G^{\mathrm{PBC}}_{ab}=\frac{2(-1)^{a-b}}{L}\,
\csc\!\left[\frac{2\pi}{L}\left(a-b+\frac12\right)\right],
\qquad a,b=1,\dots,\frac{L}{2}.
\end{equation}

To make the relation with the original matrix explicit, define the inverse map $\sigma^{-1}$ by
\[
\sigma^{-1}(m)=\begin{cases}
\dfrac{L-m+1}{2}, & m \ \text{odd},\\[3mm]
L-\dfrac{m}{2}+1, & m \ \text{even}.
\end{cases}
\]
Then
\[
R^{\mathrm{PBC}}_{mn}=\tilde R^{\mathrm{PBC}}_{\sigma^{-1}(m),\,\sigma^{-1}(n)}.
\]
In particular, if $m$ is odd and $n$ is even, we have
\[
R^{\mathrm{PBC}}_{mn}=G^{\mathrm{PBC}}_{ab},
\qquad a=\frac{L-m+1}{2},\qquad b=\frac{L-n}{2}+1.
\]
Hence
\[
a-b=\frac{n-m-1}{2},\qquad a-b+\frac12=\frac{n-m}{2},
\]
and therefore
\begin{equation}
R^{\mathrm{PBC}}_{mn}=\frac{2}{L}
(-1)^{\frac{n-m-1}{2}}\csc\!\left(\frac{\pi}{L}(n-m)\right),
\qquad m\ \text{odd},\ n\ \text{even}.
\end{equation}
By antisymmetry,
\[
R^{\mathrm{PBC}}_{nm}=-R^{\mathrm{PBC}}_{mn},
\]
while entries with $m+n$ even vanish. Thus the complete matrix is
\[
R^{\mathrm{PBC}}_{mn}=\begin{cases}
0, & m+n\ \mathrm{even},\\[2mm]
\displaystyle
\frac{2}{L}(-1)^{\frac{n-m-1}{2}}
\csc\!\left(\frac{\pi}{L}(n-m)\right),
& m+n\ \mathrm{odd},\ m<n,\\[3mm]
\displaystyle-\frac{2}{L}(-1)^{\frac{m-n-1}{2}}\csc\!\left(\frac{\pi}{L}(m-n)\right),
& m+n\ \mathrm{odd},\ m>n.
\end{cases}
\]

Now, we continue with the OBC. We use the same odd-even reversed permutation as before,
\[
(1,2,\dots,2L)\ \longrightarrow\ (2L-1,2L-3,\dots,3,1\,|\,2L,2L-2,\dots,4,2),
\]
so that the odd indices appear first in decreasing order, followed by the even indices, also in decreasing order. In the permuted basis, the antisymmetric matrix has the checkerboard form
\begin{equation}
\tilde{\mathbf{R}}^{\mathrm{OBC}}=\mathbf{P}^{\top}\mathbf{R}^{\mathrm{OBC}}\mathbf{P}=\begin{pmatrix}
0 &\mathbf{G}^{\mathrm{OBC}}\\[2mm]
-\big(\mathbf{G}^{\mathrm{OBC}}\big)^{\top} & 0
\end{pmatrix},
\end{equation}
where $G^{\mathrm{OBC}}\in\mathbb{C}^{(L/2)\times(L/2)}$ has entries
\begin{equation}
G^{\mathrm{OBC}}_{ab}=\frac{2(-1)^{a-b}}{L}\,
\csc\!\left[\frac{2\pi}{L}\left(a-b+\frac12\right)\right],
\qquad a,b=1,\dots,\frac{L}{2}.
\end{equation}

To express $R^{\mathrm{OBC}}$ in the original basis, define the inverse map
\[
\sigma^{-1}(m)=
\begin{cases}
\dfrac{2L-m+1}{2}, & m \ \text{odd},\\[3mm]
2L-\dfrac{m}{2}+1, & m \ \text{even}.
\end{cases}
\]
Equivalently, if $m$ is odd and $n$ is even, we may write
\[
a=\frac{2L-m+1}{2},\qquad b=\frac{2L-n+2}{2},
\]
so that
\[
R^{\mathrm{OBC}}_{mn}=G^{\mathrm{OBC}}_{ab},
\qquad (m\ \text{odd},\ n\ \text{even}),
\]
and similarly
\[
R^{\mathrm{OBC}}_{mn}=-G^{\mathrm{OBC}}_{ba},
\qquad (m\ \text{even},\ n\ \text{odd}).
\]
For $m$ odd and $n$ even, we have
\[
a-b=\frac{n-m-1}{2},\qquad
a-b+\frac12=\frac{n-m}{2},
\]
and
\[
a+b-\frac12
=
2L+1-\frac{m+n}{2}.
\]
Therefore,
\begin{align}
R^{\mathrm{OBC}}_{mn}
&=
\frac{(-1)^{\frac{n-m-1}{2}}}{2L+1}
\left[
\csc\!\left(\frac{\pi}{2L+1}\frac{n-m}{2}\right)
+
\csc\!\left(\frac{\pi}{2L+1}\left(2L+1-\frac{m+n}{2}\right)\right)
\right] \notag\\[2mm]
&=
\frac{(-1)^{\frac{n-m-1}{2}}}{2L+1}
\left[
\csc\!\left(\frac{\pi}{2L+1}\frac{n-m}{2}\right)
+
\csc\!\left(\frac{\pi}{2L+1}\frac{m+n}{2}\right)
\right],
\qquad (m\ \text{odd},\ n\ \text{even}),
\end{align}
where in the second line we used $\csc(\pi-x)=\csc x$.

By antisymmetry,
\[
R^{\mathrm{OBC}}_{nm}=-R^{\mathrm{OBC}}_{mn},
\]
and all entries with $m+n$ even vanish. Hence the full matrix can be written as
\[
R^{\mathrm{OBC}}_{mn}=
\begin{cases}
0, & m+n\ \mathrm{even},\\[2mm]
\displaystyle
\frac{(-1)^{\frac{n-m-1}{2}}}{2L+1}
\left[
\csc\!\left(\frac{\pi}{2L+1}\frac{n-m}{2}\right)
+
\csc\!\left(\frac{\pi}{2L+1}\frac{m+n}{2}\right)
\right],
& m\ \mathrm{odd},\ n\ \mathrm{even},\\[4mm]
\displaystyle
-\frac{(-1)^{\frac{m-n-1}{2}}}{2L+1}
\left[
\csc\!\left(\frac{\pi}{2L+1}\frac{m-n}{2}\right)
+
\csc\!\left(\frac{\pi}{2L+1}\frac{m+n}{2}\right)
\right],
& m\ \mathrm{even},\ n\ \mathrm{odd}.
\end{cases}
\]

In both cases, the matrices $\mathbf{R}$ obtained from $\mathbf{G}^{\mathrm{TFI}}$ through this construction coincide exactly with the matrices $\mathbf{R}$ listed in Table~\ref{tab:GR_matricesTFI-X}, i.e.\ the antisymmetric matrices associated with the critical XX chain.

\paragraph{Examples: Relation between $\mathbf{G}^{TFI}_{L\times L}$ and $\mathbf{R}^{XX}_{2L\times2L}$.} We start using the formulas for PBC and OBC in Table I to obtain the $\mathbf{G}^{\mathsf{TFI}}$ and $\mathbf{R}^{\mathrm{XX}}$:

For $L=2$ we have,
\[
\mathbf{G}^{\mathrm{PBC}}_{2}=
\begin{pmatrix}
0.70710678 & 0.70710678\\
-0.70710678 & 0.70710678
\end{pmatrix},
\qquad
\mathbf{G}^{\mathrm{OBC}}_{2}=
\begin{pmatrix}
0.89442719 & 0.44721360\\
-0.44721360 & 0.89442719
\end{pmatrix}.
\]

\[
\mathbf{R}^{\mathrm{PBC}}_{4}=
\begin{pmatrix}
0 & -0.70710678 & 0 & 0.70710678\\
0.70710678 & 0 & 0.70710678 & 0\\
0 & -0.70710678 & 0 & -0.70710678\\
-0.70710678 & 0 & 0.70710678 & 0
\end{pmatrix},
\]
\[
\mathbf{R}^{\mathrm{OBC}}_{4}=
\begin{pmatrix}
0 & 0.89442719 & 0 & 0.44721360\\
-0.89442719 & 0 & 0.44721360 & 0\\
0 & -0.44721360 & 0 & 0.89442719\\
-0.44721360 & 0 & -0.89442719 & 0
\end{pmatrix}.
\]

For the permutation
\[
(1,2,3,4)\mapsto(3,1,4,2),
\qquad
\widetilde{\mathbf R}=\mathbf P\,\mathbf R\,\mathbf P^{\mathsf T},
\]
we obtain

\[
\widetilde{\mathbf R}^{\mathrm{PBC}}_{4}=
\begin{pmatrix}
0 & 0 & -0.70710678 & -0.70710678\\
0 & 0 & 0.70710678 & -0.70710678\\
0.70710678 & -0.70710678 & 0 & 0\\
0.70710678 & 0.70710678 & 0 & 0
\end{pmatrix},
\]
with
\[
\mathbf B^{\mathrm{PBC}}_2=
\begin{pmatrix}
-0.70710678 & -0.70710678\\
0.70710678 & -0.70710678
\end{pmatrix},
\]
and
\[
\widetilde{\mathbf R}^{\mathrm{OBC}}_{4}=
\begin{pmatrix}
0 & 0 & 0.89442719 & -0.44721360\\
0 & 0 & 0.44721360 & 0.89442719\\
-0.89442719 & -0.44721360 & 0 & 0\\
0.44721360 & -0.89442719 & 0 & 0
\end{pmatrix},
\]
with
\[
\mathbf B^{\mathrm{OBC}}_2=
\begin{pmatrix}
0.89442719 & -0.44721360\\
0.44721360 & 0.89442719
\end{pmatrix}.
\]

We can check that
\begin{equation*}
\mathbf{G}^{\mathrm{PBC}}_{2}=\mathbf{D}^\top\mathbf B^{\mathrm{PBC}}_2\mathbf{D},\qquad\mathbf{G}^{\mathrm{OBC}}_{2}=\mathbf{D}^\top\mathbf B^{\mathrm{OBC}}_2\mathbf{D},
\end{equation*}
where $\mathbf{D}=\operatorname{diag}(1,-1)$.

For $L=4$ we have,
\begin{equation*}
\mathbf G^{\mathrm{PBC}}_4=
\begin{pmatrix}
0.65328148 & 0.65328148 & -0.27059805 & 0.27059805 \\
-0.27059805 & 0.65328148 & 0.65328148 & -0.27059805 \\
 0.27059805 & -0.27059805 & 0.65328148 & 0.65328148 \\
-0.65328148 & 0.27059805 & -0.27059805 & 0.65328148
\end{pmatrix},
\end{equation*}

\begin{equation*}
\mathbf G^{\mathrm{OBC}}_4=
\begin{pmatrix}
0.86208561 &0.49481813 & -0.10398025 &0.03393414 \\
-0.36726748 & 0.75810536 &0.52875228 & -0.10398025 \\
 0.26328723 & -0.33333333 &0.75810536 &0.49481813 \\
-0.22935309 & 0.26328723 & -0.36726748 & 0.86208561
\end{pmatrix}.
\end{equation*}

\[
\mathbf{R}^{\mathrm{PBC}}_{8}=
\begin{pmatrix}
0 & -0.65328148 & 0 & 0.27059805 & 0 & -0.27059805 & 0 & 0.65328148 \\
0.65328148 & 0 & 0.65328148 & 0 & -0.27059805 & 0 & 0.27059805 & 0 \\
0 & -0.65328148 & 0 & -0.65328148 & 0 & 0.27059805 & 0 & -0.27059805 \\
-0.27059805 & 0 & 0.65328148 & 0 & 0.65328148 & 0 & -0.27059805 & 0 \\
0 & 0.27059805 & 0 & -0.65328148 & 0 & -0.65328148 & 0 & 0.27059805 \\
0.27059805 & 0 & -0.27059805 & 0 & 0.65328148 & 0 & 0.65328148 & 0 \\
0 & -0.27059805 & 0 & 0.27059805 & 0 & -0.65328148 & 0 & -0.65328148 \\
-0.65328148 & 0 & 0.27059805 & 0 & -0.27059805 & 0 & 0.65328148 & 0
\end{pmatrix},
\]
\[
\mathbf{R}^{\mathrm{OBC}}_{8}=
\begin{pmatrix}
0 & 0.86208561 & 0 & 0.36726748 & 0 & 0.26328723 & 0 & 0.22935309 \\
-0.86208561 & 0 & 0.49481813 & 0 & 0.10398025 & 0 & 0.03393414 & 0 \\
0 & -0.49481813 & 0 & 0.75810536 & 0 & 0.33333333 & 0 & 0.26328723 \\
-0.36726748 & 0 & -0.75810536 & 0 & 0.52875228 & 0 & 0.10398025 & 0 \\
0 & -0.10398025 & 0 & -0.52875228 & 0 & 0.75810536 & 0 & 0.36726748 \\
-0.26328723 & 0 & -0.33333333 & 0 & -0.75810536 & 0 & 0.49481813 & 0 \\
0 & -0.03393414 & 0 & -0.10398025 & 0 & -0.49481813 & 0 & 0.86208561 \\
-0.22935309 & 0 & -0.26328723 & 0 & -0.36726748 & 0 & -0.86208561 & 0
\end{pmatrix}.
\]

For the permutation
\[
(1,2,3,4,5,6,7,8)\mapsto(7,5,3,1,8,6,4,2),
\qquad
\widetilde{\mathbf R}=\mathbf P\,\mathbf R\,\mathbf P^{\mathsf T},
\]
we obtain

\[
\widetilde{\mathbf{R}}^{\mathrm{PBC}}_8=
\begin{pmatrix}
0 & 0 & 0 & 0 & -0.65328148 & -0.65328148 & 0.27059805 & -0.27059805 \\
0 & 0 & 0 & 0 & 0.27059805 & -0.65328148 & -0.65328148 & 0.27059805 \\
0 & 0 & 0 & 0 & -0.27059805 & 0.27059805 & -0.65328148 & -0.65328148 \\
0 & 0 & 0 & 0 & 0.65328148 & -0.27059805 & 0.27059805 & -0.65328148 \\
0.65328148 & -0.27059805 & 0.27059805 & -0.65328148 & 0 & 0 & 0 & 0 \\
0.65328148 & 0.65328148 & -0.27059805 & 0.27059805 & 0 & 0 & 0 & 0 \\
-0.27059805 & 0.65328148 & 0.65328148 & -0.27059805 & 0 & 0 & 0 & 0 \\
0.27059805 & -0.27059805 & 0.65328148 & 0.65328148 & 0 & 0 & 0 & 0
\end{pmatrix},
\]
with

\[
\mathbf B^{\mathrm{PBC}}_4=
\begin{pmatrix}
-0.65328148 & -0.65328148 & 0.27059805 & -0.27059805 \\
0.27059805 & -0.65328148 & -0.65328148 & 0.27059805 \\
-0.27059805 & 0.27059805 & -0.65328148 & -0.65328148 \\
0.65328148 & -0.27059805 & 0.27059805 & -0.65328148
\end{pmatrix},
\]
and
\[
\widetilde{\mathbf{R}}^{\mathrm{OBC}}_8=
\begin{pmatrix}
0 & 0 & 0 & 0 & 0.86208561 & -0.49481813 & -0.10398025 & -0.03393414 \\
0 & 0 & 0 & 0 & 0.36726748 & 0.75810536 & -0.52875228 & -0.10398025 \\
0 & 0 & 0 & 0 & 0.26328723 & 0.33333333 & 0.75810536 & -0.49481813 \\
0 & 0 & 0 & 0 & 0.22935309 & 0.26328723 & 0.36726748 & 0.86208561 \\
-0.86208561 & -0.36726748 & -0.26328723 & -0.22935309 & 0 & 0 & 0 & 0 \\
0.49481813 & -0.75810536 & -0.33333333 & -0.26328723 & 0 & 0 & 0 & 0 \\
0.10398025 & 0.52875228 & -0.75810536 & -0.36726748 & 0 & 0 & 0 & 0 \\
0.03393414 & 0.10398025 & 0.49481813 & -0.86208561 & 0 & 0 & 0 & 0
\end{pmatrix},
\]
with

\[
\mathbf B^{\mathrm{OBC}}_4=
\begin{pmatrix}
0.86208561 & -0.49481813 & -0.10398025 & -0.03393414 \\
0.36726748 & 0.75810536 & -0.52875228 & -0.10398025 \\
0.26328723 & 0.33333333 & 0.75810536 & -0.49481813 \\
0.22935309 & 0.26328723 & 0.36726748 & 0.86208561
\end{pmatrix}.
\]

We can check that
\begin{equation*}
\mathbf{G}^{\mathrm{PBC}}_{4}=\mathbf{D}^\top\mathbf B^{\mathrm{PBC}}_4\mathbf{D},\qquad\mathbf{G}^{\mathrm{OBC}}_{4}=\mathbf{D}^\top\mathbf B^{\mathrm{OBC}}_4\mathbf{D},
\end{equation*}
where $\mathbf{D}=\operatorname{diag}(1,-1,1,-1)$.

\paragraph{Example: SRE(TFI)--SR(XX) correspondence.} As a pedagogical illustration of the correspondence established in this work, we consider the smallest nontrivial periodic example: the TFI chain with $L=2$ and the doubled XX chain with $2L=4$. In this case, the equality between the stabilizer R\'enyi entropy (SRE) and the Shannon--R\'enyi entropy (SR) can be checked directly from the explicit probability distributions.

\textbf{TFI chain at $L=2$}: For the critical TFI chain with periodic boundary conditions, the non-degenerate ground state in the Neveu--Schwarz sector is
\begin{equation*}
\ket{\psi^{\mathrm{TFI}}_{0}}=-\cos\!\left(\frac{\pi}{8}\right)\ket{00}-\sin\!\left(\frac{\pi}{8}\right)\ket{11}.
\end{equation*}
Its associated Pauli distribution has six nonzero weights,
\begin{equation*}
\mathcal{P}_{\mathrm{TFI}}
=
\left\{
\frac{1}{4},\frac{1}{4},
\frac{1}{8},\frac{1}{8},\frac{1}{8},\frac{1}{8}
\right\},
\end{equation*}
corresponding to the Pauli strings
\[
II,\ ZZ,\ IZ,\ ZI,\ XX,\ YY.
\]
Therefore, for any Rényi index $\alpha>0$, $\alpha\neq1$, the stabilizer entropy is
\begin{equation}
M_\alpha\!\left(\ket{\psi^{\mathrm{TFI}}_{0}}\right)=
\frac{1}{1-\alpha}\ln\!\left[2\left(\frac{1}{4}\right)^\alpha
+4\left(\frac{1}{8}\right)^\alpha\right].
\end{equation}

\textbf{XX chain at $2L=4$}: The corresponding non-degenerate ground state of the critical XX chain is
\begin{equation*}
\ket{\psi^{\mathrm{XX}}_{0}}=\frac{1}{2\sqrt{2}}\left(-\,\ket{0011}-\sqrt{2}\,\ket{0101}-\ket{0110}-\ket{1001}-\sqrt{2}\,\ket{1010}-\ket{1100}\right).
\end{equation*}
Its computational basis probabilities are
\begin{equation*}
p_{\mathrm{XX}}=\left\{\frac{1}{4},\frac{1}{4},\frac{1}{8},\frac{1}{8},\frac{1}{8},\frac{1}{8}\right\},
\end{equation*}
up to the reordering of the basis states. Hence, the Shannon--R\'enyi entropy is
\begin{equation}
H_\alpha\!\left(\ket{\psi^{\mathrm{XX}}_{0}}\right)=\frac{1}{1-\alpha}\ln\!\left[2\left(\frac{1}{4}\right)^\alpha+4\left(\frac{1}{8}\right)^\alpha\right].
\end{equation}

\textbf{Direct comparison}: Since the Pauli distribution of the TFI ground state and the computational-basis distribution of the XX ground state coincide, the two entropies are exactly equal:
\begin{equation*}
M_\alpha\!\left(\ket{\psi^{\mathrm{TFI}}_{0}}\right)=H_\alpha\!\left(\ket{\psi^{\mathrm{XX}}_{0}}\right).
\end{equation*}
This equality is the simplest explicit illustration of the general SRE--SR correspondence.

For a few standard values of $\alpha$, one finds
\begin{align}
M_{1/2} &= H_{1/2} = 2\ln(1+\sqrt{2}), \\
M_{2}   &= H_{2}   = \ln\!\left(\frac{16}{3}\right), \\
M_{4}   &= H_{4}   = \frac{1}{3}\ln\!\left(\frac{1024}{9}\right).
\end{align}
In particular, for $\alpha=2$ this gives the numerical value
\[
M_2 = H_2 \approx 1.67398.
\]

This small example makes the correspondence completely transparent: the TFI Pauli distribution and the XX computational-basis distribution contain the same weights, and therefore produce identical Rényi entropies.

\section{General Quadratic Fermionic Chains}\label{Sec5}

We consider a general one-dimensional quadratic spin-$\tfrac12$ chain with finite interaction range $R$. The model can be written with either periodic boundary conditions (PBC) or open boundary conditions (OBC). In the spin language, the periodic chain is defined by
\begin{equation}
\label{eq:Hspin}
H=\frac{1}{2}\sum_{r=1}^{R}\sum_{j=1}^{N}\Big[(A_r+B_r)\sigma_j^{x}\sigma_{j+r}^{x}+(A_r-B_r)\sigma_j^{y}\sigma_{j+r}^{y}\Big]Z_{j+1}^{\,j+r-1}
+\frac{A_0}{2}\sum_{j=1}^{N}\sigma_j^{z},
\end{equation}
where the sites are understood modulo $N$ and
\begin{equation}
\sigma_{j+N}^\alpha\equiv\sigma_j^\alpha,\qquad Z_{j+1}^{\,j+r-1}=\prod_{k=j+1}^{j+r-1}\sigma_k^{z},\qquad
Z_{j+1}^{\,j+r-1}\big|_{r=1}\equiv\mathbf{1}.
\end{equation}
The coefficients satisfy $A_{-r}=A_r$ and $B_{-r}=-B_r$, which ensures Hermiticity. We focus exclusively on the periodic case. This family includes, as special cases, the standard XY, transverse-field Ising (TFI), and XX chains, as well as longer-range generalizations.

For $R=1$, one recovers standard nearest-neighbor chains. The anisotropic XY model in a transverse field,
\begin{equation}
H^{\mathsf{XY}}=-J\sum_{j=1}^{N}\left[\frac{1+\gamma}{4}\sigma_j^{x}\sigma_{j+1}^{x}+\frac{1-\gamma}{4}\sigma_j^{y}\sigma_{j+1}^{y}
\right]-\frac{h}{2}\sum_{j=1}^{N}\sigma_j^{z},
\end{equation}
is obtained for $A_1=-\frac{J}{2}$, $B_1=-\frac{J\gamma}{2}$, $A_0=-h$, with $A_{r>1}=B_{r>1}=0$. 

The transverse-field Ising chain follows in the isotropic limit $\gamma=1$,
\begin{equation}
H^{\mathsf{TFI}}=-\frac{J}{2}\sum_{j=1}^{N}\sigma_j^{x}\sigma_{j+1}^{x}
-\frac{h}{2}\sum_{j=1}^{N}\sigma_j^{z},
\end{equation}
corresponding to $A_1=B_1=-\frac{J}{2}$, and $A_0=-h$, while the XX chain,
\begin{equation}
H^{\mathsf{XX}}=-\frac{J}{4}\sum_{j=1}^{N}\big(
\sigma_j^{x}\sigma_{j+1}^{x}+\sigma_j^{y}\sigma_{j+1}^{y}\big)-\frac{h}{2}\sum_{j=1}^{N}\sigma_j^{z},
\end{equation}
is recovered for $A_1=-\frac{J}{2}$, $B_1=0$, and $A_0=-h$.

For $R>1$, the Hamiltonian~\eqref{eq:Hspin} naturally generates long-range extensions that remain quadratic in fermionic variables after the Jordan--Wigner transformation. In particular, for $R=2$ one obtains next-nearest-neighbor generalizations of the TFI and XX chains. An extended TFI-type model is given by
\begin{equation}
H^{\mathsf{TFI}}_2=-\frac{1}{2}\sum_{j=1}^{N}\left(
J_1\sigma_j^{x}\sigma_{j+1}^{x}+J_2\sigma_j^{x}\sigma_{j+1}^{z}\sigma_{j+2}^{x}\right)-\frac{h}{2}\sum_{j=1}^{N}\sigma_j^{z},
\end{equation}
which corresponds to $A_1=B_1=-\frac{J_1}{2}$, $A_2=B_2=-\frac{J_2}{2}$, and $A_0=-h$. Similarly, a next-nearest-neighbor XX-type chain reads
\begin{equation}
H^{\mathsf{XX}}_{2}=-\frac{1}{4}\sum_{j=1}^{N}\Big[
J_1(\sigma_j^{x}\sigma_{j+1}^{x}+\sigma_j^{y}\sigma_{j+1}^{y})
+J_2(\sigma_j^{x}\sigma_{j+1}^{z}\sigma_{j+2}^{x}+\sigma_j^{y}\sigma_{j+1}^{z}\sigma_{j+2}^{y})\Big]-\frac{h}{2}\sum_{j=1}^{N}\sigma_j^{z},
\end{equation}
generated by $A_1=-\frac{J_1}{2}$, $B_1=0$, $A_2=-\frac{J_2}{2}$, $B_2=0$, and $A_0=-h$. In both cases, the $Z$-string associated with $r=2$ encodes the fermionic statistics, rendering the spin Hamiltonian apparently interacting while preserving its quadratic fermionic structure.

\paragraph{From spins to fermions.}

The general Hamiltonian~\eqref{eq:Hspin} can be mapped to a system of spinless Dirac fermions by means of the Jordan--Wigner (JW) transformation
\begin{align}\label{eq:JWtransform}
\sigma_j^{z}&=2c_j^{\dagger}c_j-1, \\
\sigma_j^{x}&=\Big[\prod_{k=1}^{j-1}(1-2c_k^{\dagger}c_k)\Big](c_j^{\dagger}+c_j), \\
\sigma_j^{y}&=-i\Big[\prod_{k=1}^{j-1}(1-2c_k^{\dagger}c_k)\Big](c_j^{\dagger}-c_j),
\end{align}
where $c_j$ and $c_j^\dagger$ are canonical fermionic annihilation and creation operators satisfying $\{c_i,c_j^\dagger\}=\delta_{ij}$. In this representation, $c_j^\dagger$ creates a spin-up excitation at site $j$, while the fermionic vacuum corresponds to the fully polarized spin state $|\downarrow\downarrow\cdots\downarrow\rangle$.

For $r$-neighbor couplings, one obtains
\begin{align}
Z_{j+1}^{\,j+r-1}&=
\begin{cases}
\displaystyle\prod_{k=j+1}^{j+r-1}(2c_k^\dagger c_k-1), & r>1,\\[6pt]
\mathbf{1}, & r=1,
\end{cases}\\
\sigma^x_j\sigma^x_{j+r}&=(-1)^{r-1}(c_j^\dagger-c_j)\,Z_{j+1}^{\,j+r-1}\,(c_{j+r}^\dagger+c_{j+r}),\\
\sigma^y_j\sigma^y_{j+r}&=(-1)^{r}(c_j^\dagger+c_j)\,Z_{j+1}^{\,j+r-1}\,(c_{j+r}^\dagger-c_{j+r}).
\end{align}
Substituting these expressions into Eq.~\eqref{eq:Hspin} yields a quadratic fermionic Hamiltonian,
\begin{equation}
\label{eq:Hpre}
H=\sum_{r=1}^{R}\sum_{j=1}^{N}\left[A_r\,c_j^{\dagger}c_{j+r}+\frac{B_r}{2}\big(c_j^{\dagger}c_{j+r}^{\dagger}-c_jc_{j+r}\big)
\right]+\text{h.c.}+A_0\sum_{j=1}^{N}c_j^{\dagger}c_{j}-\frac{A_0N}{2},
\end{equation}
where the Hermitian-conjugate terms are required, since the bracketed expression is not manifestly Hermitian.

The Hermitian conjugate contribution can be re-expressed by performing the index transformation $r\to -r$ and relabeling lattice sites using PBC. This allows one to combine positive- and negative-distance terms, as well as the on-site contribution proportional to $A_0$, into a single symmetric sum over $r=-R,\dots,R$. Adopting the conventions
\begin{equation*}
A_{-r}=A_r,\qquad B_{-r}=-B_r,
\end{equation*}
which ensures Hermiticity, the Hamiltonian can be written in the compact form
\begin{equation}
\label{eq:Hdirac}
H=\sum_{r=-R}^{R}\sum_{j=1}^{N}\left[A_r\,c_j^{\dagger}c_{j+r}+\frac{B_r}{2}\big(c_j^{\dagger}c_{j+r}^{\dagger}-c_jc_{j+r}\big)
\right]+\text{const}.
\end{equation}

For OBC, the JW mapping is strictly local. In contrast, for PBC, the JW string wraps around the chain and introduces the fermion-parity operator
\begin{equation*}
\hat P = (-1)^{\hat N_f},\qquad
\hat N_f=\sum_{j=1}^{N}c_j^\dagger c_j,
\end{equation*}
which enforces the boundary condition
\begin{equation*}
c_{N+j}=\hat P\,c_j.
\end{equation*}
Consequently, the fermionic Hilbert space decomposes into even ($\hat P=+1$) and odd ($\hat P=-1$) parity sectors, corresponding to anti-periodic (Neveu--Schwarz) and periodic (Ramond) boundary conditions, respectively. Since $[\hat P,H]=0$, these sectors do not mix, and the physical ground state is obtained by selecting the sector with the lowest energy. 
\paragraph{Discrete Fourier transform.}
Translation invariance allows for a diagonalization in momentum space. The fermionic boundary conditions depend on the fermion-parity sector. In the Neveu--Schwarz (NS) sector, corresponding to even fermion parity, the fermions satisfy anti-periodic boundary conditions, and the allowed momenta are
\begin{equation*}
\theta_k=\frac{2\pi}{N}\Big(k-\tfrac12\Big),\qquad k=1,\dots,N.
\end{equation*}
In the Ramond (R) sector, the fermions are periodic, with momenta $2\pi k/N$. In the following, we work in the NS sector, which contains the physical ground state for the models considered in this work.

We introduce the discrete Fourier transform
\begin{equation*}
c_j=\frac{1}{\sqrt{N}}\sum_{k=1}^{N} e^{i\theta_k j}\,c_{\theta_k},\qquad
c_j^\dagger=\frac{1}{\sqrt{N}}\sum_{k=1}^{N} e^{-i\theta_k j}\,c_{\theta_k}^\dagger.
\end{equation*}
Defining the mode functions
\begin{align}
\varepsilon(\theta_k)
&:=\sum_{r=-R}^{R}A_r e^{ir\theta_k}
= A_0+2\sum_{r>0}A_r\cos(r\theta_k),\\
\Delta(\theta_k)
&:=\sum_{r=-R}^{R}B_r e^{ir\theta_k}
= 2i\sum_{r>0}B_r\sin(r\theta_k),
\end{align}
the Hamiltonian~\eqref{eq:Hdirac} becomes
\begin{equation}
\label{eq:Hk}
H=\sum_{k=1}^{N}\Big[
\varepsilon(\theta_k)\,c_{\theta_k}^\dagger c_{\theta_k}
+\tfrac12\big(\Delta(\theta_k)\,c_{\theta_k}^\dagger c_{-\theta_k}^\dagger
-\Delta^*(\theta_k)\,c_{-\theta_k}c_{\theta_k}\big)
\Big]+\text{const}.
\end{equation}

\paragraph{Bogoliubov diagonalization and symbol representation.}

Introducing the complex dispersion symbol
\begin{equation*}
f(e^{i\theta}) := \varepsilon(\theta)+i\Delta(\theta),
\end{equation*}
the single-particle spectrum is
\begin{equation*}
E(\theta)=|f(e^{i\theta})|.
\end{equation*}
It is convenient to isolate the phase of the symbol,
\begin{equation}
s(\theta):=\frac{f(e^{i\theta})}{|f(e^{i\theta})|}\in\mathbb{T},
\end{equation}
which completely determines the ground-state structure of the model.

The Hamiltonian is diagonalized by the Bogoliubov transformation
\begin{equation*}
\eta_{\theta_k}
=\frac{1}{2}\left[
\left(1+s(\theta_k)\right)c_{\theta_k}
+
\left(1-s(\theta_k)\right)c_{-\theta_k}^\dagger
\right],
\end{equation*}
which brings it to the diagonal form
\begin{equation}
\label{eq:Hdiagonalform}
H=\sum_{k=1}^{N} |f(e^{i\theta_k})|\,\eta_{\theta_k}^\dagger\eta_{\theta_k}
+\text{const}.
\end{equation}
The quasiparticle vacuum is defined by
\begin{equation*}
\eta_{\theta_k}\,|0_\eta\rangle=0,\qquad \forall k.
\end{equation*}

\paragraph{Correlation \texorpdfstring{$\boldsymbol G^{(f)}$}{G(f)} matrix.}

Once the ground state is written in terms of the Bogoliubov modes, all equal-time correlations can be expressed through the phase symbol \(s(\theta)\). In momentum space, the relevant kernel is simply \(s(\theta_k)\); in real space, the corresponding correlation matrix is obtained by inverse discrete Fourier transform. This leads naturally to the Toeplitz-type matrix \(\boldsymbol G^{(f)}(N)\), whose entries are
\begin{equation}
\label{eq:G-def-section1}
G^{(f)}_{nm}
=\frac{(-1)^{n-m}}{N}\sum_{k=1}^{N}
s(\theta_k)\,e^{i\theta_k(n-m)},
\qquad n,m=1,\dots,N.
\end{equation}
Thus, \(\boldsymbol G^{(f)}\) is not introduced ad hoc: it is precisely the real-space representation of the symbol \(s(\theta)\), and it compactly encodes the ground-state correlations of the quadratic fermionic chain.

The function \(f(z)\) is a trigonometric polynomial,
\begin{equation}
f(z)=\sum_{m=-R}^{R} t_m z^m=-\sum_{m=-R}^{R}(A_m+B_m)z^m,
\end{equation}
with real-space couplings related by
\begin{equation*}
A_r=-\frac{t_r+t_{-r}}{2},\qquad
B_r=-\frac{t_r-t_{-r}}{2}.
\end{equation*}
The function \(f(z)\) fully determines the model and its excitation spectrum. 
For example, the transverse-field Ising (TFI) chain corresponds to \(f(z)=z+h\), 
whereas the critical XX chain is obtained from \(f(z)=z+z^{-1}\). 
Criticality occurs whenever \(f(z)\) has zeros on the unit circle \(|z|=1\).

More generally, one may consider two families of range-\(R\) critical models. 
For the TFI-type chain, only the couplings \(A_0\) and \(A_R\) are nonzero, while
\(A_r=B_r=0\) for \(1\le r\le R-1\). 
Likewise, for the XX-type chain, only \(A_0\) and \(A_R\) are nonzero, with
\(B_r=0\) for all \(r\). 
In this way, the range-\(R\) TFI-type model is generated by
\(f(z)=z^R+1\), while the critical XX-type model at \(h=0\) is described by
\(f(z)=z^R+z^{-R}\).

Explicitly, the corresponding Hamiltonian can be written as
\begin{equation}
H_R=\sum_{j=1}^{N}\left[
A_R\,c_j^{\dagger}c_{j+R}
+\frac{B_R}{2}\left(c_j^{\dagger}c_{j+R}^{\dagger}-c_jc_{j+R}\right)
\right]+\text{const.}
\end{equation}

A summary of the coupling structure and corresponding functions \(f(z)\) for the most relevant models is provided in Table~\ref{tab:f-functions}, where we explicitly list the nonvanishing coefficients. In particular, for the range-\(R\) generalizations, all intermediate couplings vanish, i.e. \(A_r=B_r=0\) for \(1\le r\le R-1\).

\begin{table*}[!ht]
\centering
\footnotesize
\setlength{\tabcolsep}{6pt}
\renewcommand{\arraystretch}{1.2}

\begin{tabular}{|l|l|l|l|l|}
\hline
Model & $R$ & $\{A_r\},\{B_r\}$ & $f(z,J,h)$ & $f(z)$--critical \\ 
\hline

TFI 
& 1 
& $\{A_0=-h,A_1=-\frac{J}{2}\},\{B_0=0,B_1=-\frac{J}{2}\}$ 
& $Jz+h$ 
& $z+1$ 
\\[6pt]

XX 
& 1 
& $\{A_0=-h,A_1=-\frac{J}{2}\},\{B_0=0,B_1=0\}$ 
& $\frac{J}{2}(z+z^{-1})+h$ 
& $z+z^{-1},\quad h=0$ 
\\[8pt]

XY  
& 1 
& $\{A_0=-h,A_1=-\frac{J}{2}\},\{B_0=0,B_1=-\frac{J\gamma}{2}\}$ 
& $\frac{J}{2}[(1+\gamma)z+(1-\gamma)z^{-1}]+h$ 
& $\begin{aligned}
&z^{-1}\!\left(z+\frac{h+\kappa}{1+\gamma}\right)\!
\left(z+\frac{h-\kappa}{1+\gamma}\right)\left(\frac{1+\gamma}{2}\right),\\
&\kappa=\sqrt{h^2+\gamma^2-1},\quad \gamma^2+h^2\geq1
\end{aligned}$ 
\\[10pt]

TFI-type 
& R 
& $\{A_0=-h,A_R=-J/2\},\{B_0=0,B_R=-J/2\}$ 
& $Jz^R+h$ 
& $z^R+1$ 
\\

XX-type 
& R 
& $\{A_0=-h,A_R=-J\},\{B_0=0,B_R=0\}$ 
& $J(z^R+z^{-R})+h$ 
& $z^R+z^{-R}$ 
\\ 

\hline
\end{tabular}

\caption{Function \(f(z)\) and coupling structure for representative quadratic fermionic chains. 
In the $R$-range generalizations (TFI-type and XX-type), all couplings vanish except those at distance \(R\), i.e., \(A_r=B_r=0\) for \(r\neq 0,R\).}
\label{tab:f-functions}
\end{table*}

\paragraph{Correlation \texorpdfstring{$\boldsymbol G^{(f)}$}{G(f)} matrix examples.}
By using Eq.~\eqref{eq:G-def-section1}, we construct the matrices $\mathbf{G}^{(f)}(N)$. We then apply the diagonal permutation matrix
\begin{equation*}
D=\operatorname{diag}(1,-1,1,\dots,-1),
\end{equation*}
which reorganizes the structure of $\mathbf{G}^{(f)}(N)$. We employ this extra permutation only so that the ordering is exactly the one you find if you use the formulas in Table I; this does not affect the physics of the system.

It is instructive to first examine the case of the TFI family with $R=1$. In this simplest setting, the structure of the correlation matrix becomes transparent. For higher-range models ($R>1$), both in the TFI and XX families, the corresponding matrices $\mathbf{G}^{(f)}(N)$ can be systematically rearranged into block structures composed of TFI correlation matrices. This block decomposition is precisely the content of Theorem~2.

\paragraph{TFI-type family:}
For $R=1$, $N=2,4,6,8$ and $f=z+1$ we obtain,
\begin{equation*}
\mathbf{G}^{\mathsf{TFI}}(2)=\begin{pmatrix}
0.707107 & 0.707107 \\
-0.707107 & 0.707107
\end{pmatrix},
\end{equation*}

\begin{equation*}
\mathbf{G}^{\mathsf{TFI}}(4)=\begin{pmatrix}
0.653281 & 0.653281 & -0.270598 & 0.270598\\
-0.270598 & 0.653281 & 0.653281 & -0.270598\\
0.270598 & -0.270598 & 0.653281 & 0.653281\\
-0.653281 & 0.270598 & -0.270598 & 0.653281
\end{pmatrix},
\end{equation*}

\begin{equation*}
\mathbf G^{\mathsf{TFI}}(6)=
\begin{pmatrix}
 0.643951 & 0.643951 & -0.235702 & 0.172546 & -0.172546 & 0.235702 \\
 -0.235702 & 0.643951 & 0.643951 & -0.235702 & 0.172546 & -0.172546 \\
 0.172546 & -0.235702 & 0.643951 & 0.643951 & -0.235702 & 0.172546 \\
 -0.172546 & 0.172546 & -0.235702 & 0.643951 & 0.643951 & -0.235702 \\
 0.235702 & -0.172546 & 0.172546 & -0.235702 & 0.643951 & 0.643951 \\
 -0.643951 & 0.235702 & -0.172546 & 0.172546 & -0.235702 & 0.643951
\end{pmatrix},
\end{equation*}

\begin{equation*}
\mathbf G^{\mathsf{TFI}}(8)=\scalebox{0.8}{$
\begin{pmatrix}
 0.64072886 &  0.64072886 & -0.22499406 &  0.15033622 & -0.12744889 &  0.12744889 & -0.15033622 &  0.22499406 \\
-0.22499406 &  0.64072886 &  0.64072886 & -0.22499406 &  0.15033622 & -0.12744889 &  0.12744889 & -0.15033622 \\
 0.15033622 & -0.22499406 &  0.64072886 &  0.64072886 & -0.22499406 &  0.15033622 & -0.12744889 &  0.12744889 \\
-0.12744889 &  0.15033622 & -0.22499406 &  0.64072886 &  0.64072886 & -0.22499406 &  0.15033622 & -0.12744889 \\
 0.12744889 & -0.12744889 &  0.15033622 & -0.22499406 &  0.64072886 &  0.64072886 & -0.22499406 &  0.15033622 \\
-0.15033622 &  0.12744889 & -0.12744889 &  0.15033622 & -0.22499406 &  0.64072886 &  0.64072886 & -0.22499406 \\
 0.22499406 & -0.15033622 &  0.12744889 & -0.12744889 &  0.15033622 & -0.22499406 &  0.64072886 &  0.64072886 \\
-0.64072886 &  0.22499406 & -0.15033622 &  0.12744889 & -0.12744889 &  0.15033622 & -0.22499406 &  0.64072886
\end{pmatrix}$},
\end{equation*}

For $R=2$, $N=4,8$ and $f=z^2+1$ we obtain,
\begin{equation*}
\mathbf{G}^{(z^2+1)}(4)=\begin{pmatrix}
 0.707107 & 0 & 0.707107 & 0\\
 0 & 0.707107 & 0 & 0.707107 \\
 -0.707107 & 0& 0.707107 & 0 \\
 0 & -0.707107 & 0& 0.707107 \\
\end{pmatrix},
\end{equation*}

For the permutation
\[
(1,2,3,4)\mapsto(1,3,2,4),
\qquad
\widetilde{\mathbf{G}}^{(z^2+1)}(4)=\mathbf \Pi_2\,\mathbf{G}^{(z^2+1)}(4)\,\mathbf \Pi^\top_2,
\]
we obtain
\begin{equation*}
\tilde{\mathbf{G}}^{(z^2+1)}(4)=\begin{pmatrix}
\mathbf{G}^{\mathsf{TFI}}(2)&\mathbf{0}\\
\mathbf{0}&\mathbf{G}^{\mathsf{TFI}}(2)
\end{pmatrix},
\end{equation*}

\begin{equation*}
\mathbf{G}^{(z^2+1)}(8)=\scalebox{0.8}{$\begin{pmatrix}
0.653281 & 0 & 0653281 & 0 & -0.270598 & 0 & 0.270598 & 0 \\
 0 & 0.653281 & 0 & 0.653281 & 0 & -0.270598 & 0 & 0.270598 \\
 -0.270598 & 0 & 0.653281 & 0 & 0.653281 & 0 & -0.270598 & 0 \\
 0 & -0.270598 & 0 & 0.653281 & 0 & 0.653281 & 0 & -0.270598 \\
 0.270598 & 0 & -0.270598 & 0 & 0.653281 & 0 & 0.653281 & 0 \\
 0 & 0.270598 & 0 & -0.270598 & 0 & 0.653281 & 0 & 0.653281 \\
 -0.653281 & 0 & 0.270598 & 0 & -0.270598 & 0 & 0.653281 & 0 \\
 0 & -0.653281 & 0 & 0.270598 & 0 & -0.270598 & 0 & 0.653281
\end{pmatrix}$},
\end{equation*}
For the permutation
\[
(1,2,3,4,5,6,7,8)\mapsto(1,3,5,7,2,4,6,8),
\qquad
\widetilde{\mathbf{G}}^{(z^2+1)}(8)=\mathbf \Pi_2\,\mathbf{G}^{(z^2+1)}(8)\,\mathbf \Pi^\top_2,
\]
we obtain
\begin{equation*}
\tilde{\mathbf{G}}^{(z^2+1)}(8)=\begin{pmatrix}
\mathbf{G}^{\mathsf{TFI}}(4)&\mathbf{0}\\
\mathbf{0}&\mathbf{G}^{\mathsf{TFI}}(4)
\end{pmatrix},
\end{equation*}

For $R=3$, $N=6$, and $f=z^3+1$ we have

\begin{equation*}
\mathbf G^{(z^3+1)}(6)=
\begin{pmatrix}
0.707107 & 0 & 0 & 0.707107 & 0 & 0 \\
 0 & 0.707107 & 0 & 0 & 0.707107 & 0 \\
 0 & 0 & 0.707107 & 0 & 0 & 0.707107 \\
 -0.707107 & 0 & 0 & 0.707107 & 0 & 0 \\
 0 & -0.707107 & 0 & 0 & 0.707107 & 0 \\
 0 & 0 & -0.707107 & 0 & 0 & 0.707107
\end{pmatrix},
\end{equation*}

For the permutation
\[
(1,2,3,4,5,6)\mapsto(1,4,2,5,3,6),
\qquad
\widetilde{\mathbf{G}}^{(z^3+1)}(6)=\mathbf \Pi_3\,\mathbf{G}^{(z^3+1)}(6)\,\mathbf \Pi^\top_3,
\]
we obtain
\begin{equation*}
\tilde{\mathbf{G}}^{(z^3+1)}(6)=\begin{pmatrix}
\mathbf{G}^{\mathsf{TFI}}(2)&\mathbf{0}&\mathbf{0}\\
\mathbf{0}&\mathbf{G}^{\mathsf{TFI}}(2)&\mathbf{0}\\
\mathbf{0}&\mathbf{0}&\mathbf{G}^{\mathsf{TFI}}(2)
\end{pmatrix},
\end{equation*}

For $R=4$, $N=8$, and $f=z^4+1$ we have

\begin{equation*}
\mathbf{G}^{(z^4+1)}(8)=\scalebox{0.8}{$\begin{pmatrix}
 0.707107 & 0 & 0 & 0 & 0.707107 & 0 & 0 & 0 \\
 0 & 0.707107 & 0 & 0 & 0 & 0.707107 & 0 & 0 \\
 0 & 0 & 0.707107 & 0 & 0 & 0 & 0.707107 & 0 \\
 0 & 0 & 0 & 0.707107 & 0 & 0 & 0 & 0.707107 \\
 -0.707107 & 0 & 0 & 0 & 0.707107 & 0 & 0 & 0 \\
 0 & -0.707107 & 0 & 0 & 0 & 0.707107 & 0 & 0 \\
 0 & 0 & -0.707107 & 0 & 0 & 0 & 0.707107 & 0 \\
 0 & 0 & 0 & -0.707107 & 0 & 0 & 0 & 0.707107
\end{pmatrix}$},
\end{equation*}

For the permutation
\[
(1,2,3,4,5,6,7,8)\mapsto(1,5,2,6,3,7,4,8),
\qquad
\widetilde{\mathbf{G}}^{(z^4+1)}(8)=\mathbf \Pi_4\,\mathbf{G}^{(z^4+1)}(8)\,\mathbf \Pi^\top_4,
\]
we obtain
\begin{equation*}
\tilde{\mathbf{G}}^{(z^4+1)}(8)=\begin{pmatrix}
\mathbf{G}^{\mathsf{TFI}}(2)&\mathbf{0}&\mathbf{0}&\mathbf{0}\\
\mathbf{0}&\mathbf{G}^{\mathsf{TFI}}(2)&\mathbf{0}&\mathbf{0}\\
\mathbf{0}&\mathbf{0}&\mathbf{G}^{\mathsf{TFI}}(2)&\mathbf{0}\\
\mathbf{0}&\mathbf{0}&\mathbf{0}&\mathbf{G}^{\mathsf{TFI}}(2)
\end{pmatrix},
\end{equation*}

\paragraph{XX-type family:}
For $R=1$, $N=4,8$ and $f=z+z^{-1}$ we obtain,
\begin{equation*}
\mathbf{G}^{\mathsf{XX}}(4)=\begin{pmatrix}
0& 0.707107 & 0 & -0.707107 \\
0.707107 & 0 & 0.707107 & 0 \\
0& 0.707107 & 0 & 0.707107 \\
-0.707107 & 0 & 0.707107 & 0
\end{pmatrix},
\end{equation*}

For the permutation
\[
(1,2,3,4)\mapsto(1,3,2,4),
\qquad
\widetilde{\mathbf{G}}^{\mathsf{XX}}(4)=\mathbf \Pi_2\,\mathbf{G}^{\mathsf{XX}}(4)\,\mathbf \Pi^\top_2,
\]
we obtain
\begin{equation*}
\widetilde{\mathbf{G}}^{\mathsf{XX}}(4)=\begin{pmatrix}
\mathbf{0}&\mathbf{P}^\top_2\mathbf{G}^{\mathsf{TFI}}(2)\mathbf{P}_2\\
\mathbf{P}^\top_2(\mathbf{G}^{\mathsf{TFI}}(2))^\top\mathbf{P}_2&\mathbf{0}
\end{pmatrix},\qquad \mathbf{P}_2=\begin{pmatrix}
    0&1\\
    1&0
\end{pmatrix},
\end{equation*}

\begin{equation*}
\mathbf G^{\mathsf{XX}}(8)=\scalebox{0.8}{$
\begin{pmatrix}
 0 &  0.65328148 &  0 & -0.27059805 & -0 &  0.27059805 &  0 & -0.65328148 \\
 0.65328148 &  0 &  0.65328148 &  0 & -0.27059805 & -0 &  0.27059805 &  0 \\
 0 &  0.65328148 &  0 &  0.65328148 &  0 & -0.27059805 & -0 &  0.27059805 \\
-0.27059805 &  0 &  0.65328148 &  0 &  0.65328148 &  0 & -0.27059805 & -0 \\
-0 & -0.27059805 &  0 &  0.65328148 &  0 &  0.65328148 &  0 & -0.27059805 \\
 0.27059805 & -0 & -0.27059805 &  0 &  0.65328148 &  0 &  0.65328148 &  0 \\
 0 &  0.27059805 & -0 & -0.27059805 &  0 &  0.65328148 &  0 &  0.65328148 \\
-0.65328148 &  0 &  0.27059805 & -0 & -0.27059805 &  0 &  0.65328148 &  0
\end{pmatrix}$},
\end{equation*}

For the permutation
\[
(1,2,3,4,5,6,7,8)\mapsto(1,3,5,7,2,4,6,8),
\qquad
\widetilde{\mathbf{G}}^{\mathsf{XX}}(8)=\mathbf \Pi_2\,\mathbf{G}^{\mathsf{XX}}(8)\,\mathbf \Pi^\top_2,
\]
we obtain
\begin{equation*}
\widetilde{\mathbf{G}}^{\mathsf{XX}}(8)=\begin{pmatrix}
\mathbf{0}&\mathbf{P}^\top_2\mathbf{G}^{\mathsf{TFI}}(4)\mathbf{P}_2\\
\mathbf{P}^\top_2(\mathbf{G}^{\mathsf{TFI}}(4))^\top\mathbf{P}_2&\mathbf{0}
\end{pmatrix},\qquad \mathbf{P}_2=\begin{pmatrix}
    0&0&0&1\\
    0&0&1&0\\
    0&1&0&0\\
    1&0&0&0
\end{pmatrix},
\end{equation*}

For $R=2$, $N=8$ and $f=z^2+z^{-2}$ we obtain,
\begin{equation*}
\mathbf{G}^{(z^2+z^{-2})}(8)=\scalebox{0.8}{$\begin{pmatrix}
0& 0& 0.707107 & 0& 0& 0& -0.707107 & 0\\
 0& 0& 0& 0.707107 & 0& 0& 0& -0.707107 \\
 0.707107 & 0& 0& 0& 0.707107 & 0& 0& 0\\
 0& 0.707107 & 0& 0& 0& 0.707107 & 0& 0\\
 0& 0& 0.707107 & 0& 0& 0& 0.707107 & 0\\
 0& 0& 0& 0.707107 & 0& 0& 0& 0.707107 \\
 -0.707107 & 0& 0& 0& 0.707107 & 0& 0& 0\\
 0& -0.707107 & 0& 0& 0& 0.707107 & 0& 0
\end{pmatrix}$}.
\end{equation*}

For the permutation
\[
(1,2,3,4,5,6,7,8)\mapsto(1,5,3,7,2,6,4,8),
\qquad
\widetilde{\mathbf{G}}^{(z^2+z^{-2})}(8)=\mathbf \Pi_4\,\mathbf{G}^{(z^2+z^{-2})}(8)\,\mathbf \Pi^\top_4,
\]
we obtain
\begin{equation*}
\widetilde{\mathbf{G}}^{(z^2+z^{-2})}(8)=\begin{pmatrix}
\mathbf{0}&\mathbf{P}^\top_2\mathbf{G}^{\mathsf{TFI}}(4)\mathbf{P}_2&\mathbf{0}&\mathbf{0}\\
\mathbf{P}^\top_2(\mathbf{G}^{\mathsf{TFI}}(4))^\top\mathbf{P}_2&\mathbf{0}&\mathbf{0}&\mathbf{0}\\
\mathbf{0}&\mathbf{0}&\mathbf{0}&\mathbf{P}^\top_2\mathbf{G}^{\mathsf{TFI}}(4)\mathbf{P}_2\\
\mathbf{0}&\mathbf{0}&\mathbf{P}^\top_2(\mathbf{G}^{\mathsf{TFI}}(4))^\top\mathbf{P}_2&\mathbf{0}
\end{pmatrix},\qquad \mathbf{P}_2=\begin{pmatrix}
    0&0&0&1\\
    0&0&1&0\\
    0&1&0&0\\
    1&0&0&0
\end{pmatrix},
\end{equation*}

\section{Proof of Theorem 3}\label{Sec6}

In this Section, we prove Theorem~3 by showing that the matrices
\(\bold{G}^{(f)}\) associated with the symbols \(f(z)=z^n+1\) and
\(f(z)=z^m+z^{-m}\) admits natural reductions after a suitable reordering of the lattice sites. The basic idea is that the dependence of the symbol on \(n\) or \(m\) induces a decomposition into independent residue classes, so that the full
matrix splits into simpler blocks. The proof is therefore organized in two steps, corresponding to the two families of symbols appearing in the theorem. In the first paragraph, we show that \(f(z)=z^n+1\) reduces to \(n\) identical copies of the fundamental \(z+1\) block. In the second paragraph, we show that \(f(z)=z^m+z^{-m}\) reduces to \(m\) identical off-diagonal blocks, each of which is equivalent, up to permutations and diagonal phases, to the same basic \(z+1\) building block. This makes explicit that the general cases are obtained by repeating and rearranging the elementary TFI-type structure.
\paragraph{ Block structure for $f(z)=z^{n}+1$.}
For $L\in\mathbb{N}$ and a function $f$ on the unit circle, define
\begin{equation}
G_{ij}\equiv G^{(f)}_{ij}
=\frac{(-1)^{\,i-j}}{L}\sum_{k=1}^{L}
\frac{f\!\left(e^{i\theta_k}\right)}{\big|f\!\left(e^{i\theta_k}\right)\big|}\,
e^{\,i\theta_k (i-j)},
\qquad
\theta_k:=\frac{2\pi}{L}\Big(k-\tfrac{1}{2}\Big).
\label{eq:G-def}
\end{equation}
{\bf{Theorem[Mode--pair reduction for $z^n+1$]}}:
{\it{Take $f(z)=z^{n}+1$. Assume $\,2n\mid L\,$ (equivalently, $L/n$ is even). Let $\Pi$ be the permutation that groups the indices by residue classes modulo $n$,
\[
\Pi:\ (1,2,\dots,L)\ \longmapsto\ \big(1,1+n,1+2n,\dots;\ 2,2+n,\dots;\ \dots;\ n,\dots\big).
\]
Then $\Pi^{\!\top}\bold{G}^{(z^{n}+1)}_{L}\Pi$ is block diagonal with $n$ identical blocks of size $L/n$, and each block is (up to a harmless alternating-sign conjugation when $n$ is even) the $f(z)=z+1$ matrix at half-shift on size $L/n$:
\begin{equation}
\boxed{\
\Pi^{\!\top}\bold{G}^{(z^{n}+1)}_{L}\Pi
=\bigoplus_{j=1}^{n} \bold{B},\qquad
\bold{B}=
\begin{cases}
\;\;\bold{G}^{(z+1)}_{\,L/n}, & n\ \text{odd},\\[4pt]
\;\bold{D}\,\bold{G}^{(z+1)}_{\,L/n}\,\bold{D}, & n\ \text{even},
\end{cases}
}
\qquad
\bold{D}:=\mathrm{diag}(1,-1,1,-1,\dots)\in\mathbb{R}^{(L/n)\times(L/n)}.
\label{eq:block-claim}
\end{equation}}}

We now show why \eqref{eq:block-claim} holds. For $f(z)=z^{n}+1$ we have
\[
\frac{f(e^{i\theta})}{|f(e^{i\theta})|}=e^{\,i\frac{n}{2}\theta}\,\mathrm{sgn}\!\big(\cos\tfrac{n\theta}{2}\big),
\]
with the natural continuous-phase convention at the finitely many zeros where $\cos(\tfrac{n\theta}{2})=0$. On the half-shifted grid $\theta_k=\tfrac{2\pi}{L}(k-\tfrac12)$ and under the shift $k\mapsto k+\tfrac{L}{n}$ (i.e. $\theta\mapsto\theta+\tfrac{2\pi}{n}$) one has
\[
e^{\,i\frac{n}{2}(\theta+\frac{2\pi}{n})}\,\mathrm{sgn}\!\Big(\cos\tfrac{n}{2}(\theta+\tfrac{2\pi}{n})\Big)
=\big(-e^{\,i\frac{n}{2}\theta}\big)\cdot\big(-\,\mathrm{sgn}(\cos\tfrac{n\theta}{2})\big)
= e^{\,i\frac{n}{2}\theta}\,\mathrm{sgn}(\cos\tfrac{n\theta}{2}),
\]
so the ratio $f(e^{i\theta_k})/|f(e^{i\theta_k})|$ is invariant along each length-$n$ orbit $\{k,k+\tfrac{L}{n},\dots,k+(n-1)\tfrac{L}{n}\}$. Grouping the sum in \eqref{eq:G-def} by these orbits produces the geometric factor
\[
\sum_{r=0}^{n-1} e^{\,i\theta_{k+r\frac{L}{n}}(i-j)}
=\sum_{r=0}^{n-1} e^{\,i\theta_k(i-j)}\,e^{\,i\frac{2\pi}{n}r\,(i-j)}
=e^{\,i\theta_k(i-j)}\sum_{r=0}^{n-1}\!\big(e^{\,i\frac{2\pi}{n}(i-j)}\big)^{r},
\]
which vanishes unless $n$ divides $(i-j)$. This kills all couplings between different residue classes modulo $n$ and yields $n$ independent blocks of size $L/n$. Finally, within each block, the remaining kernel is exactly the $z+1$ case on size $L/n$; the global factor $(-1)^{i-j}$ in \eqref{eq:G-def} restricts to an alternating-sign conjugation by $D$ when $n$ is even (and cancels when $n$ is odd), producing \eqref{eq:block-claim}.

\paragraph{Block reduction for $f(z)=z^m+z^{-m}$ on the half--shifted grid.}

Fix integers $m\ge 1$ and $L\ge 1$. Let
\begin{equation}
\theta_k=\frac{2\pi}{L}\Big(k-\frac{1}{2}\Big),\qquad k=1,2,\dots,L,
\end{equation}
and, for a function $f$ on the unit circle, define the $L\times L$ matrix
\begin{equation}
\label{eq:Gdef}
G^{(f)}_{nm}
=\frac{(-1)^{\,n-m}}{L}\sum_{k=1}^{L}
\frac{f\!\left(e^{i\theta_k}\right)}{\big|f\!\left(e^{i\theta_k}\right)\big|}\;
e^{\,i\theta_k (n-m)}\,,\qquad n,m\in\{1,\dots,L\}.
\end{equation}
We are interested in $f(z)=z^m+z^{-m}$, for which
\begin{equation}
\label{eq:s-symbol}
\frac{f(e^{i\theta})}{|f(e^{i\theta})|}=\operatorname{sgn}\!\big(\cos(m\theta)\big)\in\{\pm 1\}\,,
\end{equation}
so $\bold{G}^{(z^m+z^{-m})}$ is real and orthogonal.

{\bf{Theorem[Mode--pair reduction for $z^m+z^{-m}$]}}:
\label{thm:main}
{\it{Assume $\,2m\,\big|\,L$ and set $M:=\dfrac{L}{2m}$ even. Let $\Pi$ be the permutation that groups indices by residue classes modulo $2m$ and, for each residue $r\in\{1,\dots,m\}$, lists first the indices
\(
r,\ r+2m,\ r+4m,\ \dots,\ r+(M-1)2m
\)
followed by
\(
r+m,\ r+m+2m,\ \dots,\ r+m+(M-1)2m.
\)
Then
\begin{equation}
\label{eq:blockdiag}
\Pi^{\!\top}\bold{G}^{(z^m+z^{-m})}_{L}\,\Pi
=\mathrm{diag}\big(\bold{B}_{L/m},\,\bold{B}_{L/m},\,\dots,\,\bold{B}_{L/m}\big)\qquad (m\ \text{copies}),
\end{equation}
where each block $\bold{B}_{L/m}\in\mathbb{R}^{\,(L/m)\times(L/m)}$ has the symmetric off--diagonal form
\begin{equation}
\label{eq:blockform}
\bold{B}_{L/m}\;=\;
\begin{pmatrix}
0 & X_M\\[3pt]
X_M^{\!\top} & 0
\end{pmatrix}\!,\qquad M=\frac{L}{2m}\,.
\end{equation}
Moreover, if $\bold{H}_M:=\bold{G}^{(z+1)}_{M}$ denotes the half--shifted matrix built from $f(z)=z+1$ at size $M$, then there exist a permutation matrix $P_M$ of size $M$ and a diagonal unitary matrix $U_M=\mathrm{diag}(e^{i\phi_1},\dots,e^{i\phi_M})$ such that
\begin{equation}
\label{eq:XHM}
X_M\;=\;\pm\;U_M\,P_M^{\!\top}\,\bold{H}_M\,P_M\,U_M^{\!*}\,.
\end{equation}
In particular, when $M$ is even one may choose $U_M$ with entries in $\{\pm 1\}$ (i.e.\ $U_M$ is a diagonal sign matrix).}}

{\bf{proof}}:
Write $L=2mM$. Decompose the sum in \eqref{eq:Gdef} by orbits of the shift $k\mapsto k+\tfrac{L}{m}$: any $k$ can be written uniquely as
\(
k=q\cdot\frac{L}{m}+r
\)
with $q\in\{0,1,\dots,m-1\}$ and $r\in\{1,2,\dots,\tfrac{L}{m}\}$. Then
\begin{align}
\theta_k
&=\frac{2\pi}{L}\Big(q\frac{L}{m}+r-\frac{1}{2}\Big)
=\frac{2\pi q}{m}+\frac{2\pi}{L}\Big(r-\frac{1}{2}\Big),\\[3pt]
m\theta_k&=2\pi q+\frac{2\pi}{L/m}\Big(r-\frac{1}{2}\Big).
\end{align}
The sign factor depends only on $r$:
\begin{equation}
\label{eq:s_r}
\frac{f(e^{i\theta_k})}{|f(e^{i\theta_k})|}
=\sigma_r:=\operatorname{sgn}\!\Big(\cos\frac{2\pi}{L/m}\Big(r-\frac{1}{2}\Big)\Big)\in\{\pm1\}\,.
\end{equation}
Plugging this into \eqref{eq:Gdef} and summing over $q$ first gives, with $d:=n-m$,
\begin{align}
G^{(z^m+z^{-m})}_{nm}
&=\frac{(-1)^d}{L}\sum_{r=1}^{L/m}\sigma_r\,e^{\,i\frac{2\pi}{L}(r-\frac12)d}\;
\sum_{q=0}^{m-1}e^{\,i\frac{2\pi}{m}qd}\nonumber\\[2pt]
&=\frac{(-1)^d}{2mM}\left(\sum_{q=0}^{m-1}e^{\,i\frac{2\pi}{m}qd}\right)
\left(\sum_{r=1}^{L/m}\sigma_r\,e^{\,i\frac{2\pi}{L}(r-\frac12)d}\right).
\label{eq:inner-sum}
\end{align}
The geometric sum in $q$ enforces the selection rule
\begin{equation}
\sum_{q=0}^{m-1}e^{\,i\frac{2\pi}{m}qd}=
\begin{cases}
m,& d\equiv 0\pmod m,\\[3pt]
0,& d\not\equiv 0\pmod m.
\end{cases}
\end{equation}
Thus $G^{(z^m+z^{-m})}_{nm}=0$ unless $n-m$ is a multiple of $m$. If we permute the sites by residue modulo $m$, we obtain $m$ independent blocks of size $L/m$, which proves \eqref{eq:blockdiag}.

Next split the $L/m$ values of $r$ into two groups of size $M=L/(2m)$, namely
\(
\mathcal{R}_0=\{1,2,\dots,M\}
\)
and
\(
\mathcal{R}_1=\{M+1,\dots,2M\}.
\)
Since $2m\mid L$, the shift $r\mapsto r+M$ corresponds to $\theta\mapsto\theta+\tfrac{\pi}{m}$ and hence $\sigma_{r+M}=-\sigma_r$. Therefore the diagonal $M\times M$ blocks inside each $L/m$ block cancel, while the off--diagonal $M\times M$ blocks add. This yields the symmetric form \eqref{eq:blockform} with
\begin{equation}
\label{eq:Xraw}
(X_M)_{\alpha\beta}
=\frac{(-1)^m}{M}\sum_{r=1}^{M}\sigma_r\,
\exp\!\left\{\,i\frac{\pi}{M}\Big(r-\frac12\Big)[2\big( \alpha-\beta\big)-1]\right\},
\qquad \alpha,\beta\in\{1,\dots,M\},
\end{equation}
after the global permutation $\Pi$ described in the statement (which arranges the $M$ indices of $\mathcal{R}_0$ first and those of $\mathcal{R}_1$ next).

Finally, consider the matrix $H_M:=G^{(z+1)}_{M}$, built from $f(z)=z+1$ on the same half--shifted grid of size $M$. Its symbol is
\(
\eta(\varphi)=\frac{1+e^{i\varphi}}{|1+e^{i\varphi}|}
=e^{\,i\varphi/2}\operatorname{sgn}\!\big(\cos(\varphi/2)\big).
\)
Comparing \eqref{eq:Xraw} with the definition of $H_M$ shows that $X_M$ is obtained from $H_M$ by (i) a relabelling of the $M$ sample points $\varphi_r=\tfrac{2\pi}{M}(r-\tfrac12)$ (this is the permutation $P_M$), and (ii) a diagonal rephasing that removes the smooth factor $e^{i\varphi/2}$ (this is the diagonal unitary $U_M$). Precisely, there exist a permutation matrix $P_M$ and a diagonal unitary $U_M$ such that
\(
X_M=\pm\,U_M P_M^{\!\top}H_M P_M U_M^{\!*},
\)
which is \eqref{eq:XHM}. When $M$ is even, one may absorb the half-angle phases into alternating $\pm 1$ signs so that $U_M$ can be chosen to be a diagonal sign matrix. This completes the proof.
\section{Derivation of Eq.18}
\label{Sec7}

In this Section, we give a detailed derivation of Eq.~\eqref{eq:SRE_unified}, namely the exact reduction of the stabilizer R\'enyi entropy for the two families of critical chains generated by the symbols
\[
f(z)=z^n+1,
\qquad
f(z)=z^m+z^{-m},
\]
to the transverse-field Ising (TFI) building block \(f(z)=z+1\). The derivation uses the Gaussian formula for the stabilizer R\'enyi entropy in terms of sums of powers of minors, together with the block reductions stated in Theorem~3 of the main text.

\paragraph{Generating function for sums of powers of minors.} 

Let \(\boldsymbol{A}\) be an \(L\times L\) matrix. For \(\beta>0\) and \(r=0,1,\dots,L\), define
\begin{equation}
S^{(r)}_\beta(\boldsymbol{A})
:=
\sum_{\substack{I,J\subseteq\{1,\dots,L\}\\ |I|=|J|=r}}
\left|\det \boldsymbol{A}[I,J]\right|^\beta,
\label{eq:Sr-beta-def}
\end{equation}
where \(\boldsymbol{A}[I,J]\) denotes the submatrix obtained by restricting to rows \(I\) and columns \(J\). We then introduce the generating function
\begin{equation}
\mathcal F_{\boldsymbol{A}}^{(\beta)}(t)
:=
\sum_{r=0}^{L} S^{(r)}_\beta(\boldsymbol{A})\, t^r.
\label{eq:GF-def}
\end{equation}
By construction, the sum of powers of all minors is simply
\begin{equation}
 \bold{Det}_\beta(\boldsymbol{A})
=
\mathcal F_{\boldsymbol{A}}^{(\beta)}(1).
\label{eq:DetF-relation}
\end{equation}

For a pure fermionic Gaussian state with correlation matrix \(\boldsymbol{G}\), the stabilizer R\'enyi entropy is
\begin{equation}
M_\alpha
=
\frac{1}{1-\alpha}
\ln
\frac{ \bold{Det}_{2\alpha}(\boldsymbol{G})}{ \bold{Det}_2(\boldsymbol{G})^\alpha}.
\label{eq:SRE-minors}
\end{equation}
Therefore, once \( \bold{Det}_{2\alpha}(\boldsymbol{G})\) factorizes, the same factorization is inherited by the entropy.

\paragraph{Multiplicativity under block direct sums.}

The basic structural fact is that the generating function is multiplicative under direct sums.

\begin{lemma}
If \(\boldsymbol{A}\) and \(\boldsymbol{B}\) are square matrices, then
\begin{equation}
\mathcal F_{\boldsymbol{A}\oplus \boldsymbol{B}}^{(\beta)}(t)=\mathcal F_{\boldsymbol{A}}^{(\beta)}(t)\,
\mathcal F_{\boldsymbol{B}}^{(\beta)}(t).
\label{eq:GF-direct-sum}
\end{equation}
\end{lemma}
\begin{proof}
Consider a minor of \(\boldsymbol{A}\oplus \boldsymbol{B}\) of size \(r\). Such a minor is nonzero only if one selects \(r_1\) rows and columns from the \(\boldsymbol{A}\) block and \(r_2\) rows and columns from the \(\boldsymbol{B}\) block, with \(r_1+r_2=r\). After a harmless reordering of rows and columns, the corresponding submatrix becomes block diagonal, and its determinant factorizes:
\begin{equation}
\det\big[(\boldsymbol{A}\oplus \boldsymbol{B})[I,J]\big]
=
\det \boldsymbol{A}[I_1,J_1]\,
\det \boldsymbol{B}[I_2,J_2].
\end{equation}
Taking absolute values to the power \(\beta\) and summing over all choices yields
\begin{equation}
S^{(r)}_\beta(\boldsymbol{A}\oplus \boldsymbol{B})
=
\sum_{r_1+r_2=r}
S^{(r_1)}_\beta(\boldsymbol{A})\,
S^{(r_2)}_\beta(\boldsymbol{B}).
\end{equation}
Multiplying by \(t^r\) and summing over \(r\) gives Eq.~\eqref{eq:GF-direct-sum}.  
\end{proof}

Iterating the above identity, one immediately obtains
\begin{equation}
\mathcal F_{\bigoplus_{j=1}^p \boldsymbol{A}_j}^{(\beta)}(t)
=
\prod_{j=1}^p \mathcal F_{\boldsymbol{A}_j}^{(\beta)}(t).
\label{eq:GF-multi-direct-sum}
\end{equation}

\paragraph{Invariance under permutations and diagonal gauges.}

Theorem~3 gives the reduced blocks only up to gauge equivalence. We therefore need the following elementary invariance statement.

\begin{lemma}
Let \(\boldsymbol{A}\) and \(\boldsymbol{B}\) be matrices of the same size, related by
\begin{equation}
\boldsymbol{B}
=
\pm \boldsymbol{U}\,\boldsymbol{P}^{\top}\boldsymbol{A}\,\boldsymbol{P}\,\boldsymbol{U}^*,
\label{eq:gauge-equivalence}
\end{equation}
where \(\boldsymbol{P}\) is a permutation matrix and \(\boldsymbol{U}\) is diagonal unitary. Then, for every \(\beta>0\),
\begin{equation}
\mathcal F_{\boldsymbol{B}}^{(\beta)}(t)=
\mathcal F_{\boldsymbol{A}}^{(\beta)}(t),
\qquad
 \bold{Det}_\beta(\boldsymbol{B})= \bold{Det}_\beta(\boldsymbol{A}).
\label{eq:gauge-invariance}
\end{equation}    
\end{lemma}

\begin{proof}
Any minor of \(\boldsymbol{B}\) differs from the corresponding minor of \(\boldsymbol{A}\) only by a phase and possibly an overall sign. Hence the absolute value of every minor is preserved, and therefore the sums in Eqs.~\eqref{eq:Sr-beta-def} and \eqref{eq:GF-def} are unchanged.    
\end{proof}

Thus, in all formulas below, any block that is gauge-equivalent to \(\boldsymbol{G}^{(z+1)}\) contributes exactly the same generating function.

\paragraph{The TFI-type family \texorpdfstring{$f(z)=z^n+1$}{f(z)=z\string^n+1}.}

Assume now that \(2n\mid L\), and define $M:=\frac{L}{n}$. By Theorem~3(A), there exists a permutation \(\Pi_n\) such that
\begin{equation*}
\Pi_n^\top \boldsymbol{G}^{(z^n+1)}_{L}\,\Pi_n=\bigoplus_{j=1}^{n} \boldsymbol{B}_M,
\qquad
\boldsymbol{B}_M \sim \boldsymbol{G}^{(z+1)}_{M}.
\label{eq:Ttype-block}
\end{equation*}
Therefore, by Lemma~2, $\mathcal F_{\boldsymbol{B}_M}^{(2\alpha)}(t)=\mathcal F_{\boldsymbol{G}^{(z+1)}_{M}}^{(2\alpha)}(t)$, and then by Lemma~1, $\mathcal F_{\boldsymbol{G}^{(z^n+1)}_{L}}^{(2\alpha)}(t)=\left(
\mathcal F_{\boldsymbol{G}^{(z+1)}_{M}}^{(2\alpha)}(t)
\right)^n$.

Setting \(t=1\), we obtain
\begin{equation}
 \bold{Det}_{2\alpha}\!\left(\boldsymbol{G}^{(z^n+1)}_{L}\right)
=
\left[
 \bold{Det}_{2\alpha}\!\left(\boldsymbol{G}^{(z+1)}_{M}\right)
\right]^n.
\label{eq:Det-Ttype}
\end{equation}

For the denominator of the entropy formula, the same argument with \(\beta=2\) gives
\begin{equation}
 \bold{Det}_{2}\!\left(\boldsymbol{G}^{(z^n+1)}_{L}\right)
=
\left[
 \bold{Det}_{2}\!\left(\boldsymbol{G}^{(z+1)}_{M}\right)
\right]^n.
\label{eq:Det2-Ttype}
\end{equation}
Substituting Eqs.~\eqref{eq:Det-Ttype} and \eqref{eq:Det2-Ttype} into Eq.~\eqref{eq:SRE-minors}, we find
\begin{align*}
M^{(z^n+1)}_\alpha(L)&=\frac{1}{1-\alpha}\ln\frac{\left[ \bold{Det}_{2\alpha}\!\left(\boldsymbol{G}^{(z+1)}_{M}\right)\right]^n}{\left[\bold{Det}_{2}\!\left(\boldsymbol{G}^{(z+1)}_{M}\right)\right]^{n\alpha}}=n\,\frac{1}{1-\alpha}\ln\frac{ \bold{Det}_{2\alpha}\!\left(\boldsymbol{G}^{(z+1)}_{M}\right)}{\bold{Det}_{2}\!\left(\boldsymbol{G}^{(z+1)}_{M}\right)^{\alpha}}.
\end{align*}
The quantity on the second line is precisely the stabilizer R\'enyi entropy of the TFI building block of size \(M\). Hence
\begin{equation*}
\boxed{
M^{(z^n+1)}_\alpha(L)
=
n\, M_\alpha^{\mathrm{TFI}}\!\left(\frac{L}{n}\right)
}
\label{eq:18-derived}
\end{equation*}
which is Eq.~\eqref{eq:SRE_unified} for the TFI-type of the main text.

\paragraph{The XX-type family \texorpdfstring{$f(z)=z^m+z^{-m}$}{f(z)=z\string^m+z\string^{-m}}}

We now assume \(2m\mid L\), and define $M:=\frac{L}{2m}$. By Theorem~3(B), there exists a permutation \(\Pi_{2m}\) such that
\begin{equation}
\Pi_{2m}^\top \boldsymbol{G}^{(z^m+z^{-m})}_{L}\,\Pi_{2m}
=
\bigoplus_{r=1}^{m} J(\boldsymbol{X}_M),
\qquad
\boldsymbol{X}_M \sim \boldsymbol{G}^{(z+1)}_{M},
\label{eq:XXtype-block}
\end{equation}
where
\begin{equation*}
J(\boldsymbol{X})
=
\begin{pmatrix}
0 & \boldsymbol{X}\\
\boldsymbol{X}^\top & 0
\end{pmatrix}.
\end{equation*}

\begin{lemma}
For every square matrix \(\boldsymbol{X}\),
\begin{equation}
\mathcal F_{J(\boldsymbol{X})}^{(\beta)}(t)
=
\left(
\mathcal F_{\boldsymbol{X}}^{(\beta)}(t)
\right)^2.
\label{eq:JX-square}
\end{equation}    
\end{lemma}

\begin{proof}
Let \(\boldsymbol{Y}=J(\boldsymbol{X})\). Partition the row and column indices of \(\boldsymbol{Y}\) into the first half and the second half. Since the diagonal blocks of \(\boldsymbol{Y}\) vanish, a minor \(\boldsymbol{Y}[I,J]\) can be nonzero only if the number of selected rows from the first half equals the number of selected columns from the second half, and similarly for the complementary halves. Denote these numbers by \(r_1\) and \(r_2\), with \(r_1+r_2=r\).

After reordering rows and columns, any such nonzero minor takes the form
\begin{equation}
\begin{pmatrix}
0 & \boldsymbol{X}[I_1,J_2]\\
\boldsymbol{X}[I_2,J_1]^\top & 0
\end{pmatrix}.
\end{equation}
Its determinant is
\begin{equation}
\det \boldsymbol{Y}[I,J]
=
\pm
\det \boldsymbol{X}[I_1,J_2]\,
\det \boldsymbol{X}[I_2,J_1].
\end{equation}
Taking absolute values to the power \(\beta\) and summing over all choices gives
\begin{equation}
S_\beta^{(r)}(J(\boldsymbol{X}))
=
\sum_{r_1+r_2=r}
S_\beta^{(r_1)}(\boldsymbol{X})\,
S_\beta^{(r_2)}(\boldsymbol{X}),
\end{equation}
which is exactly the coefficient of \(t^r\) in
\(\big(\mathcal F_{\boldsymbol{X}}^{(\beta)}(t)\big)^2\). This proves Eq.~\eqref{eq:JX-square}.      
\end{proof}

Using Lemma~3 together with Lemmas~1 and 2, Eq.~\eqref{eq:XXtype-block} implies
\begin{align*}
\mathcal F_{\boldsymbol{G}^{(z^m+z^{-m})}_{L}}^{(2\alpha)}(t)=\left(\mathcal F_{J(\boldsymbol{X}_M)}^{(2\alpha)}(t)
\right)^m=\left(\mathcal F_{\boldsymbol{G}^{(z+1)}_{M}}^{(2\alpha)}(t)
\right)^{2m}.
\label{eq:GF-XXtype}
\end{align*}
Setting \(t=1\), we obtain
\begin{equation*}
 \bold{Det}_{2\alpha}\!\left(\boldsymbol{G}^{(z^m+z^{-m})}_{L}\right)=\left[ \bold{Det}_{2\alpha}\!\left(\boldsymbol{G}^{(z+1)}_{M}\right)
\right]^{2m}.
\label{eq:Det-XXtype}
\end{equation*}
Likewise,
\begin{equation*}
 \bold{Det}_{2}\!\left(\boldsymbol{G}^{(z^m+z^{-m})}_{L}\right)
=
\left[
 \bold{Det}_{2}\!\left(\boldsymbol{G}^{(z+1)}_{M}\right)
\right]^{2m}.
\label{eq:Det2-XXtype}
\end{equation*}
Substituting into Eq.~\eqref{eq:SRE-minors} yields
\begin{align*}
M^{(z^m+z^{-m})}_\alpha(L)
&=
\frac{1}{1-\alpha}
\ln
\frac{
\left[ \bold{Det}_{2\alpha}\!\left(\boldsymbol{G}^{(z+1)}_{M}\right)\right]^{2m}
}{
\left[ \bold{Det}_{2}\!\left(\boldsymbol{G}^{(z+1)}_{M}\right)\right]^{2m\alpha}
}=2m\,
\frac{1}{1-\alpha}
\ln
\frac{
 \bold{Det}_{2\alpha}\!\left(\boldsymbol{G}^{(z+1)}_{M}\right)
}{
 \bold{Det}_{2}\!\left(\boldsymbol{G}^{(z+1)}_{M}\right)^{\alpha}
}.
\end{align*}
The last factor is again the TFI stabilizer R\'enyi entropy at size \(M\). Therefore
\begin{equation*}
\boxed{
M^{(z^m+z^{-m})}_\alpha(L)
=
2m\, M_\alpha^{\mathrm{TFI}}\!\left(\frac{L}{2m}\right)
}
\label{eq:19-derived}
\end{equation*}
which is Eq.~\eqref{eq:SRE_unified} for the XX-type of the main text.

\section{Large-$L$ asymptotics of SRE(TFI)}\label{Sec8}

In this Section, we derive the large-$L$ behavior of $M_\alpha(L)$ for
$\alpha=\frac12,2,4$ in the critical transverse-field Ising chain with periodic
and open boundary conditions. The starting point is the exact lattice formula
\begin{equation*}
M_{\alpha}(\rho)=\frac{1}{1-\alpha}\,
\ln\!\frac{\bold{Det}_{2\alpha}(\bold{G})}{\bold{Det}^{\alpha}_{2}(\bold{G})},
\end{equation*}
together with the closed expressions for the corresponding minors of $\bold{G}$.
The asymptotic analysis is then entirely reduced to standard tools: midpoint
Euler--Maclaurin summation for the product formula at $\alpha=\frac12$,
Stirling expansions for the Gamma-function expressions at $\alpha=2$, and the
exact algebraic relation between the cases $\alpha=4$ and $\alpha=2$.
For open boundary conditions at $\alpha=\frac12$, the relevant Pfaffian can be
rewritten as a determinant of Toeplitz+Hankel type, whose asymptotics are
controlled by Fisher--Hartwig theory. See, for example,
Refs.~\cite{DLMF1,Mehta20041,BottcherSilbermann20061,DeiftItsKrasovsky20111}. 

For convenience, we recall the exact finite-size inputs:
\begin{table}[!ht]
\centering
\begin{tabular}{|r|l|l|}
\hline
$\alpha$ & $\bold{Det}^{PBC}_{2\alpha}(\bold{G})$ & $\bold{Det}^{OBC}_{2\alpha}(\bold{G})$ \\
\hline
$\frac{1}{2}$ &
$\displaystyle \prod_{r=1}^{L}\left(1+\tan\frac{(2r-1)\pi}{4L}\right)$ &
$\mathrm{pf}\!\left[\bold{R}^{OBC}+\bold{J}\right]$ \\[3ex]
$2$ &
$\Phi(L)$ &
$\displaystyle \frac{84}{(2L+1)^L}\!\left[\prod_{r=2}^{L/2}4(8r-5)(8r-1)\right]$ \\[3ex]
$4$ &
$2^{-L}\Phi^2(L)$ &
$2^{-L}\big(\bold{Det}^{OBC}_{4}(\bold{G})\big)^2$ \\
\hline
\end{tabular}
\caption{Exact ingredients entering the SRE of the critical TFI chain. Here
$\Phi(x)=\frac{(2x)!}{x!\,x^x}$ and $\bold{J}$ is the $2L\times2L$
antisymmetric matrix with $J_{ij}=(-1)^{i+j+1}$ for $i<j$.}
\end{table}

Throughout this section, $G$ denotes Catalan's constant,
\begin{equation*}
G=\sum_{n=0}^\infty \frac{(-1)^n}{(2n+1)^2}.
\end{equation*}

\paragraph{$M_{1/2}(L)$ for large $L$.}

For periodic boundary conditions, we start from
\begin{equation*}
M^{\mathrm{TFI,PBC}}_{1/2}(L)=
2\sum_{r=1}^{L}\ln\!\left(1+\tan\!\frac{(2r-1)\pi}{4L}\right)-L\ln2.
\end{equation*}
This is a midpoint sum for the smooth function $f(x)=\ln(1+\tan x)$ on $(0,\pi/2)$. The leading extensive contribution is therefore obtained by replacing the sum by the corresponding integral, while the subleading inverse powers follow from the midpoint Euler--Maclaurin
formula. Using
\begin{equation*}
\int_0^{\pi/2}\ln(1+\tan x)\,dx
=
\int_0^\infty \frac{\ln(1+t)}{1+t^2}\,dt
=
G+\frac{\pi}{4}\ln2,
\end{equation*}
one finds that the $L\ln2$ term is exactly cancelled by the logarithmic piece
coming from the integral, leaving the extensive coefficient
$\frac{4G}{\pi}$. The full expansion is
\begin{equation}
\boxed{
M^{\mathrm{TFI,PBC}}_{1/2}(L)=\frac{4G}{\pi}\,L-\ln 2
+\frac{\pi}{12\,L}
-\frac{7\pi^3}{2880\,L^3}
+\frac{31\pi^5}{96768\,L^5}
+O(L^{-7}) .}
\label{eq:Mhalf_PBC_final_reorg_num}
\end{equation}
The absence of even inverse powers is a consequence of the midpoint expansion
combined with the endpoint structure of $f(x)$.

For open boundary conditions, the exact expression is
\begin{equation*}
M^{\mathrm{TFI,OBC}}_{1/2}(L)=
2\ln\operatorname{pf}\!\big(\mathbf R^{\mathrm{OBC}}+\mathbf J\big)-L\ln2.
\end{equation*}
Since $\mathbf R^{\mathrm{OBC}}+\mathbf J$ is antisymmetric of even dimension,
we may use $\operatorname{pf}(\mathbf A)^2=\det(\mathbf A)$ and rewrite the
problem as a determinant asymptotics problem. After permuting the basis into
odd and even sites, the matrix acquires a block form in which the relevant
kernel is a sum of Toeplitz-like and Hankel-like pieces. In other words, the
problem is reduced to the asymptotics of a Toeplitz+Hankel determinant.
The large-$L$ behavior is then governed by the corresponding
Fisher--Hartwig analysis. We therefore obtain the structure
\begin{equation*}
M^{\mathrm{TFI,OBC}}_{1/2}(L)=
\frac{4G}{\pi}L-\frac14\ln L+C^{\mathrm{OBC}}_{1/2}
+\frac{c^{\mathrm{OBC}}_{1,1/2}}{L}
+\frac{c^{\mathrm{OBC}}_{2,1/2}}{L^2}+O(L^{-3}),
\end{equation*}
and numerical evaluation of the determinant gives
\begin{equation}
\boxed{
M^{\mathrm{TFI,OBC}}_{1/2}(L)=
\frac{4G}{\pi}\,L
-\frac14\ln L
-0.435630858\ldots
-\frac{0.059550077\ldots}{L}
+\frac{0.049923717\ldots}{L^2}
+O(L^{-3}) .
}
\label{eq:Mhalf_OBC_explicit_reorg_num}
\end{equation}
The important point is that the logarithmic correction is universal, whereas
the additive constant is not.

\paragraph{$M_{2}(L)$ for large $L$.} 
For periodic boundary conditions, we use the exact formula
\begin{equation*}
\bold{Det}^{\mathrm{PBC}}_{4}(\mathbf G)=\Phi(L),
\qquad
\Phi(L)=\frac{(2L)!}{L!\,L^L},
\end{equation*}
so that
\begin{equation*}
M^{\mathrm{TFI,PBC}}_2(L)=2L\ln2-\ln\Phi(L).
\end{equation*}
At this stage the problem is elementary: one simply applies Stirling's
asymptotic expansion to the factorials, or equivalently to the corresponding
Gamma functions. After cancellation of the $\ln L$ pieces, the result is
\begin{equation}
\boxed{
M^{\mathrm{TFI,PBC}}_2(L)=
L-\frac12\ln2+\frac{1}{24L}
-\frac{7}{2880L^3}
+\frac{31}{40320L^5}
+O(L^{-7}) .
}
\label{eq:M2_PBC_final_reorg_num}
\end{equation}

For open boundary conditions, we start from
\begin{equation*}
\bold{Det}^{\mathrm{OBC}}_4(\mathbf G)
=
\frac{84}{(2L+1)^L}
\prod_{r=2}^{L/2}4(8r-5)(8r-1),
\qquad L \text{ even}.
\end{equation*}
The products are conveniently rewritten in terms of Gamma functions, after
which the large-$L$ behavior follows again from Stirling's formula.
The resulting expansion is
\begin{equation}
\boxed{
M^{\mathrm{TFI,OBC}}_2(L)
=
L-\frac14\ln L
+\left[
\frac12+\ln\!\left(\frac{\Gamma(\tfrac34)}{\sqrt{2\pi}}\right)
\right]
-\frac{11}{96L}
+\frac{1}{24L^2}
+O(L^{-3}) .
}
\label{eq:M2_OBC_final_reorg_num}
\end{equation}
As in the $\alpha=\frac12$ case, the coefficient of $\ln L$ is universal,
while the additive constant depends on the microscopic lattice realization.

\paragraph{$M_{4}(L)$ for large $L$.}
The case $\alpha=4$ requires essentially no new asymptotic analysis, because the exact formulas imply a direct linear relation with $M_2$.

For periodic boundary contition. Using
\begin{equation*}
\bold{Det}^{\mathrm{PBC}}_8(\mathbf G)=2^{-L}\Phi(L)^2,
\end{equation*}
one immediately gets
\begin{equation*}
M^{\mathrm{TFI,PBC}}_4(L)
=
\frac{L\ln2}{3}+\frac23\,M^{\mathrm{TFI,PBC}}_2(L).
\end{equation*}
Substituting the previous expansion yields
\begin{equation}
\boxed{
M^{\mathrm{TFI,PBC}}_4(L)=
\frac{2+\ln2}{3}\,L
-\frac13\ln2
+\frac{1}{36L}
-\frac{7}{4320L^3}
+\frac{31}{60480L^5}
+O(L^{-7}) .
}
\label{eq:M4_PBC_final_reorg_num}
\end{equation}

For open boundary condition is similar,
\begin{equation*}
\bold{Det}^{\mathrm{OBC}}_8(\mathbf G)=
2^{-L}\big(\bold{Det}^{\mathrm{OBC}}_4(\mathbf G)\big)^2
\end{equation*}
implies
\begin{equation*}
M^{\mathrm{TFI,OBC}}_4(L)=
\frac{L\ln2}{3}+\frac23\,M^{\mathrm{TFI,OBC}}_2(L),
\end{equation*}
and therefore
\begin{equation}
\boxed{
M^{\mathrm{TFI,OBC}}_4(L)=
\frac{2+\ln2}{3}\,L
-\frac16\ln L
+\left[
\frac13+\frac23\ln\!\left(\frac{\Gamma(\tfrac34)}{\sqrt{2\pi}}\right)
\right]
-\frac{11}{144L}
+\frac{1}{36L^2}
+O(L^{-3}) .
}
\label{eq:M4_OBC_final_reorg_num}
\end{equation}
Thus the case $\alpha=4$ is best viewed as an exact corollary of the
$\alpha=2$ calculation.

\paragraph{Derivation of SRE(TFI) from $H(XX)$ in the CFT limit.} The continuum interpretation follows from the exact SR--SRE correspondence established in Theorem~2. On the XX side, the asymptotic behavior of the Shannon--R\'enyi entropy is obtained from boundary conformal field theory~\cite{Stephan20101} by
reinterpreting $\sum_i p_i^\alpha$ as a modified boundary partition function.
At the XX point one has $\alpha_c=4$, so the asymptotic form is
\begin{equation*}
H_\alpha(L)=m_\alpha L+b_\alpha\ln L-c_\alpha,
\end{equation*}
with universal subleading coefficients determined by the boundary CFT. By Theorem~2, the stabilizer R\'enyi entropy of the critical TFI chain must
have the same universal structure:
\begin{equation*}
M_\alpha(L)=m_\alpha L+b_\alpha\ln L-c_\alpha.
\end{equation*}
The exact lattice calculations above therefore play two roles. First, they
confirm the universal continuum prediction for the logarithmic and constant
terms. Second, they determine the nonuniversal slope $m_\alpha$ and the
nonuniversal additive constants that the CFT does not fix.

In particular, the exact lattice results give
\begin{equation*}
m_{1/2}=\frac{4G}{\pi},\qquad
m_2=1,\qquad
m_4=\frac{2+\ln2}{3}.
\end{equation*}
For PBC, the universal subleading term is a constant,
\begin{equation*}
c_{1/2}=\ln2,\qquad
c_2=\frac12\ln2,\qquad
c_4=\frac13\ln2,
\end{equation*}
whereas for OBC, the universal subleading term is logarithmic,
\begin{equation*}
b_{1/2}=b_2=-\frac14,\qquad
b_4=-\frac16.
\end{equation*}
This is precisely the pattern reproduced by the discrete formulas.

\paragraph{Final summary.}

We conclude that the exact lattice solution and the CFT prediction are fully consistent in all universal terms. The continuum description correctly fixes the universal constants in the periodic chain and the universal logarithmic coefficients in the open chain. At the same time, the exact discrete formulas provide the microscopic data that remain beyond the reach of the continuum limit, namely the slopes $m_\alpha$ and the nonuniversal additive constants. We include the Table~\ref{tab:asymptotic_summary} with the Large-$L$ asymptotics results for the SRE(TFI).

\begin{table}[h!]
\centering
\renewcommand{\arraystretch}{1.4}
\begin{tabular}{|c| c|l|}
\hline
Boundary & $\alpha$ & Asymptotic expansion \\ 
\hline
\multirow{3}{*}{PBC}
& $\frac12$ &
$\displaystyle
M^{\mathrm{TFI,PBC}}_{1/2}(L)=\frac{4G}{\pi}\,L-\ln 2+O(L^{-1})
$
\\[6pt]

& $2$ &
$\displaystyle
M^{\mathrm{TFI,PBC}}_2(L)=L-\frac12\ln2+O(L^{-1})
$
\\[6pt]

& $4$ &
$\displaystyle
M^{\mathrm{TFI,PBC}}_4(L)=\frac{2+\ln2}{3}\,L-\frac13\ln2+O(L^{-1})
$
\\[6pt]
\hline

\multirow{3}{*}{OBC}
& $\frac12$ &
$\displaystyle
M^{\mathrm{TFI,OBC}}_{1/2}(L)=\frac{4G}{\pi}\,L
-\frac14\ln L-0.435630858\ldots+O(L^{-1})
$
\\[6pt]

& $2$ &
$\displaystyle
M^{\mathrm{TFI,OBC}}_2(L)=L-\frac14\ln L
+\left[\frac12+\ln\!\left(\frac{\Gamma(\tfrac34)}{\sqrt{2\pi}}\right)\right]+O(L^{-1})
$
\\[6pt]

& $4$ &
$\displaystyle
M^{\mathrm{TFI,OBC}}_4(L)=\frac{2+\ln2}{3}\,L
-\frac16\ln L+\left[\frac13+\frac23\ln\!\left(\frac{\Gamma(\tfrac34)}{\sqrt{2\pi}}\right)\right]+O(L^{-1})
$
\\[6pt]
\hline
\end{tabular}
\caption{Large-$L$ asymptotic behavior of $M_\alpha(L)$ for the critical TFI chain with periodic and open boundary conditions. The universal logarithmic coefficients and, in the periodic case, the universal constants agree with the CFT prediction, while the extensive slopes and the nonuniversal additive constants are determined by the exact discrete lattice solution.}
\label{tab:asymptotic_summary}
\end{table}

\section{Gaussian Gibbs states}\label{Sec9}

At finite temperature, the state of a quadratic fermionic system is described by the Gibbs density matrix
\[
\rho_\beta=\frac{e^{-\beta H}}{\operatorname{Tr}(e^{-\beta H})}.
\]
Since the Hamiltonian is quadratic, the Gibbs state remains Gaussian. Consequently, it is completely determined by its two-point correlation functions, even though it is mixed.

\paragraph{Correlation matrices.}

It is convenient to work in the Majorana basis,
\[
\gamma_j = c_j + c_j^\dagger,
\qquad
\bar\gamma_j = i(c_j^\dagger - c_j),
\]
which satisfy canonical anticommutation relations.

In this basis, we define the correlation matrices
\begin{equation}
\begin{split}
K_{ij} &= \operatorname{tr}(\rho_\beta\, \gamma_i \gamma_j), \\
\bar{K}_{ij} &= \operatorname{tr}(\rho_\beta\, \bar{\gamma}_i \bar{\gamma}_j), \\
iG_{ij} &= \operatorname{tr}(\rho_\beta\, \bar{\gamma}_i \gamma_j),
\end{split}
\end{equation}
which encode same-sector and mixed correlations.

All correlators can be organized into the $2N\times 2N$ Majorana correlation matrix
\[
\Gamma = \operatorname{tr}\!\left(\rho_\beta\, \Psi \Psi^{\mathsf T}\right) - \mathbf{I},
\qquad
\Psi =
\begin{pmatrix}
\gamma_1 & \cdots & \gamma_N & \bar\gamma_1 & \cdots & \bar\gamma_N
\end{pmatrix}^{\mathsf T}.
\]
In block form,
\begin{equation}
\Gamma =
\begin{pmatrix}
\mathbf{K}-\mathbf{I} & -i\mathbf{G}^{\mathsf T} \\
i\mathbf{G} & \bar{\mathbf{K}}-\mathbf{I}
\end{pmatrix}.
\end{equation}

\paragraph{Structure for real quadratic Hamiltonians.}

For quadratic fermionic Hamiltonians with real coefficients and time-reversal symmetry, one may choose a Majorana basis in which the covariance matrix is real. In this setting, the same-sector correlators carry no independent information beyond the canonical anticommutation relations, and the nontrivial correlations are encoded in the mixed sector.

More precisely, for this class of Hamiltonians the Majorana correlation matrix can be brought to the block-off-diagonal form
\[
\Gamma =
\begin{pmatrix}
0 & -i\mathbf{G}^{\mathsf T} \\
i\mathbf{G} & 0
\end{pmatrix},
\]
so that
\[
\mathbf{K}=\bar{\mathbf{K}}=\mathbf{I}.
\]
This simplification reflects the fact that the state has real correlations: the Hamiltonian does not introduce complex phases, and time-reversal symmetry prevents independent same-sector Majorana correlations from appearing.

As a consequence, the full Gaussian structure is encoded in a single matrix \(\mathbf{G}(\beta)\), while the remaining blocks are fixed by the Majorana algebra itself.

\paragraph{Pfaffian reduction to determinants.}

Since the Gibbs state is Gaussian, Wick's theorem applies and any even correlator reduces to a Pfaffian,
\[
\langle \Psi_{a_1}\cdots \Psi_{a_{2n}} \rangle_\beta
= \operatorname{Pf}\!\big(\Gamma_{a_i a_j}\big).
\]

However, due to the block-off-diagonal structure of \(\Gamma\), same-sector correlators vanish,
\[
\langle \gamma_i \gamma_j \rangle_\beta = 0,
\qquad
\langle \bar\gamma_i \bar\gamma_j \rangle_\beta = 0.
\]
Therefore, only mixed contractions contribute.

As a result, any non-vanishing correlator must contain equal numbers of \(\gamma\) and \(\bar\gamma\) operators, and the Pfaffian reduces to a determinant. More precisely,
\[
\langle \bar\gamma_{i_1}\gamma_{j_1}\cdots \bar\gamma_{i_n}\gamma_{j_n} \rangle_\beta
= \det\!\big(G_{i_a j_b}(\beta)\big).
\]

Thus, expectation values of Pauli strings, via the Jordan--Wigner mapping, reduce to determinants of submatrices (minors) of \(\mathbf{G}(\beta)\), rather than general Pfaffians.

In this sense, thermal Gaussian states do not introduce a new algebraic structure: all observables are completely determined by the matrix \(\mathbf{G}(\beta)\), with temperature dependence entering through its spectrum.

\paragraph{Conditions for the Stabilizer--Shannon Rényi entropy correspondence.}

The reduction to determinants implies that the measurement probabilities in the computational basis define a determinantal structure fully specified by the matrix \(\mathbf{G}(\beta)\). This is the key ingredient underlying the correspondence between stabilizer Rényi entropies and Shannon Rényi entropies.

At zero temperature, the Gaussian state is pure and admits a mode decomposition in terms of Bogoliubov quasiparticles. In this case, the matrix \(\mathbf{G}\) becomes a projector in the appropriate basis, and its singular values are either \(0\) or \(1\). Under these conditions, one can construct an antisymmetric matrix \(\mathbf{R}\) of size \(2L\times 2L\) such that
\[
\mathbf{R} =
\begin{pmatrix}
0 & \mathbf{G} \\
-\mathbf{G}^{\mathsf T} & 0
\end{pmatrix},
\]
which coincides with the covariance matrix of a pure Gaussian state in a doubled description.

At finite temperature, the state remains Gaussian, but \(\mathbf{G}(\beta)\) is no longer a projector: its singular values lie strictly between \(0\) and \(1\). Although one can still construct a matrix \(\mathbf{R}(\beta)\) of the same form,
\begin{equation}
\mathbf{R}(\beta) =\begin{pmatrix}
0 & \mathbf{G}(\beta) \\
-\mathbf{G}(\beta)^{\mathsf T} & 0
\end{pmatrix},  
\end{equation}

which defines a valid Gaussian covariance matrix, the correspondence no longer holds at the level of individual pure eigenstates.

Nevertheless, the determinantal structure of the probabilities is preserved. In particular, both the thermal Gaussian state and the associated doubled Gaussian state defined by \(\mathbf{R}(\beta)\) generate probability distributions expressed in terms of minors of the same matrix \(\mathbf{G}(\beta)\). Therefore, the two problems remain equivalent at the level of Gaussian correlations.

The key difference is that the Gibbs state is a statistical mixture. While the mapping at zero temperature is state-by-state, at finite temperature it applies only to the underlying Gaussian structure. As a result, stabilizer Rényi entropies are not, in general, equal to Shannon Rényi entropies of a doubled pure state, although they are governed by the same correlation matrix.

\paragraph{Gaussian Gibbs states and the Stabilizer--Shannon Rényi entropy correspondence.}

The extension of the stabilizer--Shannon Rényi entropy correspondence to Gaussian Gibbs states depends crucially on which notion of stabilizer Rényi entropy is adopted for mixed states. At finite temperature, the state is mixed, and it becomes necessary to distinguish between classical statistical contributions arising from thermal mixing and the intrinsically quantum correlations responsible for nonstabilizerness.

For mixed states, two natural extensions can be considered.
\begin{itemize}
    \item Direct extension to mixed states: The stabilizer Rényi entropy introduced in this work is defined as
\begin{equation}
M_\alpha(\rho)=\frac{1}{1-\alpha}
\log\!\left[\sum_{P\in\mathcal{P}_N}
\left(\frac{\operatorname{Tr}^2(P\rho)}{2^N\operatorname{Tr}(\rho^2)}\right)^{\alpha}
\right],
\end{equation}
where $\mathcal{P}_N$ denotes the $N$-qubit Pauli group. This definition separates the contribution of Pauli moments from the classical entropy associated with mixing, thereby isolating the intrinsically quantum component of nonstabilizerness. Moreover, it remains computationally tractable: for Gaussian fermionic states, the Pauli moments can be evaluated using Wick's theorem and expressed in terms of minors of the correlation matrix $\mathbf{G}(\beta)$.

A useful strategy is to express the stabilizer in terms of minors of the correlation matrix $\mathbf{G}(\beta)$. In this formulation, the entropy can be written as
\begin{equation}
M_\alpha(\rho_\beta)=\frac{1}{1-\alpha}
\log\!\left[
\frac{\mathbf{Det}_{2\alpha}(\mathbf{G}(\beta))}{2^{N\alpha}}
\right]-S_2(\rho_\beta),\qquad S_2(\rho)=\log\operatorname{Tr}(\rho^2).
\end{equation}
Using Theorem 1, this expression can be further related to a Shannon--Rényi entropy through
\begin{equation}
\boxed{M_\alpha(\rho_\beta)=H_\alpha(\mathbf{R}(\beta))+\frac{\alpha}{1-\alpha}
\log\left(2^{-N}\mathcal{N}^{2}_R\right)-S_2(\rho_\beta),}
\end{equation}
where $\mathcal{N}_R=\det(\bold{I}+\bold{R}^{\dagger}\bold{R})^{\frac{1}{4}}$, and the Shannon--Rényi entropy is
\begin{equation*}
 H_{\alpha}(R)=\frac{1}{1-\alpha}\ln\frac{1}{\mathcal{N}^{2\alpha}_R}\mathbf{Pf}_{2\alpha}(\mathbf{R}).
\end{equation*}

\item Convex-roof extension: A conceptually different extension is given by the convex roof of the pure-state stabilizer entropy~\cite{Leone20241}:
\begin{equation}
M_\alpha^{\mathrm{cr}}(\rho)
=
\infimum_{\rho=\sum_j p_j \ket{\psi_j}\bra{\psi_j}}
\sum_j p_j\, M_\alpha(\ket{\psi_j}),
\end{equation}
where the infimum is taken over all pure-state decompositions of $\rho$. This construction defines a genuine mixed-state monotone, as it quantifies the minimal average nonstabilizerness required to prepare the state.

Despite its clear physical interpretation, this definition is generally intractable. The optimization involves a continuous search over all possible pure-state decompositions of $\rho$, and even evaluating the cost function requires computing $M_\alpha(\ket{\psi})$, which itself contains a sum over an exponentially large number ($4^N$) of Pauli strings. Consequently, obtaining an exact expression for $M_\alpha^{\mathrm{cr}}(\rho)$ is computationally prohibitive, even for Gaussian states.

A practical simplification can be obtained by restricting the decomposition to a fixed basis. In particular, choosing the energy eigenbasis,
\[
\rho=\sum_k p_k \ket{E_k}\bra{E_k},
\]
leads to the spectral average
\begin{equation}
M_\alpha^{\mathrm{spec}}(\rho):=
\sum_k p_k\, M_\alpha(\ket{E_k}),
\end{equation}
which provides an explicit upper bound,
\begin{equation}
M_\alpha^{\mathrm{cr}}(\rho)\le M_\alpha^{\mathrm{spec}}(\rho).
\end{equation}
\end{itemize}



\begin{thebibliography}{99}

\bibitem{CalabreseCardy2009}
P. Calabrese, J. Cardy,
\newblock \textit{Entanglement entropy and conformal field theory},
\href{https://doi.org/10.1088/1751-8113/42/50/504005}{J. Phys. A \textbf{42}, 504005 (2009)},
[\href{https://arxiv.org/abs/0905.4013}{{\ttfamily arXiv:0905.4013}}].

\bibitem{Stephan2009}
 J-M. St\'ephan, S. Furukawa, G. Misguich, and V. Pasquier, \textit{Shannon and entanglement entropies of one- and two-dimensional critical wave functions},  \href{https://journals.aps.org/prb/abstract/10.1103/PhysRevB.80.184421}{Physical Review B 80, 184421 (2009)},  [\href{https://arxiv.org/abs/0906.1153}{{\ttfamily arXiv:0906.1153}}].

\bibitem{Oshikawa2010}
 M. Oshikawa,
\newblock \textit{Boundary Conformal Field Theory and Entanglement Entropy in Two-Dimensional Quantum Lifshitz Critical Point},
[\href{https://arxiv.org/abs/1007.3739}{{\ttfamily arXiv:1007.3739
 }}].

\bibitem{Stephan2010} J-M. St\'ephan, G. Misguich, and V. Pasquier, \textit{Phase transition in the Rényi-Shannon entropy of Luttinger liquids },  \href{https://doi.org/10.1103/PhysRevB.84.195128}{Phys.
Rev. B 84, 195128  (2011)}, [\href{https://arxiv.org/abs/1104.2544}{{\ttfamily arXiv:1104.2544}}].

\bibitem{Alcaraz2013}
F. C. Alcaraz and M. A. Rajabpour, \textit{Universal Behavior of the Shannon Mutual Information of Critical Quantum Chains}, \href{https://journals.aps.org/prl/abstract/10.1103/PhysRevLett.111.017201}{Phys. Rev. Lett. 111, 017201 (2013)}, [\href{https://arxiv.org/abs/1305.1239}{{\ttfamily arXiv:1305.1239}}].

\bibitem{StephanPRB2014}
J.-M. St\'ephan,
\newblock \textit{Shannon and R\'enyi mutual information in quantum critical spin chains},
\href{https://doi.org/10.1103/PhysRevB.90.045424}{Phys. Rev. B \textbf{90}, 045424 (2014)},
[\href{https://doi.org/10.48550/arXiv.1403.6157}{{\ttfamily arXiv:1403.6157}}].

\bibitem{Alcaraz2014}
F. C. Alcaraz, M. A. Rajabpour,
\newblock \textit{Universal behavior of the Shannon and R{\'e}nyi mutual information of quantum critical chains},
\href{https://journals.aps.org/prb/abstract/10.1103/PhysRevB.90.075132}{Phys. Rev. B \textbf{90}, 075132 (2014)},
[\href{https://arxiv.org/abs/1405.1074}{{\ttfamily arXiv:1405.1074}}].

\bibitem{LuitzAletLaflorencie2014}
D. J. Luitz, F. Alet, N. Laflorencie,
\newblock \textit{Universal Behavior beyond Multifractality in Quantum Many-Body Systems},
\href{https://doi.org/10.1103/PhysRevLett.112.057203}{Phys. Rev. Lett. \textbf{112}, 057203 (2014)},
[\href{https://arxiv.org/abs/1308.1916}{{\ttfamily arXiv:1308.1916}}].

\bibitem{Tarighi2022} B. Tarighi, R. Khasseh, M. N. Najafi, M. A. Rajabpour, \textit{Universal logarithmic correction to Rényi (Shannon) entropy in generic systems of critical quadratic fermions}, \href{https://journals.aps.org/prb/abstract/10.1103/PhysRevB.105.245109}{Phys. Rev. B \textbf{105}, 245109 (2022)}, [\href{https://arxiv.org/abs/2203.13124}{{\ttfamily arXiv:2203.13124}}].

\bibitem{central-charge}
N. U. K\"oyl\"uo\u{g}lu, S. Majumder, M. Amico, S. Mostame, E. van den Berg, M. A. Rajabpour, Z. Minev, and K. Najafi,
\textit{Measuring central charge on a universal quantum processor},
\href{https://doi.org/10.1038/s41467-025-66775-9}{Nat. Commun. \textbf{17}, 305 (2026)},
[\href{https://doi.org/10.48550/arXiv.2408.06342}{{\ttfamily arXiv:2408.06342}}].

\bibitem{Leone2022a}
S. F. E. Oliviero, L. Leone, A. Hamma
\newblock \textit{Stabilizer Rényi entropy},
\href{https://doi.org/10.1103/PhysRevLett.128.050402}{Phys. Rev. Lett. \textbf{128}, 050402 (2022)},
[\href{https://arxiv.org/abs/2106.12587}{{\ttfamily arXiv:2106.12587}}].

\bibitem{Leone2022b}
L. Leone, S. F. E. Oliviero, A.Hamma
\newblock \textit{Magic-state resource theory for the ground state of the transverse-field Ising model},
\href{https://doi.org/10.1103/PhysRevA.106.042426}{Phys. Rev. A \textbf{106}, 042426 (2022)},
[\href{https://arxiv.org/abs/2205.02247}{{\ttfamily arXiv:2205.02247}}].

\bibitem{TarabungaDalmontePRXQ2023}
P. S. Tarabunga, E. Tirrito, T. Chanda, M. Dalmonte,
\newblock \textit{Many-Body Magic Via Pauli-Markov Chains—From Criticality to Gauge Theories},
\href{https://journals.aps.org/prxquantum/abstract/10.1103/PRXQuantum.4.040317}{PRX Quantum \textbf{4}, 040317 (2023)},
[\href{https://arxiv.org/abs/2305.18541}{{\ttfamily arXiv:2305.18541}}].

\bibitem{Guglielmo2023}
G. Lami and M. Collura,
\newblock \textit{Nonstabilizerness via Perfect Pauli Sampling of Matrix Product States},
\href{https://doi.org/10.1103/PhysRevLett.131.180401}{Phys. Rev. Lett.  \textbf{131}, 180401 (2023)},
[\href{https://arxiv.org/abs/2303.05536}{{\ttfamily arXiv:2303.05536
}}].

\bibitem{Leone2024}
L. Leone, L. Bittel,
\newblock \textit{Stabilizer entropies are monotones for magic-state resource theory},
\href{https://doi.org/10.1103/PhysRevA.110.L040403}{Phys. Rev. A \textbf{110}, L040403 (2024)},
[\href{https://arxiv.org/abs/2404.11652}{{\ttfamily arXiv:2404.11652}}].

\bibitem{FuxTirritoDalmonteFazioPRR2024}
G. E. Fux, E. Tirrito, M. Dalmonte, R. Fazio,
\newblock \textit{Entanglement–nonstabilizerness separation in hybrid quantum circuits},
\href{https://journals.aps.org/prresearch/abstract/10.1103/PhysRevResearch.6.L042030}{Phys. Rev. Research \textbf{6}, L042030 (2024)},
[\href{https://arxiv.org/abs/2312.02039}{{\ttfamily arXiv:2312.02039}}].

\bibitem{Collura2024} M. Collura, J. De Nardis, V. Alba, G. Lami,
\newblock \textit{The quantum magic of fermionic Gaussian states},
\href{https://doi.org/10.22331/q-2026-03-23-2036}{Quantum 10, 2036 (2026)},
[\href{https://arxiv.org/abs/2412.05367}{{\ttfamily arXiv:2412.05367}}].

\bibitem{HaugLeeKimPRL2024}
T. Haug, S. Lee, M. S. Kim,
\newblock \textit{Efficient quantum algorithms for stabilizer entropies},
\href{https://doi.org/10.1103/PhysRevLett.132.240602}{Phys. Rev. Lett. \textbf{132}, 240602 (2024)},
[\href{https://arxiv.org/abs/2305.19152}{{\ttfamily arXiv:2305.19152}}].

\bibitem{DingPRXQuantum2025}
Y.-M. Ding,  Z. Wang, Z. Yan,
\newblock \textit{Evaluating many-body stabilizer R\'enyi entropy by sampling reduced Pauli strings: Singularities, volume law, and nonlocal magic},
\href{https://journals.aps.org/prxquantum/abstract/10.1103/pyzr-jmvw}{PRX Quantum \textbf{6}, 030328 (2025)},
[\href{https://arxiv.org/abs/2501.12146}{{\ttfamily arXiv:2501.12146}}].

\bibitem{ViscardiDalmonteHammaTirrito2025}
M. Viscardi, M. Dalmonte, A. Hamma, E. Tirrito,
\newblock \textit{Interplay of entanglement structures and stabilizer entropy in spin models}, \href{https://scipost.org/10.21468/SciPostPhysCore.9.1.012}{SciPost Phys. Core \textbf{9}, 012 (2026)},[\href{https://arxiv.org/abs/2503.08620}{{\ttfamily arXiv:2503.08620}}].

\bibitem{Fan2025}
C. Fan, X. Qian, H.-C. Zhang, R.-Z. Huang, M. Qin, and T. Xiang,
\textit{Disentangling critical quantum spin chains with Clifford circuits},
\href{https://doi.org/10.1103/PhysRevB.111.085121}{Phys. Rev. B \textbf{111}, 085121 (2025)},
[\href{https://arxiv.org/abs/2411.12683}{{\ttfamily arXiv:2411.12683}}].

\bibitem{Frau2025}
M. Frau, P. S. Tarabunga, M. Collura, E. Tirrito, and M. Dalmonte,
\textit{Stabilizer disentangling of conformal field theories},
\href{https://doi.org/10.21468/SciPostPhys.18.5.165}{SciPost Phys. \textbf{18}, 165 (2025)},
[\href{https://arxiv.org/abs/2411.11720}{{\ttfamily arXiv:2411.11720}}].

\bibitem{Hoshino2025}
M. Hoshino, M. Oshikawa, Y. Ashida,
\newblock \textit{Stabilizer R\'enyi Entropy and Conformal Field Theory},
\href{https://doi.org/10.1103/ylsz-dm3y}{Phys. Rev. X \textbf{16}, 011037 (2026)},
[\href{https://arxiv.org/abs/2503.13599}{{\ttfamily arXiv:2503.13599}}].

\bibitem{Hoshino2025b}
M. Hoshino, Y. Ashida,
\newblock \textit{Stabilizer Rényi Entropy Encodes Fusion Rules of Topological Defects and Boundaries},
\href{https://doi.org/10.1103/1tyr-rlbb}{Phys. Rev. Lett. \textbf{136}, 080402 (2026)},
[\href{https://arxiv.org/abs/2507.10656}{{\ttfamily arXiv:2507.10656}}].

\bibitem{IslamNature2015}
R. Islam \textit{et al.},
\newblock \textit{Measuring entanglement entropy in a quantum many-body system},
\href{https://doi.org/10.1038/nature15750}{Nature \textbf{528}, 77 (2015)}.

\bibitem{ElbenPRL2018}
A. Elben, B. Vermersch, M. Dalmonte, J. I. Cirac, P. Zoller,
\newblock \textit{R\'enyi entropies from random quenches in atomic Hubbard and spin models},
\href{https://doi.org/10.1103/PhysRevLett.120.050406}{Phys. Rev. Lett. \textbf{120}, 050406 (2018)},
[\href{
https://doi.org/10.48550/arXiv.1709.05060
}{{\ttfamily arXiv:1709.05060}}].

\bibitem{BrydgesScience2019}
T. Brydges \textit{et al.},
\newblock \textit{Probing R\'enyi entanglement entropy via randomized measurements},
\href{https://doi.org/10.1126/science.aau4963}{Science \textbf{364}, 260 (2019)},
[\href{https://arxiv.org/abs/1806.05747}{{\ttfamily arXiv:1806.05747}}].

\bibitem{ElbenNRP2022}
A. Elben, R. Kueng, H.-Y. Huang, R. van Bijnen, C. Kokail, M. Dalmonte, P. Calabrese, B. Kraus, J. Preskill, P. Zoller, B. Vermersch,
\newblock \textit{The randomized measurement toolbox},
\href{https://doi.org/10.1038/s42254-022-00535-2}{Nat. Rev. Phys. \textbf{5}, 9–24 (2023)},
[\href{https://arxiv.org/abs/2203.11374}{{\ttfamily arXiv:2203.11374}}].

\bibitem{OlivieroNPJQI2022}
S. F. E. Oliviero, L. Leone, A. Hamma, S. Lloyd,
\newblock \textit{Measuring magic on a quantum processor},
\href{
https://doi.org/10.1038/s41534-022-00666-5}{npj Quantum Inf. \textbf{8}, 148 (2022)},
[\href{https://arxiv.org/abs/2204.00015}{{\ttfamily arXiv:2204.00015}}].

\bibitem{HowardCampbell2017}
M. Howard, E. Campbell,
\newblock \textit{Application of a resource theory for magic states to fault-tolerant quantum computing},
\href{https://doi.org/10.1103/PhysRevLett.118.090501}{Phys. Rev. Lett. \textbf{118}, 090501 (2017)},
[\href{https://arxiv.org/abs/1609.07488}{{\ttfamily arXiv:1609.07488}}].

\bibitem{Veitch2014}
V. Veitch, S. H. Mousavian, D. Gottesman, J. Emerson,
\newblock \textit{The resource theory of stabilizer quantum computation},
\href{https://doi.org/10.1088/1367-2630/16/1/013009}{New J. Phys. \textbf{16}, 013009 (2014)},
[\href{https://arxiv.org/abs/1307.7171}{{\ttfamily arXiv:1307.7171}}].

\bibitem{WangWildeSu2020}
X. Wang, M. M. Wilde, Y. Su,
\newblock \textit{Quantifying the magic of quantum channels},
\href{
https://doi.org/10.1088/1367-2630/ab451d}{New J. Phys. \textbf{21}, 103002 (2019)},
[\href{
https://doi.org/10.48550/arXiv.1903.04483}{{\ttfamily arXiv:1903.04483}}].

\bibitem{SeddonPRXQ2021}
J. R. Seddon, E. Campbell,
\newblock \textit{Quantifying Magic for Multi-Qubit Operations},
\href{https://doi.org/10.1098/rspa.2019.0251}{Proc. A \textbf{475} (2227): 20190251 (2019)},
[\href{
https://doi.org/10.48550/arXiv.1901.03322}{{\ttfamily arXiv:1901.03322}}].

\bibitem{Haug2023}
T. Haug and L. Piroli,
\newblock \textit{Quantifying nonstabilizerness of matrix product states},
\href{https://doi.org/10.1103/PhysRevB.107.035148}
{Phys. Rev. B \textbf{107}, 035148 (2023)},
[\href{https://arxiv.org/abs/2207.13076}
{{\ttfamily arXiv:2207.13076}}].

\bibitem{Hallam2026}
A. Hallam, R. Smith, and Z. Papi\'c,
\newblock \textit{Spectral signatures of nonstabilizerness and criticality in infinite matrix product states},
\href{https://doi.org/10.1103/77wh-qv29}
{Phys. Rev. B \textbf{113}, 245113 (2026)},
[\href{https://arxiv.org/abs/2602.15116}
{{\ttfamily arXiv:2602.15116}}].

\bibitem{LIEB1961407}
E.~Lieb, T.~Schultz and D.~Mattis,
\textit{{Two soluble models of an antiferromagnetic chain}},
\href{https://doi.org/10.1016/0003-4916(61)90115-4}{Annals of Physics \textbf{16}(3), 407 (1961)}.

\bibitem{Peschel2003}
I. Peschel,
\newblock \textit{Calculation of reduced density matrices from correlation functions},
\href{https://doi.org/10.1088/0305-4470/36/14/101}{J. Phys. A \textbf{36}, L205 (2003)},
[\href{https://arxiv.org/abs/cond-mat/0212631}{{\ttfamily arXiv:cond-mat/0212631}}].

\bibitem{Kitaev2003}G. Vidal, J. I. Latorre, E. Rico, and A. Kitaev
\textit{{Entanglement in Quantum Critical Phenomena}}, \href{https://doi.org/10.1103/PhysRevLett.90.227902}{Phys. Rev. Lett. \textbf{90}, 227902 (2003)}, [\href{https://arxiv.org/abs/quant-ph/0211074
}{{\ttfamily arXiv:quant-ph/0211074
}}].

\bibitem{JinKorepin2004}
B.-Q. Jin, V. E. Korepin,
\newblock \textit{Quantum Spin Chain, Toeplitz Determinants and the Fisher–Hartwig Conjecture},
\href{https://doi.org/10.1023/B:JOSS.0000037230.37166.42}{J. Stat. Phys. \textbf{116}, 79 (2004)}.

\bibitem{Its:2008}
   A. R. Its, F. Mezzadri and M. Y. Mo,
   \textit{{Entanglement Entropy in Quantum Spin Chains with Finite Range
Interaction}}
  , \href{https://doi.org/10.1007/s00220-008-0566-6}{ Communications in Mathematical Physics {\bfseries 284}, 117 (2008)},
  [\href{https://arxiv.org/abs/0708.0161}{{\ttfamily
  arXiv:0708.0161}}].

\bibitem{PeschelEisler2009}
I. Peschel, V. Eisler,
\newblock \textit{Reduced density matrices and entanglement entropy in free lattice models},
\href{https://doi.org/10.1088/1751-8113/42/50/504003}{J. Phys. A \textbf{42}, 504003 (2009)},
[\href{https://arxiv.org/abs/0906.1663}{{\ttfamily arXiv:0906.1663}}].

\bibitem{IvanovAbanovCheianov2013}
D. A. Ivanov, A. G. Abanov, V. V. Cheianov,
\newblock \textit{Counting free fermions on a line: Toeplitz determinants and Fisher–Hartwig asymptotics},
\href{https://doi.org/10.1088/1751-8113/46/8/085003}{J. Phys. A \textbf{46}, 085003 (2013)},
[\href{
https://doi.org/10.48550/arXiv.1112.2530
}{{\ttfamily arXiv:1112.2530}}].

\bibitem{GrohaEsslerCalabrese2018}
S. Groha, F. H. L. Essler, P. Calabrese,
\newblock \textit{Full counting statistics in the transverse field Ising chain},
\href{https://scipost.org/10.21468/SciPostPhys.4.6.043}{SciPost Phys. \textbf{4}, 043 (2018)},
[\href{
https://doi.org/10.48550/arXiv.1803.09755
}{{\ttfamily arXiv:1803.09755}}].

\bibitem{Verresen:2018}
  R. Verresen, N. G. Jones, and F. Pollmann,
  \textit{{Topology and edge modes in quantum critical chains}}
  , \href{https://doi.org/10.1103/PhysRevLett.120.057001}{ Phys. Rev.
Lett. {\bfseries 120}, 057001 (2018)},
  [\href{https://arxiv.org/abs/1709.03508}{{\ttfamily
  arXiv:1709.03508}}].

\bibitem{Verresen:2019}
  N. G. Jones, R. Verresen,
  \textit{{Asymptotic correlations in gapped and critical topological phases of 1D quantum systems}}
  , \href{https://doi.org/10.1007/s10955-019-02257-9}{ J. Stat. Phys. \textbf{175}, 1164--1213 (2019)},
  [\href{https://arxiv.org/abs/1805.06904}{{\ttfamily
  arXiv:1805.06904}}].

\bibitem{Ares2021}
F. Ares, M. A. Rajabpour, J. Viti,
\newblock \textit{Exact full counting statistics for the staggered magnetization and the domain walls in the XY spin chain},
\href{https://doi.org/10.1103/PhysRevE.103.042107}{Phys. Rev. E \textbf{103}, 042107 (2021)},
[\href{https://arxiv.org/abs/2012.14012}{{\ttfamily arXiv:2012.14012
}}].

\bibitem{Surace2022}  J. Surace, and L. Tagliacozzo, \textit{Fermionic Gaussian states: an introduction to numerical approaches}, \href{
https://doi.org/10.21468/SciPostPhysLectNotes.54}{ SciPost Phys. Lect. Notes \textbf{54} (2022)}, [\href{https://doi.org/10.48550/arXiv.2111.08343}{{\ttfamily arXiv: 2111.08343
}}].

\bibitem{TKR2024}
B. Tarighi, R. Khasseh, and M. A. Rajabpour, \textit{{Efficient Representation of Gaussian Fermionic Pure States in Non-Computational
Bases}},
\href{https://doi.org/10.1103/PhysRevA.109.062214}{Phys. Rev. A \textbf{109}, 062214 (2024)},
[\href{https://arxiv.org/abs/2403.03289}{{\ttfamily arXiv:2403.03289}}].
%

\bibitem{Rajabpour2025a}
M. A. Rajabpour, M. A. Seifi Mirjafarlou, R. Khasseh, \textit{{Explicit Pfaffian Formula for Amplitudes of Fermionic Gaussian Pure States in Arbitrary Pauli Bases}}, \href{https://doi.org/10.1103/PhysRevB.111.235102}{Phys. Rev. B. \textbf{111},  235102 (2025)},
[\href{https://doi.org/10.48550/arXiv.2502.04857}{{\ttfamily arXiv:2502.04857}}].

\bibitem{Supplement}
See Supplemental Material for: Sec.~I, the proof of Theorem~1 and a short derivation of Eq.~(10); Sec.~II, the permutation freedom and invariance of the \(\mathbf R \leftrightarrow \mathbf G\) mapping and recall the Pfaffian relation between $\bold{R}$ and $\bold{R}'$; Sec.~III, the Gaussian doubling map, its physical interpretation, and an explicit \(L=2\) example; Sec.~IV, the proof of Theorem~2 and explicit relations between \(\mathbf G^{\mathrm{TFI}}\) and \(\mathbf R^{\mathrm{XX}}\); Sec.~V, the formulation of general quadratic fermionic chains, their representation in terms of the complex symbol \(f(z)\), and the construction of the corresponding correlation matrix \(\mathbf G^{(f)}\); Sec.~VI, the proof of Theorem~3: the block reduction for \(f(z)=z^n+1\), and the block reduction for \(f(z)=z^m+z^{-m}\); In Sec.~VII we derive eq.~\eqref{eq:SRE_unified}; Sec.~VIII, the derivation of the large-\(L\) asymptotics for \(\alpha=\frac12,2,4\), together with their comparison to the CFT-limit formulas obtained through the exact SR--SRE correspondence; and Sec.~IX, the extension to Gaussian Gibbs states for stabilizer--Shannon R\'enyi correspondence, and additional Refs.~\cite{Heyraud2025, DLMF, Mehta2004,BottcherSilbermann2006, DeiftItsKrasovsky2011}.

\bibitem{Heyraud2025}
V.~Heyraud, H.~Chomet, and J.~Tilly,
\textit{Unified framework for matchgate classical shadows},
\href{https://doi.org/10.1038/s41534-025-01015-y}{npj Quantum Inf. \textbf{11}, 65 (2025)},
[\href{https://doi.org/10.48550/arXiv.2409.03836}{{\ttfamily arXiv:2409.03836}}].

\bibitem{DLMF}
NIST Digital Library of Mathematical Functions,
\textit{§2.10 Asymptotic Approximations of Sums and Sequences; §5.11 Asymptotic Expansions of the Gamma Function},
Release 1.2.4 (2025),
edited by F. W. J. Olver, A. B. Olde Daalhuis, D. W. Lozier, et al.,
\href{https://dlmf.nist.gov}{dlmf.nist.gov}.

\bibitem{Mehta2004}
M. L. Mehta,
\textit{Random Matrices},
3rd ed. (Elsevier, Amsterdam, 2004),
[\href{https://shop.elsevier.com/books/random-matrices/lal-mehta/978-0-12-088409-4}{Elsevier}].

\bibitem{BottcherSilbermann2006}
A. B\"ottcher and B. Silbermann,
\textit{Analysis of Toeplitz Operators},
2nd ed. , \href{https://doi.org/10.1007/3-540-32436-4}{(Springer, Berlin, 2006)}.

\bibitem{DeiftItsKrasovsky2011}
P. Deift, A. Its, and I. Krasovsky,
\textit{Asymptotics of Toeplitz, Hankel, and Toeplitz+Hankel determinants with Fisher--Hartwig singularities},
\href{https://doi.org/10.4007/annals.2011.174.2.12}{Ann. Math. \textbf{174}, 1243 (2011)},
[\href{https://arxiv.org/abs/0905.0443}{{\ttfamily arXiv:0905.0443}}].
%

\bibitem{pfaffinho}
The \textit{Pfaffinho} is a Pfaffian of a submatrix from a skew-symmetric matrix.

\bibitem{TerhalDiVincenzo2002}
B.~M.~Terhal and D.~P.~DiVincenzo,
\textit{{Classical simulation of noninteracting-fermion quantum circuits,}}
\href{https://doi.org/10.1103/PhysRevA.65.032325}{{Phys.\ Rev.\ A \textbf{65}, 032325 (2002)}}, [\href{https://doi.org/10.48550/arXiv.quant-ph/0108010}{{\ttfamily 	arXiv:quant-ph/0108010}}].

\bibitem{JozsaMiyake2008}
R.~Jozsa and A.~Miyake,
\textit{{Matchgates and classical simulation of quantum circuits,}}
\href{https://doi.org/10.1098/rspa.2008.0189}{{Proc.\ R.\ Soc.\ A \textbf{464}, 3089--3106 (2008)}}, [\href{https://doi.org/10.48550/arXiv.0804.4050}{{\ttfamily arXiv:0804.4050}}].

\bibitem{Langer2026}
M.~Langer, R.~Morral-Yepes, A.~Gammon-Smith, F.~Pollmann, and B.~Kraus,
\textit{{Matchgate circuit representation of fermionic Gaussian states: optimal preparation, approximation, and classical simulation,}}
[\href{https://doi.org/10.48550/arXiv.2603.05675}{{\ttfamily arXiv:2603.05675}}].

\bibitem{Lieb2016}
A. Giuliani, I. Jauslin,  E. H. Lieb,  \textit{{A Pfaffian Formula for Monomer–Dimer Partition Functions}}, \href{https://doi.org/10.1007/s10955-016-1484-1}{J Stat Phys 163, 211–238 (2016)},

\bibitem{Lieb1968}
E. H. Lieb, \textit{{A theorem on Pfaffians}}, \href{https://doi.org/10.1016/S0021-9800(68)80078-X}{J. Comb. Theory 5, 313–319 (1968)}.

\bibitem{chiral}
Here, “chiral reduction” refers to an algebraic block decomposition and does not imply physical chirality.

\bibitem{commentHoshino}

Refs.\cite{Hoshino2025,Hoshino2025b} develop a BCFT framework for stabilizer R\'enyi entropy (SRE) and present formulas that are described as holding for arbitrary R\'enyi index~$\alpha$. Using the exact SRE\(\leftrightarrow\)Shannon--R\'enyi (SR) mapping established here  the SR entropy is known to exhibit a boundary phase transition at
\(\alpha_c=4\) (free-fermion point), with replica-based expressions valid only for \(\alpha\leq\alpha_c\) and different universal behavior for \(\alpha>\alpha_c\)~\cite{Stephan2010}.

\bibitem{NajafiRamezanpourRajabpour2025}
M. Nattagh Najafi, A. Ramezanpour, and M. A. Rajabpour,
\textit{A field theory representation of sum of powers of principal minors and physical applications},
\href{https://doi.org/10.21468/SciPostPhysCore.8.3.051}{SciPost Phys. Core \textbf{8}, 051 (2025)},
[\href{https://arxiv.org/abs/2403.09874}{{\ttfamily arXiv:2403.09874}}].
\end{thebibliography}

\begin{thebibliography}{99}

\bibitem{TKR20241}
B. Tarighi, R. Khasseh, and M. A. Rajabpour, \textit{{Efficient Representation of Gaussian Fermionic Pure States in Non-Computational
Bases}},
\href{https://doi.org/10.1103/PhysRevA.109.062214}{Phys. Rev. A. \textbf{109}, 062214 (2024)},
[\href{https://arxiv.org/abs/2403.03289}{{\ttfamily arXiv:2403.03289}}].

\bibitem{Leone2022b1}
L. Leone, S. F. E. Oliviero, A.Hamma
\newblock \textit{Magic-state resource theory for the ground state of the transverse-field Ising model},
\href{https://doi.org/10.1103/PhysRevA.106.042426}{Phys. Rev. A 106, \textbf{106}, 042426 (2022)},
[\href{https://arxiv.org/abs/2205.02247}{{\ttfamily arXiv:2205.02247 }}].

\bibitem{Rajabpour2025a1}
M. A. Rajabpour, M. A. Seifi Mirjafarlou, R. Khasseh, \textit{{Explicit Pfaffian Formula for Amplitudes of Fermionic Gaussian Pure States in Arbitrary Pauli Bases}}, \href{https://doi.org/10.1103/PhysRevB.111.235102}{Phys. Rev. B. \textbf{111},  235102 (2025)},
[\href{https://doi.org/10.48550/arXiv.2502.04857}{{\ttfamily arXiv:2502.04857}}].


\bibitem{TerhalDiVincenzo20021}
B.~M.~Terhal and D.~P.~DiVincenzo,
\textit{{Classical simulation of noninteracting-fermion quantum circuits,}}
\href{https://doi.org/10.1103/PhysRevA.65.032325}{Phys.\ Rev.\ A \textbf{65}, 032325 (2002)},[\href{https://doi.org/10.48550/arXiv.quant-ph/0108010}{{\ttfamily arXiv:quant-ph/0108010}}].


\bibitem{JozsaMiyake20081}
R.~Jozsa and A.~Miyake,
\textit{{Matchgates and classical simulation of quantum circuits,}}\href{https://doi.org/10.1098/rspa.2008.0189}{Proc. R. Soc. A \textbf{464}, 3089-3106 (2008)},[\href{https://doi.org/10.48550/arXiv.0804.4050}{{\ttfamily arXiv: 0804.4050}}].

\bibitem{Heyraud20251}
V.~Heyraud, H.~Chomet, J.~Tilly,
\textit{{Unified framework for matchgate classical shadows,}} [\href{https://doi.org/10.48550/arXiv.2409.03836}{\ttfamily arXiv: 2409.03836}]

\bibitem{Langer20261}
M.~Langer, R.~Morral-Yepes, A.~Gammon-Smith, F.~Pollmann, and B.~Kraus,
\textit{{Matchgate circuit representation of fermionic Gaussian states: optimal preparation, approximation, and classical simulation,}}
[\href{https://doi.org/10.48550/arXiv.2603.05675}{{\ttfamily arXiv:2603.05675}}].

\bibitem{Stephan20101} J-M. St\'ephan, G. Misguich, and V. Pasquier, \textit{Phase transition in the Rényi-Shannon entropy of Luttinger liquids },  \href{https://doi.org/10.1103/PhysRevB.84.195128}{Phys.
Rev. B 84, 195128  (2011)}, [\href{https://arxiv.org/abs/1104.2544}{{\ttfamily arXiv:1104.2544}}].


\bibitem{DLMF1}
NIST Digital Library of Mathematical Functions,
\textit{§2.10 Asymptotic Approximations of Sums and Sequences; §5.11 Asymptotic Expansions of the Gamma Function},
Release 1.2.4 (2025),
edited by F. W. J. Olver, A. B. Olde Daalhuis, D. W. Lozier, et al.,
\href{https://dlmf.nist.gov}{dlmf.nist.gov}.

\bibitem{Mehta20041}
M. L. Mehta,
\textit{Random Matrices},
3rd ed. (Elsevier, Amsterdam, 2004),
[\href{https://shop.elsevier.com/books/random-matrices/lal-mehta/978-0-12-088409-4}{Elsevier}].

\bibitem{BottcherSilbermann20061}
A. B\"ottcher and B. Silbermann,
\textit{Analysis of Toeplitz Operators},
2nd ed. , \href{https://doi.org/10.1007/3-540-32436-4}{(Springer, Berlin, 2006)}.

\bibitem{DeiftItsKrasovsky20111}
P. Deift, A. Its, and I. Krasovsky,
\textit{Asymptotics of Toeplitz, Hankel, and Toeplitz+Hankel determinants with Fisher--Hartwig singularities},
\href{https://doi.org/10.4007/annals.2011.174.2.12}{Ann. Math. \textbf{174}, 1243 (2011)},
[\href{https://arxiv.org/abs/0905.0443}{{\ttfamily arXiv:0905.0443}}].


\bibitem{Leone20241}
L. Leone, L. Bittel,
\textit{{Stabilizer entropies are monotones for magic-state resource theory}},
\href{https://doi.org/10.1103/PhysRevA.110.L040403}{Phys. Rev. A \textbf{110}, L040403 (2024)}.
\end{thebibliography}
\end{document}